\documentclass[
	fontsize=12pt,         
	numbers=noenddot,    	
    parindent=half,        
    listof=totoc,        	
    bibliography=totoc,  	
	headsepline=true,       
	footsepline=false, 		
    DIV=12                	
]{scrartcl}
\addtokomafont{disposition}{\boldmath}

\usepackage[utf8]{inputenc}
\usepackage[T1]{fontenc}
\usepackage[english]{babel}

\usepackage{graphicx}
\usepackage{subfigure}
\usepackage{color}

\usepackage{amsmath, amsfonts, amssymb, bbm, bm, mathabx}

\usepackage[citestyle = authoryear,
bibstyle = authoryear, dashed = false, giveninits = true,
backend=biber, maxnames = 2, maxbibnames = 10, uniquename=false, uniquelist=false]{biblatex} 
\usepackage[defaultlines = 2, all]{nowidow}

\usepackage[a4paper, left = 2.5cm, right = 2.5cm, top = 2.5cm, bottom = 2.5cm]{geometry}

\usepackage{booktabs} 
\usepackage{longtable}
\usepackage{tabularx}

\usepackage[hidelinks]{hyperref} 

\usepackage{lscape}
\usepackage{nowidow}

\usepackage{bbm, mathabx}

\usepackage{rotating}
\usepackage{placeins}
\usepackage{fontawesome5}
\usepackage{float}

\newcommand{\dif}{\mathop{}\!\mathrm{d}}
\DeclareMathOperator{\E}{\mathbb{E}}
\DeclareMathOperator{\Prob}{\mathbb{P}}

\AtEveryBibitem{\clearfield{month}}
\AtEveryCitekey{\clearfield{month}}
\AtEveryBibitem{\clearlist{language}}

\renewbibmacro*{volume+number+eid}{%
	\printfield{volume}%
	\setunit*{\addnbthinspace}
	\printfield{number}%
	\setunit{\addcomma\space}%
	\printfield{eid}}
\DeclareFieldFormat[article]{number}{\mkbibparens{#1}}
\DeclareFieldFormat[article,periodical]{volume}{\mkbibbold{#1}}

\title{Methods for Quantifying Dataset Similarity: a Review, Taxonomy and Comparison}
\author{Marieke Stolte\thanks{Corresponding author, e-mail: \texttt{stolte@statistik.tu-dortmund.de}}, Franziska Kappenberg, \\
	Andrea Bommert, and Jörg Rahnenführer}
\date{Department of Statistics,
	TU Dortmund University}

\begin{document}
\definecolor{timescolor}{rgb}{1,0.5,0}%
\definecolor{condcheckcolor}{rgb}{0.3372549, 0.7058824, 0.9137255}
\definecolor{checkcolor}{rgb}{0, 0.6196078, 0.4509804}

\newcommand{\mycheck}{\textcolor{checkcolor}{\faCheck}}
\newcommand{\mytimes}{\textcolor{timescolor}{\faTimes}}
\newcommand{\condcheck}{\textcolor{condcheckcolor}{(\faCheck)}}

\maketitle
\section*{Abstract}	
Quantifying the similarity between datasets has widespread applications in statistics and machine learning. The performance of a predictive model on novel datasets, referred to as generalizability, depends on how similar the training and evaluation datasets are. Exploiting or transferring insights between similar datasets is a key aspect of meta-learning and transfer-learning. In simulation studies, the similarity between distributions of simulated datasets and real datasets, for which the performance of methods is assessed, is crucial. In two- or $k$-sample testing, it is checked, whether the underlying distributions of two or more datasets coincide.

Extremely many approaches for quantifying dataset similarity have been proposed in the literature. We examine more than 100 methods and provide a taxonomy, classifying them into ten classes. In an extensive review of these methods the main underlying ideas, formal definitions, and important properties are introduced. 

We compare the 118 methods in terms of their applicability, interpretability, and theoretical properties, in order to provide recommendations for selecting an appropriate dataset similarity measure based on the specific goal of the dataset comparison and on the properties of the datasets at hand. An online tool facilitates the choice of the appropriate dataset similarity measure.

\tableofcontents

\section{Introduction}
Quantifying how similar or different two or more datasets are is a crucial subtask in various applications of statistics and machine learning. Examples of applications include but are not limited to (i) the assessment of the generalizability of a predictive model to a broader context, (ii) the transfer of knowledge from one task to another task in transfer-learning or meta-learning, (iii) the comparison of distributions of simulated data and data from the true data generating process when planning and implementing simulation studies, and (iv) checking whether the underlying distributions of two or more datasets coincide via two- or $k$-sample testing.

In statistics and machine learning, generalizability is a measure of a model's performance for a broader context, compared to the data on which it was fitted. Generalizability is a useful property since conclusions drawn from the present study can be transferred to a more general set of study objects. The performance of the model on a new or unseen dataset depends on the similarity between the dataset that was used for fitting the model and the new dataset. Quantifying this similarity with a univariate measure thus can help to assess whether generalizability is given without fitting the model on the new dataset.

In meta-learning and transfer-learning, a central component is to exploit or transfer insights between different datasets. For example, some meta-learning models try to find the most suitable datasets to train specific models. Likewise, in transfer-learning, a common approach is to pre-train a model on a large (source) dataset and then fine-tune the model on the (target) dataset of interest. Also in this situation, it is crucial to understand and measure similarities between datasets in order to select appropriate source datasets for the first step of the process.

Further, when transferring insights of simulation studies to given data, the similarity between distributions of simulated datasets and the distribution of a (target) dataset, for which the performance of methods is assessed, is critical. If assumptions are made about the underlying distribution, such as that it is a normal distribution, and if these assumptions are not met, the conclusions from the simulation study for the target dataset may be fundamentally flawed.

In the statistical learning and machine learning literature, a vast number of approaches for quantifying dataset similarity have been proposed. However, to the best of our know\-ledge, there is no comprehensive comparison between many of these different approaches. In the following, we refer to some publications in which at least comparisons of subsets of the methods have been carried out.

In \textcite{thas_comparing_2010}, many methods for comparing distributions for univariate data, including graphical methods as well as hypothesis tests, are explained and discussed, but the multivariate case is not covered.
In \textcite{rachev_probability_1991} and in \textcite{liese_convex_1987}, properties of several probability metrics and divergences, respectively, are discussed, i.e.\ distance measures between probability distributions.

In general, many articles that present new methods for measuring dataset similarity include brief summaries of competing methods \parencite[e.g.][]{rosenbaum_exact_2005, biswas_nonparametric_2014, chen_new_2017, sarkar_graph-based_2020,  deb_multivariate_2021, kim_classification_2021, li_measuring_2022, huang_kernel_2022}.
Most of these include only a small number of competing methods and, in most cases, only methods based on the same principle.
Further, some articles provide comprehensive reviews of single methodological classes (e.g.\ \cite{muandet_kernel_2017} for kernel mean embeddings or \cite{szekely_energy_2017} for the energy distance).

Simulation studies comparing the new method with some previous methods are often presented additionally \parencite[e.g.][]{biswas_nonparametric_2014, chwialkowski_fast_2015, mondal_high_2015, jitkrittum_interpretable_2016, petrie_graph-theoretic_2016, chen_new_2017, lopez-paz_revisiting_2017, liu_classifier_2018, liu_learning_2020, sarkar_graph-based_2020}. 

We do not know of any comparison of methods belonging to many of the different approaches, in particular methods based on different principles. In this paper, we give an extensive review providing characterization and classification of dataset similarity methods and their properties.
In total, we examine more than 100 methods for quantifying dataset similarity and provide a taxonomy dividing the methods into ten classes, based on the underlying principles we identified. The methods were selected from an extensive literature search by using the following criteria: 
\begin{itemize}
	\item \emph{The method is applicable for multivariate data}. This excludes the vast literature on methods for comparing univariate distributions. For example, a comprehensive overview of methods for one-dimensional data can be found in \textcite{thas_comparing_2010}.
	\item \emph{The method requires no specific parametric or distributional assumptions on the underlying distributions of the datasets} (e.g.\ normal distribution). The general assumptions of discrete or continuous data are allowed since they can be easily verified in practice.
	\item \emph{The method does not focus on a particular property of the data} (e.g.\ means), but on the entire dataset or its entire distribution. This particularly excludes tests based solely on location or scale differences.
\end{itemize}
The classes into which the methods are divided are (i) comparison of cumulative distribution functions, density functions, or characteristic functions; (ii) methods based on multivariate ranks; (iii) discrepancy measures for distributions; (iv) graph-based methods; (v) methods based on inter-point distances; (vi) kernel-based methods; (vii) methods based on binary classification; (viii) distance and similarity measures for datasets; (ix) comparison based on summary statistics; and (x) testing approaches. The division is based on the fundamental underlying ideas that we identified in the set of analyzed dataset similarity methods. This taxonomy is not strict, but helpful for structuring the set of methods. Some methods could be classified into several classes. In those cases, we put them into the class matching their main idea.

Moreover, we present a comprehensive comparison of the methods. We introduce 22 criteria to judge the applicability, interpretability, and theoretical properties of dataset similarity measures. For each method, we check which of these criteria are fulfilled to provide guidance for the choice of a suitable method for quantifying the similarity of given datasets. To further facilitate the comparison of methods we implement an online tool that allows for interactive filtering and sorting of the methods (\url{https://shiny.statistik.tu-dortmund.de/data-similarity}).

In Section \ref{sec:notation}, we present the notation and assumptions that are used throughout the article. In Section \ref{sec:methods}, a detailed description of all methods can be found. The methods of each class are presented in a separate subsection. Within each class, methods are ordered chronologically. An overview of all methods is given in Section \ref{sec:overview}. 
In Section~\ref{sec:summ.methods}, we present a summary of the methods in the ten classes. For each group of methods, we describe its general concept and explain some prototypical example methods. This summary points out the main ideas of each class and can be understood without reading the detailed method descriptions.
In Section~\ref{sec:comp}, we introduce the list of criteria for rating the dataset similarity measures. These criteria are organized according to three main categories: applicability, interpretability, and theoretical properties.
In Section~\ref{sec:results}, we provide the results of the method comparison based on the criteria presented before. 
In Section \ref{sec:summary}, a brief summary of the review and comparison and an outlook are given.

\section{Notation and general assumptions}\label{sec:notation}
In general, a dataset can be viewed in two different ways, which impacts the approach to measure similarity. First, a dataset can be viewed simply as a collection of points in space. Second, and this is the more common view in the methods we examined, a dataset can be viewed as a sample of random variables that follow a true underlying distribution. In the second case, it is often of interest to estimate the similarity of these underlying distributions rather than the similarity of the datasets themselves. Therefore, many of the methods presented below focus on comparing multivariate distributions rather than directly comparing datasets.

In the following we assume at least two different datasets $\mathcal{D}_1$ and $\mathcal{D}_2$ consisting of $n_1$ and $n_2$, respectively, samples $X_1,\dots, X_{n_1} \sim F_1$ and $Y_1,\dots, Y_{n_2} \sim F_2$.
We assume $X_i, Y_j\in \mathbb{R}^p \,\forall i\in\{1,\dots,n_1\}, j\in \{1,\dots,n_2\}$ and call the $p$ components of each sample \textit{features}.
We denote the pooled sample as $\{Z_1,\dots, Z_N\} = \{X_1,\dots, X_{n_1}, Y_1,\dots, Y_{n_2}\}$, where $N = n_1+n_2$ is the total sample size.
For most of the methods, we assume that all $Z_i$ are distributed independently. For asymptotics, it is assumed that $N\to\infty$ such that $\frac{n_1}{N}\to const\in(0,1)$ if not explicitly stated otherwise.
We define the two-sample problem as the testing problem  
\begin{equation}
	H_0: F_1 = F_2 \text{ vs. } H_1: F_1\ne F_2.\label{two.sample.problem}
\end{equation}
This testing problem is sometimes also called testing for homogeneity of the two distributions.

In some cases, we also assume that there are $n_i$ observations of a target variable in each dataset, but most methods only require the feature variables and cannot deal with a target variable in a meaningful way.
Analogously to the two-sample problem, we define the $k$-sample problem for $k\ge 2$ datasets $\mathcal{D}_1,\dots,\mathcal{D}_k$ with sample sizes $n_i, i= 1, \dots, k$ as
\[
H_0: F_1 = F_2 = \dots = F_k \text{ vs. } H_1: \exists i\ne j\in\{1,\dots,k\}: F_i\ne F_j,
\]
where $F_i$ denotes the distribution of each sample in the $i$th dataset. We use the notation $F_i$ to denote the distribution as well as its cumulative distribution function.
By $f_i$ we denote the corresponding density functions if they exist.
If not explicitly stated otherwise we refer to the special case of the two-sample problem (\ref{two.sample.problem}).

In general, we denote random variables in uppercase letters and the corresponding observations in lowercase letters.
We use the hat symbol to denote estimators.
$\hat{F}_i$ and $\hat{f}_i$ denote the empirical distribution and density functions, respectively.
We use $T$ as the symbol for (test) statistics and $d(\cdot, \cdot)$ to denote distance measures.

\section{Detailed description of data similarity methods}\label{sec:methods}
In the following, we describe all selected dataset similarity methods. The methods are sorted according to the classes we identified. The first subsection gives an overview of all methods. In Subsection \ref{sec:cdf} to \ref{sec:testing}, the methods belonging to each class are described.

\subsection{Overview of all methods}\label{sec:overview}
Table \ref{tab:method.list} gives an overview of the dataset similarity methods included in this review. The first column gives the name of the method, if available, and otherwise, the reference where the method is defined. In all cases, the reference is linked. If the article defining the method is published online, the online publication can be accessed by clicking on the link that is given in parentheses after the method name. The classes to which the methods are assigned are given in subheadings within the body of the table. The subclasses are listed in the second column. The third column shows the section and page where the method is described within this article. 
\begin{longtable}{p{0.4\linewidth}p{0.35\linewidth}p{0.15\linewidth}}
	\bottomrule
	\endfoot
	\caption{List of all dataset similarity methods.}
	\label{tab:method.list}\\
	\toprule    
	Method/ Article & Subclass & Section, Page \\ 
	\midrule\endhead
	\multicolumn{3}{l}{\textbf{Comparison of CDFs, density or characteristic functions}}\\[1ex]
	\textcite{bickel_distribution_1969} (\href{https://www.jstor.org/stable/2239194}{link}) &  Comparison of CDFs & \ref{sec:cdf}, p.\ \pageref{bickel_distribution_1969}\\ 
	\textcite{biau_asymptotic_2005} (\href{https://doi.org/10.1109/TIT.2005.856979}{link}) &  Comparison of CDFs & \ref{sec:cdf}, p.\ \pageref{biau_asymptotic_2005}\\ 
	\textcite{boeckel_multivariate_2018} (\href{http://arxiv.org/abs/1809.04090}{link}) & Comparison of CDFs &  \ref{boeckel_multivariate_2018}, p.\ \pageref{boeckel_multivariate_2018} \\ 
	\textcite{ntoutsi_general_2008} (\href{https://doi.org/10.1137/1.9781611972788.73}{link}) &  Comparison of density functions & \ref{ntoutsi_general_2008}, p.\ \pageref{ntoutsi_general_2008} \\ 
	\textcite{ganti_framework_1999} (\href{https://doi.org/10.1006/jcss.2001.1808}{link}) &  Comparison of density functions &  \ref{ganti_framework_1999}, p.\ \pageref{ganti_framework_1999} \\
\end{longtable}

\subsection{Comparison of cumulative distribution functions}\label{sec:cdf}

In this subsection, methods based on the comparison of cumulative distribution functions (cdf) will be presented. Comparing distributions by their cumulative distribution functions is an intuitive approach since a distribution is fully characterized by its cumulative distribution function. In the one-dimensional case, methods of the Kolmogorov-Smirnov (KS) type that compare the maximal absolute difference of the (empirical) cumulative distribution functions are particularly popular. Still, their extension to the multivariate case is not straightforward \parencite{ramdas_wasserstein_2017}. 

\paragraph{Extension of the Kolmogorov-Smirnov test via permutation}\phantomsection
\label{bickel_distribution_1969}
\textcite{bickel_distribution_1969} gives a generalization of the Kolmogorov-Smirnov test to multivariate data based on applying a permutation procedure to the classical Kolmogorov-Smirnov test. It is distribution-free for continuous distributions and consistent against all alternatives. \textcite{bickel_distribution_1969} states that the asymptotic value of the cut-off point depends on the distribution $F$ if $F$ is not continuous. No details are given on the practical implementation of the test. \textcite{chen_new_2017} claim that the required sample size is exponential in the dimension $p$ and \textcite{mondal_high_2015} note that the test cannot be used for $p > n_i$.

\paragraph{Extension of the Kolmogorov-Smirnov test via partitioning}\phantomsection
\label{biau_asymptotic_2005}
\textcite{biau_asymptotic_2005} design a test using the $L^1$ distance between empirical distributions restricted to a finite partition of the support of the two distributions. For this test, $n_1 = n_2$ is required and a finite partition of $\mathbb{R}^p$ is needed. The authors themselves state that the ``choice of the partition in [the test statistic] is a difficult one''. \textcite{biau_asymptotic_2005} assume for this partition that for $N \to \infty$ the maximum of measures over each part goes to zero. A rectangle partition is said to be a good choice if cell probabilities are approximately equal. The resulting test is distribution-free and strongly consistent.
In addition, an asymptotic version of the test is given, which is not distribution-free but is consistent. According to \textcite{gretton_kernel_2006}, performing the test becomes difficult or impossible for high-dimensional problems due to the partitioning that becomes increasingly difficult in higher dimensions.

\paragraph{Multivariate distribution functions based on measure transportation}\phantomsection\label{boeckel_multivariate_2018}
\textcite{boeckel_multivariate_2018} define a new generalization of distribution functions to the multivariate case, called $\nu$-Brenier Distribution Functions (BDF), and their empirical counterparts. The $\nu$-BDF of a distribution is defined as the push-forward (measure-preserving transformation) of a continuous measure to a reference measure $\nu$ that has compact, convex support. To be more precise, the push forward of the mixture $t\mu_X + (1-t)\mu_Y, t \in[0,1]$ to a uniform distribution in the unit ball, where $t$ is the asymptotic ratio of sample sizes $n_1 / N$. \textcite{boeckel_multivariate_2018} assume that both measures belong to the family of absolutely continuous measures with finite second moments and compact support. They show that an analog of the Glivenko-Cantelli theorem holds for the empirical $\nu$-BDF. For their testing procedure, \textcite{boeckel_multivariate_2018} choose $\nu$ as the uniform measure on the unit sphere in $\mathbb{R}^p$. The test statistic is the 2-Wasserstein distance (\ref{eq:wasserstein}) between image measures of the distributions of $X$ and $Y$ generated by the push-forward of the mixture distribution of $X$ and $Y$ to $\nu$. In practice, the empirical counterparts are used. The procedure works by generating a uniform partition of the unit ball into $N$ parts, then calculating the optimal transport of both samples to this partition and taking the 2-Wasserstein distance of the empirical distributions of these optimal transports. The critical value is obtained by using the (1-$\alpha$)-quantile of the empirical distribution of 2-Wasserstein distances of empirical distributions of $M$ random permutations of a uniform partition of the unit ball into $N$ parts. An asymptotic upper bound for the type II error is derived. According to \textcite{deb_multivariate_2021}, this test is one of two tests for the multivariate two-sample problem that is exactly distribution-free, computationally feasible, and consistent against all alternatives. However, they criticize that the test statistic is random given the data due to external randomization in the construction of the test statistic, that strong assumptions on underlying distributions are needed, such as that the data generating distribution is compactly supported and absolutely continuous, and that there is no asymptotic null distribution theory.
The method of \textcite{boeckel_multivariate_2018} could also be seen as a method based on multivariate ranks or as a test based on the Wasserstein distance.

\subsection{Comparison of density functions}
The comparison of density functions follows a similar idea as the comparison of cumulative distribution functions. Different approaches are presented below.

\subsubsection{Comparison of probability densities based on partitions} 
\paragraph{Partitions based on decision trees I}\phantomsection
\label{ganti_framework_1999}
\textcite{ganti_framework_1999} propose measuring the deviation between datasets based on criteria derived from decision tree models.
Let $\mathcal{D}_1$ and $\mathcal{D}_2$ denote two datasets that include a categorical target variable each.

\textcite{ganti_framework_1999} calculate a decision tree model for each dataset $\mathcal{D}_1$ and $\mathcal{D}_2$ and calculate the \textit{greatest common refinement (GCR)} induced by these trees. That is the intersection of the partitions of the sample space induced by each tree. They then compare the distribution of both datasets over this GCR. Let $n_r$ denote the number of segments of the GCR, $p_i$ the proportion of observations of $\mathcal{D}_1$ that map to the $i$-th segment, and $q_i$ the respective proportion of observations of $\mathcal{D}_2$ mapping to the $i$-th segment. Then \textcite{ganti_framework_1999} compare the vector $p$ and $q$ by a difference function $f: \mathbb{R}^{n_r} \to \mathbb{R}^{n_r}$ and aggregate the results from that by an aggregate function $g: \mathbb{R}^{n_r} \to \mathbb{R}$ to obtain a measure of distance between the two datasets
\[
\text{GAN1} = g(f(p,q)).
\]
Large values then indicate differences between the datasets. They propose the absolute difference function
\begin{align*}
	f_a(p, q)_i = |p_i - q_i|,
\end{align*}
and the scaled difference function 
\begin{align*}
	f_s(p, q)_i = \begin{cases}
		\frac{|p_i - q_i|}{(p_i + q_i)/2}, & \text{if } (p_i + q_i) > 0\\
		0, \text{otherwise}
	\end{cases}.
\end{align*}
For the aggregate function, they propose the sum or maximum of the values from the difference function. For using the sum as the aggregate function together with either $f_a$ or $f_s$, it can be shown that the GCR is optimal in the sense that it gives the lowest value over all common refinements. For using the maximum, this property is not fulfilled. As in general, for different combinations of difference and aggregate functions, there might not be an upper bound for the difference given by the proposed measure GAN1, \textcite{ganti_framework_1999} propose using a Bootstrap test procedure for assessing whether or not the two datasets are generated by the same data-generating process. The lower bound for GAN1 for all proposed difference and aggregate functions is 0. 

Typical measures for monitoring change and assessing how much a new tree model differs can be seen as a special case of the measure proposed by \textcite{ganti_framework_1999}. For these, they calculate a decision tree model for dataset $\mathcal{D}_1$.

Then, for the first measure of change between models, they use this model fit on the first dataset to make predictions $\widehat{Y}$ for the target variable in dataset $\mathcal{D}_2$.
Finally, they calculate the misclassification rate
$$ \text{GAN2} = \frac{|\{i: \widehat{Y}_i \neq Y_i\}|}{|\mathcal{D}_2|},$$
which is the proportion of observations in $\mathcal{D}_2$ whose target value is predicted incorrectly by the model fitted on $\mathcal{D}_1$.

For the second measure of change between models, they consider the partition of the feature space induced by the decision tree calculated on $\mathcal{D}_1$.
Let $n_r$ again denote the number of segments of this partition, $p_i$ the proportion of observations of $\mathcal{D}_1$ that map to the $i$-th segment, and $q_i$ the respective proportion of observations of $\mathcal{D}_2$ mapping to the $i$-th segment.
If the datasets $\mathcal{D}_1$ and $\mathcal{D}_2$ come from the same data generating process, the number of observations of $\mathcal{D}_2$ that are expected to map to the $i$-th segment can be estimated by $|\mathcal{D}_2| \cdot p_i$.
The number of observations mapping to the $i$-th segment equals $|\mathcal{D}_2| \cdot q_i$.
Let $c \in \mathbb{R}$ denote a small positive constant, for example $c = 0.5$.
\textcite{ganti_framework_1999} propose calculating the $\chi^2$-statistic
$$ \text{GAN3} = \sum\limits_{i=1}^{n_r} \chi(p_i, q_i)$$
with $$\chi(p_i, q_i) = \begin{cases}
	\frac{(|\mathcal{D}_2| \cdot p_i - |\mathcal{D}_2| \cdot q_i)^2}{|\mathcal{D}_2| \cdot p_i}, & \text{if } p_i > 0,\\
	c, & \text{otherwise.}\\
\end{cases}$$

For both measures, low values indicate similar datasets.
With respect to bounds, $0 \leq \text{GAN1} \leq 1$ holds because GAN1 is a proportion.
Also, $0 \leq \text{GAN2}$ holds because it is a sum of squared values.
A dataset-independent upper bound cannot be specified for GAN2 because $\chi$ can attain higher values with more observations in $\mathcal{D}_2$. Again, a Bootstrap test is proposed to assess the significance of the $\chi^2$ value since usual asymptotics for the classical $\chi^2$ test typically will not apply here due to low expected counts. 

\textcite{ganti_framework_1999} do not address the choice of hyperparameters for creating the decision tree model.
As discussed in the previous paragraph, this can have a non-negligible impact on the resulting data similarity measures.

\paragraph{Partitions based on probability binning I}\phantomsection
\label{roederer_probability_2001}
\textcite{roederer_probability_2001} suggest probability binning for comparing the multivariate distributions of two datasets.
Their method only considers the feature space $\mathcal{X}$ and is only applicable to numeric features.
First, one dataset is chosen for defining a partition of the feature space.
To do so, for each feature, the median value and the variance are computed.
Two bins (segments) are created by splitting the feature space at the median value of the feature with the largest variance.
Then, the calculation of median and variance as well as the splitting is continued recursively for both subspaces, until a predefined minimum number of observations per bin is reached.

Having obtained the partition, the proportions of observations falling into each bin are calculated for both datasets.
Let $p_{1,i}$ denote the proportion of observations of the first dataset that fall into the $i$-th bin, and let $p_{2,i}$ denote the respective proportion for the second dataset.
Then, \textcite{roederer_probability_2001} propose the measure
$$\text{ROE} = \sum_{i=1}^{n_b} \frac{\left(p_{1,i} - p_{2,i}\right)^2}{p_{1,i} + p_{2,i}}$$
with $n_b$ denoting the number of bins to quantify the difference between the two datasets.
As stated in \textcite{roederer_probability_2001}, the measure is bounded by $0 \leq \text{ROE} \leq 2$ with low values corresponding to high similarity.

\textcite{roederer_probability_2001} explain that the minimum number of observations per bin should not be smaller than 10.
Also, it should be chosen appropriately by the user, such that a good coverage of a potentially high-dimensional space can be achieved.
With a predefined minimum number of observations per bin, the number of observations in the first dataset determines the number of bins.
This might lead to problems when the number of observations is very different for the two datasets.

\paragraph{Partitions based on probability binning II}\phantomsection
\label{wang_random_2005}
\textcite{wang_random_2005} propose a measure to quantify changes between two datasets with class labels for the application of fraud and intrusion detection. Their dataset distance measure quantifies concept drifts. It uses a universal model that has minimal learning cost. \textcite{wang_random_2005} aim to use a quantification of change to improve the current prediction model directly instead of refitting the model after change gets detected. They note that their approach of using arbitrary partition structures instead of learning a tree structure on one of the datasets as a simple alternative serves the same purpose but eliminates the cost of learning the structure. They propose to calculate the class distribution for each dataset for a certain partition of the feature space, which they call the \emph{signature} of the data, by randomly partitioning the multi-dimensional space into a set of bins. This partition can be achieved either by recursive partitioning (random decision tree) or by iteratively creating new bins (random histogram). \textcite{wang_random_2005} give explicit recommendations on how to create the partition. They recommend the use of random histograms over random decision trees. If $n_{x,j,K}$ and $n_{y,j,K}$ denote the number of observations of the $j$th part for the $K$th class for the first and second dataset, respectively, and $n_c$ is the total number of classes and $n_r$ the total number of parts in the partition, the distance function between the signatures of both datasets is defined as 
\[
\text{Dist}_s(X, Y) = \frac{1}{2} \sum_{j = 1}^{n_r}\sum_{K = 1}^{n_c} \left|\frac{n_{x,j,K} }{n_1} - \frac{n_{y,j,K} }{n_2}\right| \in [0, 1].
\]
The distance can be calculated efficiently and can be used to make predictions. The calculation is repeated to create $B$ random structures resulting in $B$ random signatures. Then, the distance measure as above is calculated for each random structure and the mean over the values is taken as the final distance between the two datasets. The random signatures can be used for classification as well.

\paragraph{Partitions based on decision trees II}\phantomsection
\label{ntoutsi_general_2008}
\textcite{ntoutsi_general_2008} propose measuring dataset similarity based on probability density estimates derived from decision trees.
Consider two classification datasets $\mathcal{D}_1$ and $\mathcal{D}_2$.
For each of them, construct a decision tree for the target variable $Y$.
Then, derive a partition of the feature space $\mathcal{X}$ based on the split rules such that each leaf node corresponds to one segment in the partition.
Next, overlay the two partitions resulting in smaller hyper rectangles.

Based on the joint partition, the probability densities $P_D(\mathcal{X})$ and $P_D(Y,\mathcal{X})$ are estimated for $D \in \{\mathcal{D}_1, \mathcal{D}_2, \mathcal{D}_1 \cup \mathcal{D}_2\}$.
Let $n_r$ denote the number of segments in the joint partition and $n_c$ the number of classes in $\mathcal{D}_1$ and $\mathcal{D}_2$.
To estimate $P_D(\mathcal{X})$, assess the proportion of observations in $D$ that fall into each segment of the joint partition, $\hat{P}_D(\mathcal{X}) \in \mathbb{R}^{n_r}$.
For the estimation of the joint density $P_D(Y,\mathcal{X})$, determine the proportion of observations that fall into each segment of the joint partition and belong to each class, $\hat{P}_D(Y, \mathcal{X}) \in \mathbb{R}^{n_r \times n_c}$.
Estimate the conditional density $P_D(Y|\mathcal{X})$ by calculating the proportion of observations belonging to each class separately for each segment, $\hat{P}_D(Y|\mathcal{X}) \in \mathbb{R}^{n_r \times n_c}$.

\textcite{ntoutsi_general_2008} consider the similarity index $$s(p, q) = \sum_{i} \sqrt{p_i \cdot q_i}$$
for vectors $p$ and $q$.
If $p$ and $q$ are $(n_r \times n_c)$-matrices, they are interpreted as $(n_r \cdot n_c)$-dimensional vectors.
For the conditional distribution, the similarity vector $S(Y|\mathcal{X}) \in \mathbb{R}^{n_r}$ is computed with $S(Y|\mathcal{X})_i = s(\hat{P}_{\mathcal{D}_1}(Y|\mathcal{X})_{i \bullet}, \hat{P}_{\mathcal{D}_2}(Y|\mathcal{X})_{i \bullet})$ and index $i \bullet$ denoting the $i$-th row.
Three similarity measures for datasets are suggested:

\begin{enumerate}
	\item NTO1 = $s(\hat{P}_{\mathcal{D}_1}(\mathcal{X}), \hat{P}_{\mathcal{D}_2}(\mathcal{X}))$
	\item NTO2 = $s(\hat{P}_{\mathcal{D}_1}(Y, \mathcal{X}), \hat{P}_{\mathcal{D}_2}(Y, \mathcal{X}))$
	\item NTO3 = $S(Y|\mathcal{X})^{T} \hat{P}_{\mathcal{D}_1 \cup \mathcal{D}_2}(\mathcal{X})$
\end{enumerate}

For probability estimates $p$ and $q$, $s(p, q)$ is bounded by $0 \leq s(p, q) \leq 1$.
Therefore, measures NTO1 and NTO2 are bounded by $0 \leq \text{NTO1, NTO2} \leq 1$.
Measure NTO3 is  bounded by $0 \leq \text{NTO3} \leq 1$ because $S(Y|\mathcal{X})_i \leq 1$ for all $i \in \{1, \ldots, n_r\}$ and $\sum_{i=1}^{n_r} (\hat{P}_{\mathcal{D}_1 \cup \mathcal{D}_2}(\mathcal{X}))_i = 1$.
For the three measures, high values correspond to high similarity.

\textcite{ntoutsi_general_2008} do not specify how to choose the hyperparameters for the decision tree computation.
Especially for decision trees with many leaf nodes, it is likely that the joint partition contains empty or very sparse segments.

\subsubsection{Comparison of probability densities based on kernel density estimation}\label{sec:char.fun}

\paragraph{Comparison of kernel density estimates in $L^2$-norm I}\phantomsection
\label{ahmad_goodness_1993}
\textcite{ahmad_goodness_1993} define a test statistic based on the $L^2$-norm of the difference between kernel density estimates. It has to be assumed that densities exist and are differentiable up to second order with bounded derivatives. Also, assumptions on the kernel function are needed. Under these assumptions, an asymptotic test is proposed. The test statistic has an asymptotic normal distribution under both null and alternative hypothesis, with less restrictive assumptions than in related articles. Different sequences of weights have to be chosen to calculate the estimators of the test statistic and its variance. 

\paragraph{Comparison of kernel density estimates in $L^2$-norm II}\phantomsection
\label{anderson_two-sample_1994}
\textcite{anderson_two-sample_1994} present a test based on the integrated square distance between kernel-based density estimates as well as asymptotic distributional results and power calculations for this test. For the calculation, a kernel and the bandwidth for kernel density estimation must be chosen. \textcite{anderson_two-sample_1994} use a bandwidth of $h = 1$ in their derivation of theoretical results. They show that the minimum distance at which the statistic can discriminate between $f_1$ and $f_2$ can be expressed by $f_2 = f_1 + N^{-1/2}h^{-p/2} \cdot g$ under the condition that $h \to 0$ as $n_1, n_2 \to \infty$, where $n_1$ and $n_2$ are assumed to be of the same order of magnitude. For this, they use the assumption that $f_1 = f_2$ has two continuous, square-integrable derivatives. Additionally, regularity conditions on the kernel are made: it has to be bounded, absolutely integrable, and its Fourier transform must not vanish on any interval. These conditions are e.g.\ fulfilled for $p$-variate uniform, standard normal densities, and $p$-variate forms of Epanechnikovs kernel. The test statistic is not asymptotically normally distributed. The resulting test is consistent.

\paragraph{Comparison of kernel density estimates based on empirical likelihood}\phantomsection
\label{cao_empirical_2006}
\textcite{cao_empirical_2006} propose a test based on comparing kernel estimators of the two density functions for continuous distributions based on an empirical likelihood criterion. They state that ``for high-dimensional distributions, the curse of dimensionality implies that the method will not be applicable in practice''. An alternative presented is to use a model-based approach that only concentrates on elliptically contoured distributions.

\subsection{Comparison of characteristic functions}
The idea of comparing characteristic functions is that they fully characterize the distribution.
\textcite{meintanis_review_2016} reviews tests based on empirical characteristic functions, including tests for the two- and $k$-sample problem and an interpretation in terms of moments of a general test statistic, which is based on integrating over the weighted squared difference of empirical characteristic functions.
Different tests can be derived from this general test statistic via different weight functions (e.g.\ \textcite{alba-fernandez_test_2008}, \textcite{lindsay_kernels_2014}, \textcite{huskova_tests_2008}).

\paragraph{Comparison of empirical characteristic functions in $L^2$-norm}\phantomsection
\label{alba-fernandez_bootstrap_2004}
\textcite{alba-fernandez_bootstrap_2004} consider the $L^2$ norm of the difference between empirical characteristic functions (ecf). Their test statistic is the integral of the weighted absolute difference between the ecfs of the two samples.
They use a trigonometric Hermite interpolant to obtain a numerical integration formula to approximate the test statistic.
The $p$-value of the test is estimated by a Bootstrap algorithm similar to \textcite{alba_homogeneity_2001}.
The test can be applied to continuous and discrete data.
The assumption of existing second moments is made for all theoretic results and additional assumptions on the approximation of the test statistic are required to show consistency against a wide range of fixed alternatives.

\paragraph{Comparison of empirical characteristic functions based on weighted integrals}\phantomsection
\label{alba-fernandez_test_2008}
\textcite{alba-fernandez_test_2008} propose a class of tests based on the weighted integral of empirical characteristic functions.
The tests are not asymptotically distribution-free and neither the null distribution nor the asymptotic null distribution of the test statistic are known. Instead, a permutation or Bootstrap procedure yields asymptotically consistent approximations of the null distribution. The weight function must be chosen, but a choice of the weight function that is not very restrictive already yields consistency against any fixed alternative. Specifically, if the weight function is chosen such that the distance between populations is larger than zero for any $F_1 \ne F_2$, the test is consistent against any fixed alternative. This is for example fulfilled for weight functions with positive density for almost all points in $\mathbb{R}^p$. The weight function also influences the computing time. There are no conditions assumed on the populations. The method can be applied to continuous as well as discrete data of any arbitrary fixed dimension. The test is a generalization of tests in \textcite{alba_homogeneity_2001} and \textcite{alba-fernandez_bootstrap_2004}. According to \textcite{li_measuring_2022}, the choice of the weight function is a difficult problem.

\paragraph{Comparison of empirical characteristic functions based on jackknife empirical likelihood}\phantomsection
\label{liu_test_2015}
\textcite{liu_test_2015} develop a jackknife empirical likelihood (JEL) test by incorporating characteristic functions. The test statistic reduces to a two-sample U-statistic, which simplifies the estimation. For fixed dimension, the authors derive a nonparametric Wilks's theorem. For $p\to \infty$, $p = o(N^{1/3})$, under some mild conditions the normalized JEL ratio statistic has a standard normal limit. For $p>N$ an alternative version of the JEL test is proposed that has an asymptotical $\chi^2_2$ distribution under the null. For computing the test statistic, a range over which the statistic is calculated has to be chosen. According to \textcite{li_measuring_2022}, the choice of the weight function is a difficult problem. \textcite{liu_test_2019} claim that the test works well in the case of small samples and also for asymmetric data.

\paragraph{Comparison of empirical characteristic functions based on characteristic distance}\phantomsection
\label{li_measuring_2022}
\textcite{li_measuring_2022} introduce the characteristic distance, which does not need any assumptions on moments and parameters and fully characterizes the homogeneity of two distributions since it is nonnegative and equal to zero if and only if the distributions are equal. The characteristic distance relies on the equivalence of almost surely equal characteristic functions and the equality
\[
\E\left(\exp\left(i\langle X, X^{\prime}\rangle\right)\arrowvert X^{\prime}\right) = \E\left(\exp\left(i\langle Y, X^{\prime}\rangle\right)\arrowvert X^{\prime}\right), \, a.s.,
\]
where $i$ denotes the imaginary unit.
Let $X^{\prime}, X^{\prime\prime}$ and $Y^{\prime}, Y^{\prime\prime}$ denote independent copies of $X$ and $Y$, respectively. Then, the characteristic distance is defined as 
\begin{align*}
	\text{CD} (X, Y) &= \E\left[\| \E\left(\exp\left(i\langle X^{\prime\prime}, X - X^{\prime}\rangle\right)\arrowvert X - X^{\prime}\right)\right.\\
	&\qquad ~~- \left.\E\left(\exp\left(i\langle Y, X - X^{\prime}\rangle\right)\arrowvert X - X^{\prime}\right)\|^2\right]\\
	&\quad + \E\left[\| \E\left(\exp\left(i\langle X, Y - Y^{\prime}\rangle\right)\arrowvert Y-Y^{\prime}\right)\right. \\
	&\qquad \quad ~-\left. \E\left(\exp\left(i\langle Y^{\prime\prime}, Y - Y^{\prime}\rangle\right)\arrowvert Y - Y^{\prime}\right)\|^2\right].
\end{align*}
An empirical version is obtained by replacing the conditional expectations with empirical means. \textcite{li_measuring_2022} derive the distribution of the characteristic distance under the null and alternative hypothesis. They call the resulting test distribution-free, but the asymptotic distribution depends on an unknown true distribution, so a permutation test is used instead. According to \textcite{li_measuring_2022}, the test has a clear and intuitive probabilistic interpretation and its estimator is easy to calculate. They derive the asymptotic distribution and show that the test is consistent against any generic alternative. The test is robust since no moment assumptions are needed and it is free of any tuning parameters.

\subsection{Methods based on multivariate ranks}\label{sec.rank}

For the univariate two-sample problem, tests based on ranks are popular methods. Since $\mathbb{R}^p$ has no natural ordering, the generalization of these methods to the multivariate case is not straightforward. Different approaches to multivariate rank procedures are presented below. 

\paragraph{Ranks based on projections obtained by binary classification}\phantomsection
\label{ghosh_distribution-free_2016}
The first procedures based on multivariate ranks, which are often only applicable to the location or scale problem instead of the general two-sample problem, were proposed by  \textcite{puri_nonparametric_1971}, \textcite{randles_multivariate_1990} (only location problem), \textcite{hettmansperger_affine_1994} (also only for location problem), \textcite{choi_approach_1997} (theory only for location problem), \textcite{hettmansperger_affine_1998} (also only for location problem). According to \textcite{ghosh_distribution-free_2016}, these procedures usually yield poor results for high-dimensional data, none of them can be used for $p>N$, and none is distribution-free in finite sample situations, and although some of them are asymptotically distribution-free and for some one can implement conditional versions using permutation type techniques. The test from \textcite{liu_quality_1993} (can only detect location and/or additional dispersion differences) is based on a quality index based on ranks. The test is distribution-free but computationally infeasible in high dimensions and also cannot be used when $p>N$. \textcite{ghosh_distribution-free_2016} instead present a general procedure for multivariate generalizations of univariate distribution-free tests based on ranks of real-valued linear functions of multivariate observations. The linear function is obtained by solving a classification problem between the two distributions. The procedure is exactly distribution-free in finite samples under very general conditions and applicable even when the dimension exceeds the sample size.\\
The idea behind the test procedure is that $H_0: F_1 = F_2$ can be expressed as $F_{\beta,1} = F_{\beta,2} \forall\beta\in\mathbb{R}^p$, where $\beta^TX\sim F_{\beta, i}$ if $X\sim F_i$. This means, if the distributions differ, it is expected that for some $\beta$ values $F_{\beta,1}$ differs from $F_{\beta,2}$. \textcite{ghosh_distribution-free_2016} choose a projection such that the separation between observations of different samples is maximized by using the direction vector of a linear classifier that discriminates between the samples. They train a support vector machine (SVM) or distance-weighted discrimination \parencite{marron_distance-weighted_2007} on a training set. Then projections are calculated on a test set and a one-sided KS test or Wilcoxon test is performed on these projections. This procedure is repeated for several train/test splits and the test statistics are averaged. Then either a randomized test is used or tests with Bonferroni correction/control of FDR for each split are performed, and $H_0$ is rejected if any of the tests reject. \textcite{ghosh_distribution-free_2016} derive asymptotic results for $p\to\infty$ under the assumption of uniformly bounded fourth moments, weak dependence among component variables, and convergence of variances and the squared differences between expectations (the last two hold automatically for i.i.d. components with bounded second moments). These assumptions ensure that under $H_1$ the amount of information for discrimination grows to infinity as the dimension increases. Under these assumptions, consistency is shown. The same holds even if $p\to\infty$ and $N\to\infty$ such that $N/p^2\to0$ and also for general one-sided linear rank statistics that are linear combinations of a monotonically increasing function applied to the ranks. The test remains distribution-free if data is transformed by a real-valued measurable function chosen on the training set (instead of linear transformation as before). Power is maximized if this transformation is chosen as the likelihood ratio (LR), but this is hard to estimate due to the curse of dimensionality. An alternative is to use a nonlinear SVM with a suitable kernel choice. 
The method can also be seen as a method based on binary classification.

\paragraph{Ranks based on optimal transport I}\phantomsection
\label{ghosal_multivariate_2021}
\textcite{ghosal_multivariate_2021} propose multivariate nonparametric tests free of tuning parameters based on a new notion of multivariate quantiles/ranks, which were introduced by \textcite{chernozhukov_mongekantorovich_2017} and use optimal transport theory (see \ref{eq:OT}). The idea behind this approach is based on the insight that in the univariate case, ranks can be understood as the solution of transporting the data distribution to the uniform distribution. Therefore, the multivariate ranks are defined as the solution to the corresponding multivariate optimal transport problem. The test statistic then consists of the integral (w.r.t.\ a reference distribution) over the squared distance of the rank map of the pooled sample applied to the quantile maps of the individual samples. The authors show asymptotic consistency under fixed alternatives and derive rates of convergence of the test statistics under null and alternative hypothesis. \textcite{ghosal_multivariate_2021} make the assumption of absolutely continuous distributions. It is not known if the test is (asymptotically) distribution-free. Instead, a permutation test is performed. The test statistic tends to zero under $H_0$ for $n_1, n_2\to\infty$ such that $n_1/N\to $ const. The test is implemented in the \texttt{R} \parencite{R_4_1_2} package \texttt{testOTM} \parencite{testOTM}.

\paragraph{Ranks based on optimal transport II}\phantomsection
\label{deb_efficiency_2021}
\textcite{deb_efficiency_2021} propose distribution-free analogues of Hotelling's $T^2$ test based on optimal transport. The test statistic is the squared difference of the mean of multivariate ranks between two samples. Here, ranks are assigned for the pooled sample based on optimal transport. They claim consistency for general alternatives and efficiency under location shift alternatives. \textcite{deb_efficiency_2021} aim to design multivariate nonparametric distribution-free tests that attain similar asymptotic relative efficiency (ARE) values compared to Hotelling's $T^2$ test. The resulting test statistic follows a limiting $\chi^2_p$ distribution under the null and a non-central $\chi^2$ distribution under contiguous alternatives. The test is consistent against large classes of natural alternatives including a location shift model and a contamination model. \textcite{deb_efficiency_2021} present numerous lower bounds on the ARE for multiple subfamilies of multivariate probability distributions, e.g.\ distributions with independent components (for multiple subfamilies there is no loss of efficiency). Their test is exactly distribution-free and generalizes the two-sided Wilcoxon rank-sum test and the van der Waerden score test for $p> 1$. The test can be further generalized by calculating scores from ranks. Then assumptions on the score function and its covariance matrix are required. A reference distribution is needed to define the ranks. The choice of this reference distribution affects the ARE of the resulting test. Throughout, the assumption of Lebesgue absolutely continuous probability measures is used. Under this assumption, weak convergence of the rank distribution to a reference distribution is shown. No moment assumptions are made for the distributions that are compared, only for the reference distribution and the score function. The moment assumptions are for example always satisfied for the identity as score function and $U[0,1]^p$ as reference distribution. A basic version of the statistic is presented in \textcite{hallin_efficient_2022} for the special case that the reference distribution is spherical uniform and for a specific choice of rank set, but the authors did not study theoretical properties as consistency and asymptotic efficiency of the resulting test. The results of \textcite{deb_efficiency_2021} prove consistency and can be used to derive ARE for this special case as well.

\paragraph{Ranks based on similarity graphs}\phantomsection\label{zhou_new_2023}
	\textcite{zhou_new_2023} define ranks based on similarity graphs and use these for constructing a two-sample test. 
	They define a sequence of simple similarity graphs $\{G_l\}_{l=0}^K$ on the pooled sample via 
	\[
	G_{l+1} = G_l \cup G_{l+1}^\ast
	\]
	with
	\[
	G_{l+1}^\ast = \arg\max_{G^\prime\in\mathcal{G}_{l+1}}\sum_{(i,j)\in G^\prime} S(Z_i, Z_j),
	\]
	where $G_0$ has no edges, $\mathcal{G}_{l+1} = \{G^\prime\in\mathcal{G}:G^\prime\cap G_l = \emptyset \}$ with $\mathcal{G}$ the set of graphs that fulfill specific user-defined constraints, and $S(\cdot,\cdot)$ a similarity measure, e.g.\ the negative Euclidean distance for Euclidean data, and $Z_1,\dots,Z_N$ denoting the pooled sample. This construction scheme includes as special cases the $K$-nearest neighbor graph, the $K$-minimum spanning tree, the $K$-minimum distance non-bipartite pairing, and the $K$-shortest Hamiltonian path. Based on the sequence of similarity graphs, \textcite{zhou_new_2023} define the following two graph-based rank matrices $R = (R_{ij})_{i,j=1}^N$. The \emph{graph-induced ranks} are defined as 
	\[
	R_{ij} = \sum_{l=1}^K \mathbbm{1}\left(\left(i,j\right)\in G_l\right)
	\]
	and the \emph{overall ranks} are defined as
	\[
	R_{ij} = \text{rank}\left(S\left(Z_i, Z_j\right), G_K\right),
	\]
	where for $(i,j)\in G_k$, $\text{rank}\left(S\left(Z_i, Z_j\right), G_K\right)$ denotes the rank of $S\left(Z_i, Z_j\right)$ among the values $\{S\left(Z_u, Z_v\right)\}_{(u,v)\in G_K}$ and otherwise it is zero. The graph-induced rank $R_{ij}$ can be interpreted as the number of graphs that contain the edge $(i,j)$ in the sequence of graphs. The overall rank can be interpreted as the rank of the similarity of edges in the graph $G_K$. Both depend on the choice of $K$. For the test, the symmetrized rank matrix $1/2(R+R^T)$ is used. For convenience, it is also denoted by $R$.

For the test statistic, the within-sample rank sums of the first and second sample are defined as 
	\[
	U_x = \sum_{i,j=1}^{n_1} R_{ij}, U_y = \sum_{i,j=n_1 + 1}^{N} R_{ij}.
	\]
	Using these, the \emph{Rank In Similarity graph Edge-count two-sample test (RISE)} statistic is defined as 
	\[
	T_R = (U_x - \mu_x, U_y - \mu_y)\Sigma^{-1}(U_x - \mu_x, U_y - \mu_y)^T,
	\]
	where $\mu_x = \E(U_x)$, $\mu_y = \E(U_y)$, and $\Sigma = \mathbb{C}\text{ov}((U_x, U_y)^T)$. The quantities $\mu_x$, $\mu_y$, and $\Sigma$ are explicitly calculated under the permutation null hypothesis, and sufficient conditions under which $\Sigma$ is invertible and therefore $T_R$ is well-defined are given. The test statistic can be decomposed into two quantities $Z_w$ and $Z_{\text{diff}}$, which can be related to the graph-based tests of \textcite{chen_new_2017}, \textcite{chen_weighted_2018}, and \textcite{zhang_graph-based_2019}. For small samples, the exact permutation null distribution can be used for testing. For large samples and under several assumptions on the similarity graphs, the asymptotic $\chi^2_2$-distribution of $T_R$ can be used for testing. For continuous distributions and the $K$-MST or $K$-NN graph based on the Euclidean distance and with $K = \mathcal{O}(1)$, the test is consistent for $n_1, n_2\to\infty$ and $n_1/N\to\pi\in(0,1)$. Under several other assumptions, consistency of the test using the graph-induced ranks for the $K$-NN graph or for using the overall ranks for the $K$-minimum distance non-bipartite pairing is shown. Extensions of the RISE test using kernel functions for the similarity measure and using another graph-based rank definition are briefly presented. The RISE test can also be classified as a graph-based approach.

\subsection{Discrepancy measures for distributions} 

There are two main classes of discrepancy measures for distributions: probability metrics and divergences. The best-known subclasses are \textit{Integral Probability Metrics} (IPM, also called probability metrics with a $\xi$-structure \parencite{zolotarev_metric_1976, zolotarev_probability_1984}) as introduced by \textcite{muller_integral_1997} and $f$-\textit{Divergences} (sometimes also called Ali-Silvey distances, going back to \textcite{ali_general_1966}, or Csiszár's $\Phi$-divergences, going back to \textcite{csiszar_informationstheoretische_1963}).
The latter were introduced in the two aforementioned articles. These two classes of Integral Probability Metrics and $f$-divergences only intersect at the total variation distance as shown by \textcite{sriperumbudur_empirical_2012}. \\
For probability metrics, \textcite{zolotarev_probability_1984} distinguishes between probability metrics with a $\Lambda$-structure, probability metrics with a $\xi$-structure (= IPMs), metrics with a Hausdorff structure, and metrics with an Integral structure. He also notes that there are other metrics that do not have any of these structures, e.g.\ the Hellinger metric. 
Other classes of divergences are Bregman-divergences \parencite{bregman_relaxation_1967} and the Burbea-Rao divergences \parencite{burbea_convexity_1982}, which will not be discussed here since they require a parametric model of the distributions. Another class is formed by $H$-divergences, which overlap with both $f$-divergences and IPMs and were recently introduced by \textcite{zhao_comparing_2021}.  \\
A detailed overview of the theory on probability metrics as well as a comprehensive list of examples can be found in \textcite{rachev_probability_1991}. A more applied description is given in \textcite{rachev_advanced_2008, rachev_probability_2011}. An overview of inequalities specifying the relationships between different $f$-divergences is given in \textcite{sason_f_2016}.
In the following, we will focus on the main ideas and important examples. We will again only look at discrepancy measures for comparing two distributions. There are often versions of the measures discussed here for the case of comparing an empirical distribution to a known distribution in a goodness-of-fit context. This case is for example discussed in \textcite{basu_statistical_2011} in more detail.\\
Note that many of the methods presented below can also be seen as methods based on inter-point distances or methods based on cumulative distribution functions or density functions, e.g.\ all $f$-divergences are based on density functions.

\subsubsection{Probability (semi-)metrics}\phantomsection
\label{probability_metrics}
\textcite{zolotarev_probability_1984} reviews probability metrics and identifies four classes, which are defined via a functional $\mu(X, Y)$ on the space of bivariate distributions. This functional takes values on $[0,\infty]$, with the following three properties: 
\begin{enumerate}
	\item $\Prob(X = Y) = 1 \Rightarrow \mu(X, Y) = 0$
	\item $\mu(X, Y) = \mu(Y, X)$
	\item $\mu(X, Y) \le \mu(X, Z) + \mu(Z, Y)$
\end{enumerate}
Note that \textcite{zolotarev_probability_1984} argues to use the term probability \emph{metric} although the first condition is only analogous to conditions from functional analysis that characterizes a semimetric and not a metric.

In contrast, \textcite{muller_integral_1997} defines a probability metric $d$ with the following properties: 
\begin{enumerate}
	\item $d(F_1, F_2) = 0 \Leftrightarrow F_1 = F_2$ (positive definite)
	\item $d(F_1, F_2) = d(F_2, F_2)$ (symmetry)
	\item $d(F_1, F_3) \le d(F_1, F_2) + d(F_2, F_3)$ (triangle inequality).
\end{enumerate}
If the first property is replaced by  
\[
d(F, F) = 0\text{ for all distributions } F,
\]
$d$ is called a probability semimetric.\\
The latter way to distinguish between probability metric and semimetric is more common in the literature \parencite[e.g.][]{rachev_probability_1991} and more precise and is therefore used in the following. \\
\textcite{zolotarev_probability_1984} propose a distinction between \textit{simple} and \textit{compound} metrics, \textcite{rachev_probability_1991} also distinguishes \textit{primary} metrics from the former. He defines three types as follows: 
Let $P$ be the joint distribution of $F_1$ and $F_2$ and let $Q$ be the joint distribution of two distributions $G_1, G_2$ defined on the same sample spaces as $F_1$ and $F_2$. Denote the space of joint distributions to which $P$ and $Q$ belong by $\mathcal{P}$. A (semi)metric $d:\mathcal{P}\to [0,\infty)$ is called \textit{primary (semi)metric}, if it is a probability (semi)metric and there exists a function $h: \mathcal{P}_1 \to \mathbb{R}^J, J\in \mathbb{N},$ such that
\[
(h(F_1) = h(G_1) \wedge h(F_2) = h(G_2)) \Leftrightarrow d(P) = d(Q).
\]
$\mathcal{P}_1$ denotes the set of Borel probability measures for some separable metric space.
Examples of primary metrics are distances between moments of distributions as well as the  $L_q$-\textit{Engineer metric}
\[
\text{EN}(X, Y; q) = \left[ \sum_{i = 1}^{p} \left\arrowvert \E\left(X_i\right) - \E\left(Y_i\right)\right\arrowvert^q\right]^{\min(q, 1/q)} \text{ with } q> 0,
\]
where $X_i, Y_i$ denote the $i$th component of the $p$-dimensional random vectors $X\sim F_1$ and $Y\sim F_2$.\\
A probability (semi)metric $d:\mathcal{P}\to [0,\infty)$ is called a \textit{simple semimetric}, if for each $P\in\mathcal{P}$ with marginals $F_1, F_2$:
\[
F_1 = F_2 \Rightarrow d(P) = 0
\]
and a \textit{simple metric} if the converse implication also holds. Simple metrics as well as primary metrics only depend on the marginals $F_1, F_2$ instead of their joint distribution and can therefore also be denoted by $d(F_1, F_2)$ instead of $d(P)$.
Examples of simple (semi)metrics are the Kantorovich metric \parencite{kantorovich_mathematical_1960}, Prokhorov metric \parencite{prokhorov_convergence_1956}, Birnbaum-Orlicz metric \parencite{birnbaum_uber_1931}, and  Zolotarev's semimetric \parencite{zolotarev_probability_1984}.\\
In the sense of \textcite{rachev_probability_1991}, every probability metric is a \textit{compound metric}. In other papers, the term compound metric is used only for metrics that are not simple \parencite{rachev_probability_1991}. Examples of compound metrics are the Ky Fan metrics \parencite{fan_entfernung_1943}.

Probability metrics can also be classified based on their structure, according to \textcite{zolotarev_probability_1984}. The different structures are presented below, mainly following the overview of \textcite{rachev_probability_1991}. We will denote the sample space by $S$ and assume that it is a metric space with a corresponding metric that we denote by $d^{\prime}$ to avoid confusion with the probability metrics denoted by $d$.

\paragraph{Hausdorff Structure}
Following \textcite{rachev_probability_1991}, a probability semimetric $d$ is said to have a \textit{Hausdoff structure} or a \textit{h-structure} if it can be represented in the following form: 
\[
d(X, Y) = h_{\lambda, \phi, \mathfrak{B}_0}(X, Y) := \max\left\{h^{\prime}_{\lambda, \phi, \mathfrak{B}_0}(X, Y), h^{\prime}_{\lambda, \phi, \mathfrak{B}_0}(Y, X)\right\}, 
\]
where 
\[
h^{\prime}_{\lambda, \phi, \mathfrak{B}_0}(X, Y) = \sup_{A\in \mathfrak{B}_0} \inf_{B\in \mathfrak{B}_0} \max\left\{\frac{1}{\lambda} r(A, B), \phi(X, Y; A, B)\right\}
\]
and 
\[
r(A, B) = \inf\{\varepsilon>0: A^{\varepsilon} \supseteq B, B^{\varepsilon} \supseteq A\}
\] 
is the Hausdorff semimetric in the Borel $\sigma$-algebra $\mathfrak{B}(S)$ with $A^{\varepsilon}$ denoting the open $\varepsilon$-neighborhood of $A$, $\lambda > 0$, $\mathfrak{B}_0 \subseteq \mathfrak{B}(S)$ and $\phi$ such that 
\begin{enumerate}
	\item $\Prob(X = Y) = 1 \Rightarrow \phi(X, Y; A, B) = 0 \,\forall A, B\in \mathfrak{B}_0$, and
	\item $\exists \text{ constant } K_{\phi} \ge 1: \forall A, B, C \in \mathfrak{B}_0 \text{ and random variables } X, Y, Z $\[
	\phi(X, Z; A, B) \le K_{\phi} [\phi(X, Y; A, C) + \phi(Y, Z; C, B)].
	\] 
\end{enumerate}
Examples are the Lévy metric for univariate distributions
\[
L(X, Y) = L(F_1, F_2) = \inf\{\varepsilon>0:F_1(x - \varepsilon) - \varepsilon \le F_2(x) \le F_1(x + \varepsilon) + \varepsilon, x\in \mathbb{R}\}
\]
and the Prokhorov metric $\pi_{\lambda}, \lambda > 0$
\begin{align*}
	\pi_{\lambda}(F_1, F_2) :=& \inf\{\varepsilon > 0: \Prob_1(C)\le\Prob_2(C^{\lambda\varepsilon}) + \varepsilon \text{ for any } C \in \mathcal{C}\}, 
\end{align*}
where $\mathcal{C}$ denotes the set of all nonempty closed subsets of $S$. Every semimetric has the trivial Hausdorff representation $h_{\lambda, \phi, \mathfrak{B}_0} = \mu$ with $\mathfrak{B}_0$ a singleton, e.g.\ $\mathfrak{B}_0 \equiv A_0$ for some set $A_0$, and $\phi(X, Y; A_0, A_0) = d(X, Y)$ \parencite[pp. 51-68]{rachev_probability_1991}. The original definition of \textcite{zolotarev_probability_1984} explicitly excludes the trivial representation.

\paragraph{$\Lambda$-structure}
A semimetric $d$ has a $\Lambda$-structure if there exists a non-negative function $\nu$ that satisfies the conditions 
\begin{enumerate}
	\item $\Prob(X = Y) = 1 \Rightarrow \nu(X, Y; t) = 0 \;\forall t\ge 0$
	\item $\nu(X, Y; t) = \nu(Y, X; t) \;\forall t\ge 0$
	\item $0\le t^{\prime} < t^{\prime\prime} \Rightarrow \nu(X, Y; t^{\prime}) \ge \nu(X, Y; t^{\prime\prime})$
	\item $\nu(X, Z; t^{\prime} + t^{\prime\prime}) \le \nu(X, Y; t^{\prime}) + \nu(Y, Z; t^{\prime\prime}) \;\forall t^{\prime}, t^{\prime\prime} \ge 0$
\end{enumerate}
such that $d$ can be represented as
\[
d(X, Y) = \Lambda_{\lambda, \nu}(X, Y) := \inf\{\varepsilon > 0: \nu(X, Y; \lambda\varepsilon) < \varepsilon\}
\]
for some $\lambda > 0$. \\
It can be shown that each semimetric has a trivial $\Lambda$-structure \parencite[pp. 68-69]{rachev_probability_1991}. Again, this trivial representation is excluded in the original definition by \textcite{zolotarev_probability_1984}. Examples of (semi-)metrics with a $\Lambda$-structure in the strong sense are the Ky Fan metric and the generalized Lévy-Prokhorov metrics. In general, each semimetric with a Hausdoff structure also has a $\Lambda$-structure \parencite[p. 69]{rachev_probability_1991}.

\paragraph{Integral structure}
\textcite{zolotarev_probability_1984} additionally defines (semi)metrics with an integral structure which comprise those (semi)metrics that can be represented as 
\[
d(X, Y) = \psi(\E(\phi(h(X, Y)))),
\]
where $h$ is a metric in $(S, d^{\prime})$ which is a measurable function, $\phi$ is a strictly increasing convex function on $(0,\infty)$ vanishing at zero and $\psi$ is the superposition of a non-decreasing concave function vanishing at zero and the inverse function of $\phi$. All metrics with an integral structure are compound metrics. An example are metrics of the form
\[
\gamma_q(X, Y) = \left[\E\left((d^{\prime})^q(X, Y)\right)\right]^{\min(1, 1/q)}, q > 0.
\]

\paragraph{Integral Probability Metrics / Probability Metrics with a $\xi$-structure}\label{sec.ipm}
Probability Metrics with a $\xi$-structure are better known as Integral probability metrics, going back to \textcite{muller_integral_1997}. They are based on the idea that if two distributions are identical, any function should have the same expectation under both distributions.
Let $\mathcal{F}$ be a set of functions $f: \mathcal{X}\to \mathbb{R}$. Then, an integral probability metric is given by
\[
\text{IPM}_{\mathcal{F}}(F_1, F_2) = \sup_{f\in\mathcal{F}} \left|\int f \dif F_1 - \int f \dif F_2\right|.
\]
All IPMs are probability metrics. \\ 
An example for an IPM is the \textit{Dudley metric $\beta$}, which is generated by $\mathcal{F}_{\beta} = \{f: \|f\|_{\infty} \le 1, \|f\|_{L} \le 1\}$, where $\|\cdot\|_L$ denotes the Lipschitz-norm defined on a metric space $(S, d^{\prime})$ as 
\[
\|f\|_L := \sup_{x\ne y\in S} \frac{|f(x) - f(y)|}{d^{\prime}(x, y)}
\]
and $\|\cdot\|_{\infty}$ denotes the supremum norm. 
Another special case is the \textit{Total Variation Metric} \parencite{zolotarev_probability_1984} 
\[
\sigma(F_1, F_2) := |F_1 - F_2|(S), 
\]
where 
\begin{equation}
	\|\mu\| := |\mu|(S) \label{eq:tot.var}
\end{equation}
denotes the total variation norm on the set of all signed measures on an arbitrary measure space $(S, \mathcal{S})$ with total variation 
\[
|\mu| = \mu^- + \mu^+.
\]
Here $\mu^-$ and $\mu^+$ denote the negative and positive parts of $\mu$, respectively. The total variation metric has generator $\mathcal{F}_{\sigma}:=\{2\cdot\mathbbm{1}_B: B\in \mathcal{S}\}$ since 
\[
\|\mu\| = 2\sup_{A\in \mathcal{S}} |\mu(A)| \, \text{for all signed measures }\mu \text{ with } \mu(S) = 0.
\] It also fulfills the property 
\begin{equation}
	d(F_1 * G, F_2 * G) \le d(F_1, F_2) \,  \text{ for all probability measures } G. \label{C}
\end{equation}
The \textit{stop-loss metric}
\[
d_{\text{SL}}(F_1, F_2) = \sup_{t\in\mathbb{R}} |\E_{F_1}(X - t)^+ - \E_{F_2}(X - t)^+|
\]
is motivated by risk-theoretic considerations \parencite{gerber_introduction_1979, rachev_approximation_1990} and has the generator $\mathcal{F}_{\text{SL}} = \{s\to\Phi_t(s) = (s-t)^+, t\in\mathbb{R}\}$. It fulfills the condition (\ref{C})
and additionally, it holds
\begin{equation}
	d_{\mathcal{F}}(\delta_a, \delta_b) = d^{\prime}(a, b), \label{R}
\end{equation}
where $\delta_x$ denotes the Dirac measure on $x$. More properties for the univariate case are given in \textcite{rachev_approximation_1990}. The last example introduced by \textcite{muller_integral_1997} is the \textit{Kantorovich-Rubinstein metric $\xi_1$} \parencite{zolotarev_probability_1984, dudley_real_1989}, which is generated by the set of Lipschitz functions $\mathcal{L}_1 = \{ \text{Lipschitz functions } f: \|f\|_L\le 1\}$. For $S = \mathbb{R}$ it reduces to 
\[
\xi_1(F_1, F_2) = \ell_1(F_1, F_2) := \int |F_1(t) - F_2(t)| \dif t.
\]
Additional to the conditions (\ref{C}) and (\ref{R}), it fulfills the condition 
\begin{equation}
	d_{\mathcal{F}}(aF_1, aF_2) = a\,d_{\mathcal{F}}(F_1, F_2).
\end{equation}
If $S$ is separable, the Kantorovich-Rubinstein metric is the dual representation of the $L_1$\textit{-Wasserstein distance} 
\[
W_1(F_1, F_2) = \inf_{\pi\in\Pi(F_1, F_2)} \int d(x, y) \dif\pi(x, y),
\]
where $\Pi(F_1, F_2)$ is the set of joint distributions with marginal distributions $F_1$ and $F_2$ (for general Wasserstein distance, see (\ref{eq:wasserstein})). There is a connection between the Kantorovich-Rubinstein metric and optimal transport (see \ref{sec.rank}, \ref{sec.ot}) via the Kantorovich–Rubinstein duality. For details see, e.g.\, \textcite{rachev_mass_1998}.
Other examples are all $L^q$-metrics
\[
\theta_p(F_1, F_2) := \|F_1 - F_2\|_q,
\]
where $q\in[1,\infty]$, as well as the engineer metric \parencite[p. 73]{rachev_probability_1991}.

\textcite{sriperumbudur_empirical_2012} define empirical estimates for the Kantorovich metric, Fortet-Mourier metric, dual-bounded Lipschitz distance (Dudley metric), total-variation distance and kernel distance (Mean Maximum Discrepancy, MMD, see Section \ref{sec.mmd}) that are easily computable and strongly consistent, except for the total-variation distance. They are motivated by their observation that homogeneity tests as an important application are often based on estimates of distances as test statistics. Further, while the MMD and the total variation metric are already successfully applied in this context, most other IPMs are not, due to the lack of good estimates for continuous random variables, especially in the multivariate case. For the application, it is crucial that the statistics have a consistent estimator exhibiting fast convergence behavior and low computational complexity. They show that the estimate for the kernel distance (MMD) in comparison is computationally cheaper, converges at a faster rate to the population value, and its rate of convergence is independent of the dimension $p$ of the space. 
The additional IPMs considered by \textcite{sriperumbudur_empirical_2012} are defined as follows. The \textit{Fortet-Mourier metric} is a generalization of the Kantorovich metric with $\mathcal{F} = \|f\|_c\le 1$, where 
\[
\|f\|_c := \sup\{\frac{|f(x) - f(y)|}{c(x, y)}: x\ne y \in S\}
\]
and $c(x, y) = d^{\prime}(x, y)\max(1, d^{\prime}(x, a)^{q-1}, d^{\prime}(y, a)^{q-1})$ for $q\ge1$ and for some $a\in S$. For $q = 1$, this yields the definition of the Kantorovich metric. The kernel (MMD) distance $\gamma_{\mathcal{F}}$ is defined by setting $\mathcal{F} = \{f:\|f\|_{\mathcal{H}}\le 1\}$, where $\|\cdot\|_{\mathcal{H}}$ denotes the norm on the reproducing kernel Hilbert space (RKHS) $\mathcal{H}$ induced by the kernel function. For details see Section \ref{sec.mmd}.\\
The general empirical estimator for an IPM given samples 
\[
\{X_1, \dots X_{n_1}\} \sim F_1 \text{ and } \{Y_1, \dots Y_{n_2}\} \sim F_2
\]
is defined as 
\begin{equation}
	\gamma_{\mathcal{F}}(F_{1,n_1}, F_{2, n_2}) = \sup_{f\in\mathcal{F}} \left|\sum_{i = 1}^N \tilde{Z}_i f(Z_i)\right|, \label{est.ipm}
\end{equation}
where $F_{1,n_1} := \frac{1}{n_1}\sum_{i = 1}^{n_1} \delta_{X_i}$ and $F_{2,n_2} := \frac{1}{n_2}\sum_{j = 1}^{n_2} \delta_{Y_j}$ denote the empirical distributions of $F_1$ and $F_2$, $N = n_1 + n_2$, with \[Z = \{Z_1,\dots,Z_N\} = \{X_1\dots,X_{n_1}, Y_1,\dots Y_{n_2}\}\] denoting the pooled sample and $\tilde{Z}_i = \frac{1}{n_1}$ when $Z_i =  X_i$ for $i = 1,\dots, n_1$ and  $\tilde{Z}_i = -\frac{1}{n_2}$ when $Z_i =  Y_{i - n_1 + 1}$ for $i = n_1+1,\dots, N$. The computation is not straightforward for arbitrary $\mathcal{F}$, so \textcite{sriperumbudur_empirical_2012} define the empirical estimators for special cases and show how to calculate them via solving linear programs or using a closed-form expression. 

The resulting estimator of the kernel distance can be given as a closed-form expression and therefore is easy to implement compared to the other estimators. Moreover, it is the only presented estimator for which (\ref{est.ipm}) has a unique solution. For the estimators for the Kantorovich and total variation metric, strong consistency is shown under the assumption that $(S, d^\prime)$ is a totally bounded metric space. To show strong consistency of the estimator for the kernel distance, the following assumptions are made: For any $r\ge 1$ and probability measure $F$, define the $L^r$-norm $\|f\|_{F, r} := \left(\int |f|^r \dif F\right)^{1/r}$ and let $L^r(F)$ denote the metric space induced by this norm. The covering number $\mathcal{N}(\varepsilon, \mathcal{F}, L^r(F))$ is the minimal number of $L^r(F)$ balls of radius $\varepsilon$ needed to cover $\mathcal{F}$. $\mathcal{H}(\varepsilon, \mathcal{F}, L^r(F)) := \log\mathcal{N}(\varepsilon, \mathcal{F}, L^r(F))$ is called the entropy of $\mathcal{F}$ using the $L^r(F)$ metric. Define the minimal envelope function as $G(x) := \sup_{f\in\mathcal{F}} |f(x)|$. Under the assumptions
\begin{enumerate}
	\item $\int_S G\dif F_1 < \infty$,
	\item $\int_S G\dif F_2 < \infty$,
	\item $\forall \varepsilon> 0, \frac{1}{n_1} \mathcal{H}(\varepsilon, \mathcal{F}, L^r(F))\stackrel{F_1}{\to} 0 $ as $n_1\to \infty$,
	\item $\forall \varepsilon> 0, \frac{1}{n_2} \mathcal{H}(\varepsilon, \mathcal{F}, L^r(F))\stackrel{F_2}{\to} 0 $ as $n_2\to \infty$,
\end{enumerate}
strong consistency can be shown for the kernel distance estimator. \\
\textcite{sriperumbudur_empirical_2012} also derive convergence rates of the estimators to their population values. For the estimators for the Kantorovich and total variation metric, these convergence rates depend on the dimension $p$, and thus in large dimensions, more samples are needed to obtain useful estimates. The rate for the kernel distance is independent of the dimension $p$. The authors also show how these convergence rates can be used to derive critical values for tests for $H_0: F_1 = F_2$ vs. $H_1: F_1 \ne F_2$. Moreover, the theoretical results on convergence and dependence on $p$ are confirmed by simulations for cases in which the measures can be computed exactly.\\
For the total variation distance, it is shown that the estimator resulting from (\ref{est.ipm}) is not consistent.

\paragraph{Tests based on Wasserstein distances}

\textcite{ramdas_wasserstein_2017} present an overview of tests based on Wasserstein distance (\ref{eq:wasserstein}) and their relationships with each other, as well as with the energy and MMD test (see Section \ref{sec.mmd}) in the multivariate case and the Kolmogorov-Smirnov (KS) test, probability-probability (PP) and quantile-quantile (QQ) plots, and receiver operating (ROC) or ordinal dominance (ODC) curves in the univariate case. They show a connection between the Wasserstein distance and the energy distance via entropic smoothing and then use the connection between energy distance and MMD noted by \textcite{sejdinovic_equivalence_2013}. The tests presented are derived only for one-dimensional data and are therefore not discussed further here.

\textcite{wang_two-sample_2021} propose a test based on the Wasserstein distance (\ref{eq:wasserstein}) for an optimal linear projection of the data. The idea behind this is to circumvent the curse of dimensionality for the Wasserstein distance through a projection of data into a lower-dimensional space. \textcite{wang_two-sample_2022} build on this idea but try to improve the test by using non-linear projections, based on their observation that the original test of \textcite{wang_two-sample_2021} cannot efficiently capture features from data with non-linear patterns. They compare the new test to the MMD (\textcite{gretton_kernel_2006}; Section \ref{sec.mmd}) and ME (\textcite{chwialkowski_fast_2015}; Section \ref{sec:kernel}) test as well as to the old version. Both tests of \textcite{wang_two-sample_2021} and \textcite{wang_two-sample_2022} rely on a train/test split of the data since an optimal projection of the data needs to be learned before performing the test. The projection of \textcite{wang_two-sample_2021} also relies on a kernel and the choice of the kernel is crucial for the test to perform well. No explicit guidelines for the choice of the kernel are given. Moreover, there is no theoretical guarantee of finding the global optimum for the projection in the nonlinear case.

\subsubsection{Divergences}\label{sec:f.div}
There are different definitions of divergences in the literature. Most have in common that a divergence is a discrepancy measure that does not fulfill all criteria for distances or metrics. It is usually required to be a non-negative function.
E.g.\, \textcite{sugiyama_direct_2013} define a divergence $d$ as a pseudo-distance, i.e.\ it acts like a distance but may violate some of the conditions 
\begin{enumerate}
	\item Non-negativity: $\forall X,Y: d(X, Y)\ge 0 $
	\item Non-degeneracy: $d(X, Y) = 0 \Leftrightarrow X = Y$ 
	\item Symmetry: $d(X, Y) = d(Y, X)$
	\item Triangle inequality: $\forall X, Y, Z: d(X, Z)\le d(X, Y) + d(Y, Z)$.
\end{enumerate}
In \textcite{zhao_comparing_2021}, given a finite set or finite-dimensional vector space $\mathcal{X}$ and a set $\mathcal{P}(\mathcal{X})$ of probability distributions on $\mathcal{X}$ with density, a probability divergence is defined as a function $D: \mathcal{P}(\mathcal{X})\times\mathcal{P}(\mathcal{X}) \to \mathbb{R}$ that satisfies 
\begin{align*}
	D(F_1, F_2) &\ge 0\\
	D(F_1, F_1) &= 0 \,\forall F_1, F_2 \in \mathcal{P}(\mathcal{X}).
\end{align*}
$D$ is called \textit{strict} if $D(F_1, F_2) > 0 \,\forall F_1 \ne F_2$ and \textit{non-strict} otherwise.
\paragraph{$f$-Divergences}
$f$-divergences use the idea that two identical distributions assign the same likelihood to every point and thus measure how far the likelihood ratio is from one \parencite{zhao_comparing_2021}. Given a convex continuous function $f:\mathbb{R}_+\to\mathbb{R}$ such that $f(1) = 0$, an $f$-divergence is defined as
\[
D_f(F_1, F_2) = \E_{F_1}\left[f(f_1(X) / f_2(X))\right],
\]
where $f_1, f_2$ are the density functions of distributions $F_1$ and $F_2$ \parencite{csiszar_informationstheoretische_1963, ali_general_1966}. \\
In the following, we will discuss the most important properties and examples of $f$-diver\-gen\-ces. There is extensive literature on $f$-divergences in general and also on more details for special $f$-divergences. For a general discussion of $f$-divergences see e.g.\ \textcite{liese_convex_1987}. For a discussion of special $f$-divergences see the literature cited below and the references therein.

\textcite{vajda_metric_2009} discusses metric properties of $f$-divergences. In general, $f$-divergences do not fulfill the conditions of a metric. Especially the triangle inequality is violated for all $f$-divergences except for the total variation metric (or multiples of it). On the other hand, positive powers of $f$-divergences are probability metrics, if the $f$-divergence itself is symmetric and bounded. For a general $f$-divergence $D_f$, it holds
\[
0\le D_f(F_1, F_2) \le f(0) + f^*(0),
\]
where 
\begin{align*}
	f(0) &= \lim_{t\downarrow0} f(t),\\
	f^*(0) &= \lim_{t\downarrow0}, f^*(t)\\
	f^*(t) &= t f\left(\frac{1}{t}\right), t > 0.
\end{align*}
It holds $D_f(F_1, F_2) = 0$ if and only if $F_1 = F_2$. If $F_1 \perp F_2$ (orthogonal, i.e.\ disjoint supports), then it holds $D_f(F_1, F_2) = f(0) + f^*(0)$. The reverse conclusion holds if the right-hand is finite. An $f$-divergence is symmetric if and only if 
\[
\exists c\in \mathbb{R}: f^*(t) = f(t) + c (t-1).
\]
For symmetric $f$-divergences, the finiteness of $f$ implies the finiteness of $f^*$. For any $f$-divergence, the relation 
\[
D_f(F_1, F_2) = D_{f^*}(F_2, F_1)
\]
holds for any two distributions $F_1, F_2$. The $f$-divergence between the restrictions of two distributions to a sub-$\sigma$-algebra is always smaller or equal to the $f$-divergence of the unrestricted distributions, with equality if this sub-$\sigma$-algebra is sufficient. In the case of a strictly convex function $f$ and a corresponding finite $f$-divergence, the equality is equivalent to sufficiency of the sub-$\sigma$-algebra. There exist representations of each $f$-divergence based on finite, measurable partitions of the sample space for the general case, and in case of a $\sigma$-algebra that is generated by an at most countable partition, it exists a representation in terms of a measurable partition of the sample space.

\textcite{vajda_metric_2009} also gives examples of $f$-divergences and checks if they fulfill the above properties. \\
The \textit{squared Hellinger distance} is defined as
\[
H^2(F_1, F_2) = 2 \int\left(\sqrt{f_1} - \sqrt{f_2}\right)^2 \dif \mu,
\]
where $\mu$ is a $\sigma$-finite measure dominating $F_1$ and $F_2$ w.r.t which the densities $f_1$ and $f_2$ exist. It satisfies all metric conditions in the power $1/2$. The same holds for the \textit{squared Le Cam distance} \textit{(Vincze-Le Cam distance)} \parencite{vincze_concept_1981, cam_asymptotic_1986}
\[
\text{LC}^2(F_1, F_2) = \frac{1}{2}\int \frac{(f_1 - f_2)^2}{f_1 + f_2} \dif \mu.
\]
In contrast, no power of the \textit{Kullback-Leibler divergence} (\textit{information divergence}) \parencite{kullback_information_1951}
\[
\text{KL}(F_1, F_2) := \int \log\left(\frac{f_1(x)}{f_2(x)}\right) f_1(x) \dif x, 
\]
where $f_1, f_2$ denote the density functions of $F_1$ and $F_2$, is a metric, since symmetry is never fulfilled. In addition, powers of the \textit{symmetrized Kullback-Leibler divergence}, which is also known as \textit{Jeffrey's divergence} 
\[
J(F_1, F_2) = \text{KL}(F_1, F_2) + \text{KL}(F_2, F_1),
\]
also do not fulfill the triangle inequality.\\
\textcite{vajda_metric_2009}  introduces the \textit{extended $\phi_{\alpha}$-divergences} $D_{\phi_{\alpha}}(F_1, F_2)$ with
\[
\phi_{\alpha}(t) = \begin{cases}
	\frac{\alpha}{|\alpha|(1-\alpha)}\left[\left(t^{1/\alpha} + 1\right)^{\alpha} - 2^{\alpha - 1} \left(t + 1\right)\right] & \text{ if }\alpha(1-\alpha)\ne0,\\
	t\log(t) + (t + 1)\log\left(\frac{2}{t + 1}\right) & \text{ if }\alpha = 1,\\
	\frac{|t - 1|}{2} & \text{ if }\alpha = 0.
\end{cases}
\]
These are symmetric $f$-divergences with $f = \phi_{\alpha}$ strict convex on $(0,\infty)$ unless $\alpha = 0$. Powers $D(F_1, F_2) = D_{\phi_{\alpha}}(F_1, F_2)^{\pi(\alpha)}$ for 
\[
\pi(\alpha) = \begin{cases}
	\frac{1}{2} & \text{ if } -\infty < \alpha \le 2,\\
	\frac{1}{\alpha} & \text{ if } \alpha > 2,\\
\end{cases}
\]
fulfill all metric properties. Special cases include the total variation, the Hellinger distance, the Le Cam distance, and the Jensen-Shannon divergence \parencite{lin_divergence_1991} (or scaled versions of them).\\
\textcite{liese_convex_1987} define a different class of $f$-divergences, called \textit{$I_{\alpha}$-divergences}. They are generated by the functions 
\[
I_{\alpha}(x) = \begin{cases}
	-\log(x) + x - 1 & \text{ if }\alpha = 0, \\
	\frac{x^{\alpha}- \alpha x + \alpha - 1}{\alpha (\alpha - 1)} & \text{ if } \alpha \ne 0, \alpha \ne 1,\\
	x\log(x) - x + 1, & \text{ if }\alpha = 1,
\end{cases}
\]
with $-\log(0) := \infty$ and $0\log(0) := 0$. A special case is the KL-divergence for $\alpha = 1$. Moreover, the $I_{\alpha}$-divergence is equal to the Rényi divergence of order $\alpha$ for $\alpha\in\{0,1\}$ and $D_{\alpha}(F_1, F_2) = \frac{1}{\alpha(\alpha - 1)} \log\left(1 + \alpha(\alpha - 1) D_{I_{\alpha}}(F_1, F_2)\right)$ for $\alpha > 0, \alpha\ne 1$, where $D_{\alpha}(F_1, F_2)$ denotes the Rényi divergence of order $\alpha$ (see definition below). \\
\textcite{sugiyama_direct_2013} provide a review of recent advances in direct divergence approximation for some $f$-divergences. They define a divergence $d$ as a pseudo-distance, i.e.\ it acts like a distance but may violate some of the conditions.\\
The first divergence considered is the Kullback-Leibler (KL) divergence. It is almost positive definite and is additive for independent random events, i.e.\ the divergence for the joint distributions equals the sum of the divergences for the marginal distributions of both variables. Moreover, the KL divergence is invariant for non-singular transformations \parencite{kullback_information_1951}. Advantages of the KL divergence according to \textcite{sugiyama_direct_2013} are that it is compatible with Maximum Likelihood (ML) estimation, invariant under input metric change, that its Riemannian geometric structure is well studied, and that it can be approximated accurately via direct density ratio estimation. However, it is not symmetric, does not fulfill the triangle inequality, its approximation is computationally expensive due to the $\log$ function, it is sensitive to outliers, and it is numerically unstable because of the strong non-linearity of the $\log$ function and the possible unboundedness of the density-ratio function. \\
The \textit{Pearson (PE) divergence} \parencite{pearson_criterion_1900}, also known as \textit{$\chi^2$ divergence}
\[
\text{PE}(F_1, F_2) := \int f_2(x) \left(\frac{f_1(x)}{f_2(x)} - 1\right)^2 \dif x
\]
is a squared-loss variant of the KL divergence. Since it is also an $f$-divergence like the KL divergence, both share similar theoretical properties. Advantages of the PE divergence according to \textcite{sugiyama_direct_2013} are again invariance under input metric change, that it can be accurately estimated via direct density-ratio estimation, that its estimator can be obtained analytically, so it is computationally much more efficient due to the compatibility of the quadratic function with least squares (LS) estimation. Also, it is more robust against outliers. But it is still not symmetric, also violates the triangle inequality, and the density ratio is possibly unbounded. \\
One way to overcome the possible unboundedness is to use the \textit{Relative Pearson (rPE) divergence} \parencite{yamada_relative_2013}
\[
\text{rPE}(F_1, F_2) := \text{PE}(F_1, F_{\alpha}) = \int f_{\alpha}(x)\left(\frac{f(x)}{f_{\alpha}(x)} - 1\right)^2 \dif x, 
\]
where for $\alpha\in[0,1)$, $f_{\alpha}$ is defined as the density function of the $\alpha$-mixture 
\[
F_{\alpha} = \alpha F_1 + (1 - \alpha) F_2
\]
of $F_1$ and $F_2$. For $\alpha= 0$ this yields the Pearson divergence. The ratio $\frac{f(x)}{f_{\alpha}(x)}$ is called the \textit{relative density-ratio} and is always upper-bounded by $\frac{1}{\alpha}$ for $\alpha>0$. The advantages of the relative Pearson divergence are that it overcomes the unboundedness, is still compatible with LS estimation, and can be approximated in almost the same way as the PE divergence via direct relative density-ratio estimation. Its approximation can still be obtained analytically in an accurate and computationally efficient manner and the rPE is still invariant under input metric change. Its disadvantages are that it violates symmetry and the triangle inequality and that the choice of $\alpha$ may not be straightforward. \\
Lastly, a divergence presented by \textcite{sugiyama_direct_2013} is the \textit{$L^2$-distance} 
\[
L^2(F_1, F_2) := \int \left(f_1(x) - f_2(x)\right)^2 \dif x
\]
which is a standard distance measure between probability measures. It does not belong to the class of $f$-divergences but rather to the class of IPMs as it is the special case of the $L^q$-distances presented above (Section \ref{sec.ipm}) for $q = 2$. Advantages according to \textcite{sugiyama_direct_2013} are that it is a proper distance measure that the density difference is always bounded as long as each density is bounded. Therefore the $L^2$-distance is stable without the need for tuning any control parameter. It is also compatible with LS estimation and can be accurately and analytically approximated in a computationally efficient and numerically stable way via direct density-difference estimation \parencite{sugiyama_density-difference_2013}. In contrast to the aforementioned divergences, the $L^2$-distance is not invariant under input metric changes. \\
Due to the advantages listed before, \textcite{sugiyama_direct_2013} argue that Pearson divergence, relative Pearson divergence, and $L^2$-distance are more useful in practice than the ``overwhelmingly popular'' KL divergence.\\
A naive way to approximate the divergences, given samples $X := \{X_i\}_{i = 1}^{n_1}\sim F_1$ and $Y := \{Y_j\}_{j = 1}^{n_2}\sim F_2$, would be to first obtain estimators for the densities $f_1, f_2$ and then compute a plug-in approximator similar to the methods for comparing density functions. This violates Vapnik's principle of never trying to solve a more general problem as an intermediate step when having a restricted amount of information, so \textcite{sugiyama_direct_2013} argue for the use of direct density-ratio or direct density-difference estimation as an alternative. ``Direct divergence approximators theoretically achieve optimal convergence rates [\dots] and compare favorably with the naive density-estimation counterparts'' \parencite{sugiyama_direct_2013}. They still suffer from the curse of dimensionality. The key idea behind these techniques is to estimate $\frac{f_1}{f_2}$ or $f_1 - f_2$ without explicitly estimating $f_1$ and $f_2$. Therefore, a density-ratio or density-difference model is used. \textcite{sugiyama_direct_2013} make use of the Gaussian density-ratio model 
\[
r(x) = \sum_{l= 1}^n \theta_l \exp\left(-\frac{\|x - X_l\|^2}{2\sigma^2}\right)
\] 
with parameters $\theta_1,\dots,\theta_n$, or of the Gaussian density-difference model 
\[
f(x) = \sum_{l = 1}^{N} \xi_l \exp\left(-\frac{\|x - c_l\|^2}{2\sigma^2}\right), 
\]
where $(c_1,\dots,c_n, c_{n+1},\dots,c_N) = (X_1,\dots,X_{n_1},Y_1,\dots, Y_{n_2})$ are Gaussian centers and $\xi_1,$ $\dots,$ $\xi_N$ parameters of the model. 

There have been different proposals for $f$-divergence estimation before. \\
\textcite{wang_divergence_2005} give an estimator for $f$-divergences for continuous distributions under certain regularity conditions that is based on estimating the density functions using a data-dependent partition of the observation space. Later, \textcite{wang_nearest-neighbor_2006} improved on this method by using nearest neighbor distances instead. This method again only works for continuous distributions. It is shown that the bias and variance of the estimator converge to zero for $n_1, n_2\to \infty$.\\
\textcite{nguyen_estimating_2010} define $M$-estimators for $f$-divergences and likelihood ratios. They make use of an equivalent reformulation of $f$-di\-ver\-gen\-ces, where an $f$-divergence can be seen as the solution to a Bayes decision problem which is a convex optimization problem. \textcite{nguyen_estimating_2010} propose a kernel-based implementation for estimation. They assume equal sample sizes and the concrete form of the estimator is only given for the KL-divergence, although the ideas for generalization to general $f$-divergences with differentiable and strictly convex $f$ are later also presented. Under several assumptions, mainly on the density ratio, consistency and convergence rates can be shown for the estimators.

\textcite{kanamori_f_2012} define tests based on $f$-divergences using density-ratio models for estimation. They need many assumptions for their theory. \textcite{kanamori_f_2012} derive an optimal estimator for $f$-divergences (regarding asymptotic variance) based on a semiparametric density-ratio model. They use this estimator as a test statistic. The critical value is calculated based on the asymptotic $\chi^2$ distribution of the test statistic. The choice of a specific $f$-divergence is left open, but up to first order, the local power does not depend on the chosen $f$-divergence. The choice of the parametric model for density ratio is left open as well. Conditions on the model to obtain optimality of the $f$-divergence estimator are given. Moreover, different examples for choosing the model and $f$-divergence are presented that fulfill the conditions for optimality. Under several additional assumptions, it is shown that the local asymptotic power of the test is equal to that of the empirical likelihood score test of \textcite{fokianos_semiparametric_2001} (not described here due to restrictive assumptions on distributions) if the density ratio model is correctly specified and that the power is larger or equal under certain misspecifications of the density ratio model. To calculate ratios of densities, implicit assumptions are required.

A generalization of $f$-divergences also known as $f$-\textit{dissimilarity} for simultaneously comparing multiple distributions can be found in \textcite{gyorfi_f-dissimilarity_1975}. This extension to the $k$-sample case is discussed in more detail by \textcite{garcia-garcia_divergences_2012}. It is shown to fulfill all properties as the two-sample version, i.e.\ the information processing property, reflexivity, invariance to affine terms, uniqueness, change of order, and bounds hold as for the two-sample case. Other extensions did not keep these properties. Define $\boldsymbol{P}_{[k]} = (P_,\dots,P_k)^T$, $t^j = \frac{1}{\dif P_j} \boldsymbol{P}_{[k]}$ and $\tilde{t}^j = \left(\frac{\dif P_1}{\dif P_j} , \dots, \frac{\dif P_{j-1}}{\dif P_j}, \frac{\dif P_{j+1}}{\dif P_j}, \dots, \frac{\dif P_k}{\dif P_j}\right)^T$. Then, the multi-distribution $f$-divergence or $f$-dissimilarity is defined as 
\[
\mathbb{I}_{\phi, j}(\boldsymbol{P}_{[k]}) = \E_{P_j}[\phi(t^j)] = \E_{P_j}[f_j(\tilde{t}^j)],
\]
where the $j$th distribution is chosen as a reference measure and $f_j \in \mathcal{C}^{k-1}_1$ is a convex function, $\mathcal{C}^k_1 := \{\phi: [0, \infty)^k\to\mathbb{R}, \phi~\text{convex}, \phi(\boldsymbol{1}_k) = 0\}$ such that $f_j(\tilde{t}^j) = \phi(t^j)$. The two expressions are equivalent. The second notation with $k-1$ terms matches the usual $f$-divergence definition for two distributions. The multi-distribution $f$-divergence can be seen as a two-step procedure. First, the probability distributions are relativized by taking Radon-Nikodym derivatives with respect to the chosen reference distribution. Second, the dispersion of the resulting likelihood ratio is measured using the convex function.

\paragraph{Rényi divergence}\phantomsection\label{renyi}
Properties of the \textit{Rényi divergence (of order $\alpha$)} \parencite{renyi_measures_1961} 
\[
D_{\alpha}(F_1, F_2) = \begin{cases} \frac{1}{\alpha - 1}\log\left(\int f_1^{\alpha} f_2^{1-\alpha} \dif \mu\right) & \alpha < 1\\
	\frac{1}{\alpha - 1}\log\left(\int f_1^{\alpha} / f_2^{\alpha - 1} \dif \mu\right) & \alpha >  1,	
\end{cases}
\]
are described in \textcite{van_erven_renyi_2014}. Here, the conventions $0/0 := 0$ and $x / 0 := \infty$, $x > 0$ are applied. Like the Kullback-Leibler divergence, the Rényi divergence is particularly popular in information theory since it can be motivated by coding. The KL divergence is a special case of Rényi divergence for order $\alpha = 1$. The Rényi divergence is only symmetric for order $\alpha = 1/2$, and in that case, it is connected to the squared Hellinger distance. In general, it fulfills the so-called skew symmetry: $D_{\alpha}(F_1, F_2) = \frac{\alpha}{1 - \alpha} D_{1-\alpha}(F_2, F_1)$ for $\alpha\in(0,1)$. For $\alpha= 2$, it is a function of the $\chi^2$-divergence. The Rényi divergence is nondecreasing in its order and it is an upper bound for $\alpha/2$ times the total variation for $\alpha\in(0,1]$. For $F_{1;1}, F_{1;2},\dots$ and $F_{2;1}, F_{2;2},\dots$ and $F_1^N = \bigtimes_{i=1}^N F_{1;i}$, $F_2^N = \bigtimes_{i=1}^N F_{2;i}$ it holds 
\[
\sum_{i = 1}^N D_{\alpha}(F_{1;i}, F_{2;i}) = D_{\alpha}(F_1^N, F_2^N)
\]
for any $\alpha\in[0,\infty]$ and any $N\in\mathbb{N}$ as well as for any $\alpha\in(0,\infty]$ and $N\in\mathbb{N}\cup\{\infty\}$ (additivity). Furthermore, the dominance of measures can be characterized by $F_1 \ll F_2$ if and only if $D_0(F_1, F_2) = 0$. On the other hand it holds that $F_1\perp F_2$ if and only if $D_{\alpha}(F_1, F_2) = \infty$ for some $\alpha\in[0,1)$ or (equivalently) all $\alpha\in[0,\infty]$. Rényi divergence is nonnegative for all $\alpha\in[0,\infty]$, but $D_{\alpha}$ is neither a metric nor the square of a metric for any order. Similar to $f$-divergences, the Rényi divergence between the restrictions of two distributions to a sub-$\sigma$-algebra is always smaller or equal to the Rényi divergence of the unrestricted distributions. For $\alpha >0$, the Rényi divergence is equal to 0 if and only if the distributions are the same. For $\alpha = 0$, $D_{\alpha}(F_1, F_2) = 0$ if and only if $F_1\ll F_2$. \textcite{van_erven_renyi_2014} extend the Rényi divergence to negative orders. The results for positive orders carry over to negative orders with reversed properties in most cases.

\paragraph{Relative information of type $s$}\phantomsection\label{taneja_relative_2004}
\textcite{taneja_relative_2004} give a generalization of the KL-divergence similar to the Rényi divergence that is defined only for discrete distributions. In addition, they present an overview of inequalities between $f$-divergences as well as Rényi divergences and their class of \textit{relative information of type $s$}.

\paragraph{$H$-divergence}\phantomsection\label{zhao_comparing_2021}
In \textcite{zhao_comparing_2021}, given a finite set or finite-dimensional vector space $\mathcal{X}$ and a set $\mathcal{P}(\mathcal{X})$ of probability distributions on $\mathcal{X}$ that have a density, a probability divergence is defined as a function $D: \mathcal{P}(\mathcal{X})\times\mathcal{P}(\mathcal{X}) \to \mathbb{R}$ that satisfies 
\begin{align*}
	D(F_1, F_2) &\ge 0\\
	D(F_1, F_1) &= 0 \;\forall F_1, F_2 \in \mathcal{P}(\mathcal{X}).
\end{align*}
$D$ is called \textit{strict} if $D(F_1, F_2) > 0 \;\forall F_1 \ne F_2$ and \textit{non-strict} otherwise. Different probability divergences are presented. 
The class of $H$-divergences is introduced by \textcite{zhao_comparing_2021}. It makes use of $H$-entropies. The idea is that distributions are different if the optimal decision loss is higher on their mixture than on each individual distribution, so the generalized entropy of the mixture distribution $(F_1 + F_2) /2$ is compared to the generalized entropy of $F_1$ and $F_2$. If $F_1$ and $F_2$ are different, it is more difficult to minimize the expected loss under the mixture, hence it should have higher generalized entropy. If the distributions are identical, the mixture is identical to $F_1$ and to $F_2$ and they all have the same generalized entropy. \\
For an action space $\mathcal{A}$ and loss function $\ell:\mathcal{X}\times \mathcal{A}\to \mathbb{R}$ a corresponding $H$-entropy
\[
H_\ell (F) = \inf_{a\in\mathcal{A}} \E_{F}[\ell(X, a)]
\]
is the Bayes optimal loss of a decision maker who must select some action $a$ not for a particular $x$, but in expectation for a random $X$ drawn from $F$. Examples are the Shannon Entropy, where $\mathcal{A} = \mathcal{P}(\mathcal{X})$ is the set of probabilities and  $\ell(x, a) = -\log a(x)$, the variance with $\mathcal{A} = \mathcal{X}$ and $\ell(x, a) = \|x - a\|_2^2$, and the predictive V-entropy with $\mathcal{A} \subset \mathcal{P}(\mathcal{X}) $ some subset of distributions and $\ell(x, a) = -\log a(x)$.\\
Using $H$-entropies, a new class of discrepancies based on optimal loss for decision tasks can be defined. For two distributions $F_1$ and $F_2$ on $\mathcal{X}$ and a continuous function $\phi:\mathbb{R}^2\to\mathbb{R}$ such that $\phi(\theta, \lambda) > 0$ whenever $\theta + \lambda > 0$ and $\phi(0, 0) = 0$
\[
D_{\ell}^{\phi}(F_1, F_2) = \phi\left(H_{\ell}\left(\frac{F_1 + F_2}{2}\right) - H_{\ell}(F_1), H_{\ell}\left(\frac{F_1 + F_2}{2}\right) - H_{\ell}(F_2)\right)
\]
is a $H$-divergence. The term $H_{\ell}\left(\frac{F_1 + F_2}{2}\right) - H_{\ell}(F_i)$ measures how much more difficult it is to minimize loss on the mixture distribution than on $F_i$, and $\phi$ maps the differences to a scalar divergence. Special cases of the general definition corresponding to particular $H$-entropies are the $H$-Jensen Shannon divergence, where $\phi(\theta, \lambda) = \frac{\theta + \lambda}{2}$ such that \[D_{\ell}^{\text{JS}} (F_1, F_2) = H_{\ell}\left(\frac{F_1 + F_2}{2}\right) - \frac{1}{2}\left(H_{\ell}(F_1) + H_{\ell}(F_2)\right),\] 
and the $H$-Min divergence, where $\phi(\theta, \lambda) = \max(\theta, \lambda)$ such that 
\[
D_{\ell}^{\text{Min}} = H_{\ell}\left(\frac{F_1 + F_2}{2}\right) - \min\left(H_{\ell}(F_1), H_{\ell}(F_2)\right).
\] 
Also, all squared MMD distances (\textcite{gretton_kernel_2006}; see Section \ref{sec.mmd}) are $H$-di\-ver\-gen\-ces.\\
Each $H$-divergence is a probability divergence. Whether the $H$-divergence is strict depends on the choice of $\ell$.\\
Given $n$ i.i.d. samples $\{X_1,\dots,X_n\}$ drawn from $F_1$ and $\{Y_1,\dots,Y_n\}$ drawn from $F_2$, an empirical estimator is given by
\begin{align*}
	\hat{D}_{\ell}^{\phi}(\hat{F}_1, \hat{F}_2) = \phi&\left(\inf_a \frac{1}{n} \sum_{i = 1}^{n} \ell\left(\tilde{Z}_i, a\right) - \inf_a \frac{1}{n} \sum_{i = 1}^{n} \ell\left(X_i, a\right),\right.\\
	& ~~~\left.\inf_a \frac{1}{n} \sum_{i = 1}^{n} \ell\left(\tilde{Z}_i, a\right) - \inf_a \frac{1}{n} \sum_{i = 1}^{n} \ell\left(Y_i, a\right)\right),
\end{align*}
with $\tilde{Z}_i = X_i b_i + Y_i(1 - b_i)$ and $b_i$ i.i.d. uniformly sampled from $\{0,1\}$ such that $\tilde{Z_i}$ is a sample from the mixture $(F_1 + F_2) / 2$. This estimator is consistent under certain regularity assumptions.\\
The $H$-divergence can be used in a permutation test for $H_0: F_1 = F_2$. The same holds for other divergences. A simulation study is performed by \textcite{zhao_comparing_2021} with $\phi(\theta, \lambda) = \left(\frac{\theta^s + \lambda^s}{2}\right)^{1/s}$, $s > 1$ and $\ell(x, a)$ the negative log-likelihood of $x$ under distribution $a$, $a\in\mathcal{A}$ with $\mathcal{A}$ a certain model family (mixture of Gaussian distributions, Parzen density estimator, Variational Autoencoder). $s$ and $\mathcal{A}$ are tuned on a training dataset and power is evaluated on test data. The test is compared to the ones based on the deep kernel MMD \parencite{liu_learning_2020}, the optimized kernel MMD \parencite{gretton_optimal_2012} as well as the ME and SCF test \parencite{chwialkowski_fast_2015, jitkrittum_interpretable_2016} and the tests using optimized frequencies \parencite{lopez-paz_revisiting_2017, cheng_classification_2022}. The test based on $H$-divergence shows the highest power under the same experimental setup that was used by \textcite{liu_learning_2020} in their experiments.

\paragraph{Distance for Probability Measures Based on Level Sets}\phantomsection\label{munoz_new_2012}
\textcite{munoz_new_2012} consider a vector space of test functions where each distribution is seen as a continuous linear functional on this space $\mathcal{D}$. In this setting, they view a probability measure as a Schwartz distribution (generalized function) $F: \mathcal{D} \to \mathbb{R}$ by setting $F(\phi) = \langle F, \phi\rangle = \int \phi \dif F = \int \phi(x) f(x) \dif\mu(x) = \langle \phi, f\rangle$, where $f$ is the density function w.r.t. the ambient measure $\mu$. Then, two probability distributions viewed as linear functionals are the same (similar) if they behave identically (similarly) on all $\phi\in\mathcal{D}$. Thus, the distance between two distributions can be measured as the differences between functional evaluations for an appropriately chosen set of test functions. \textcite{munoz_new_2012} use indicator functions of $\alpha$-level sets and define a distance of distributions by weighting the distances between the integrals of these functions w.r.t. the distributions. An $\alpha$-level set is defined as $S_{\alpha}(f) = \{x\in\mathcal{X} | f(x) \ge \alpha\}$ such that $\Prob(S_{\alpha}(f)) = 1-\nu$ for $\nu\in (0,1)$. Consider sets of the type $A_i(F) = S_{\alpha_i}(f)\setminus S_{\alpha_{i + 1}}(f), i = 1,\dots, n-1$, for a sequence $0 < \alpha_1 <\dots< \alpha_n < 1$. Then it holds for $n\to\infty$ that $A_i(F_1) = A_i(F_2) ~\forall i \Rightarrow F_1 = F_2$. \textcite{munoz_new_2012} consider $\phi_{1i} = \mathbbm{1}_{A_i(F_1)\setminus A_i(F_2)}$ and $\phi_{2i} = \mathbbm{1}_{A_i(F_2)\setminus A_i(F_1)}$ and $d_i(F_1, F_2) = |\langle F_1, \phi_{1i}\rangle - \langle F_2, \phi_{1i}\rangle| + |\langle F_1, \phi_{2i}\rangle - \langle F_2, \phi_{2i}\rangle|$ which is approximately equal to $\mu(A_i(F_1)\triangle A_i(F_2))$, where $A\triangle B = (A\setminus B) \cup (B\setminus A)$ denotes the symmetric difference of two sets. Then, the \textit{weighted level-set distance} is defined as 
\[
d_{\boldsymbol{\alpha}}(F_1, F_2) = \sum_{i = 1}^{n - 1} \alpha_i \frac{\mu(A_i(F_1)\triangle A_i(F_2))}{\mu(A_i(F_1)\cup A_i(F_2))},
\] 
where $\boldsymbol{\alpha} = \{\alpha_{(i)}\}_1^n$ and $\mu$ is the ambient measure. An estimator for the weighted level-set distance is given by 
\[
\hat{d}_{\boldsymbol{\alpha}}(F_1, F_2) = \sum_{i = 1}^{n - 1} \alpha_i \frac{ \#\left(\hat{A}_i(F_1)\triangle^S \hat{A}_i(F_2)\right)}{ \#\left(\hat{A}_i(F_1)\cup \hat{A}_i(F_2)\right)},
\] 
where $\hat{A}_i(F) = \hat{S}_{\alpha_i}(f)\setminus \hat{S}_{\alpha_{i + 1}}(f)$ are estimators of the sets $A_i$, $\#A$ denotes the number of points in $A$ and $\triangle^S$ denotes the set estimate of the symmetric difference. $d_{\boldsymbol{\alpha}}(F_1, F_2)$ and its estimator $\hat{d}_{\boldsymbol{\alpha}}(F_1, F_2)$ are both semimetrics. Estimation of level sets by using a Support Neighbor Machine \parencite{munoz_estimation_2006} and estimation of the symmetric difference between sets by using a covering of the points with closed balls to circumvent that the intersection of the observed sets is empty are described in \cite{munoz_new_2012}.\\
Building up on this, \cite{munoz_new_2013} propose another weighting in the weighted level-set distance.

\paragraph{Dataset distance based on reproducing kernel Hilbert spaces (RKHS)}\phantomsection\label{munoz_new_2013}
\textcite{munoz_new_2013} make use of a kernel to define a dataset distance. Consider a set $A$ of points generated from a distribution $F$ with sample space $\mathcal{X}$. Then, a point $y\in2^{\mathcal{X}}$ (power set of $\mathcal{X}$) is called \textit{indistinguishable} from $x\in A$ with respect to $F$ in the set $A$, $y \stackrel{A(F)}{=} x$, if $d(x, y) \le r_A$, where $r_A = \min d(x_l, x_s), x_l, x_s\in A$, denotes the minimum resolution for the dataset $A$. The idea is to build kernel functions for two datasets $A$ and $B$ from distributions $F_1$ and $F_2$ such that the kernel takes the value one for points that are indistinguishable w.r.t. $F_1$ in $A$ or w.r.t. $F_2$ in $B$, and zero otherwise. This is fulfilled for the \textit{distributional indicator kernel}. Given datasets $A$ sampled from $F_1$ and $B$ sampled from $F_2$, define $K_{A,B}: \mathcal{X}\times \mathcal{X} \to [0,1]$ as 
\[
K_{A, B}(x, y) = f_{x, r_A,\gamma}(y) + f_{y, r_B,\gamma}(x) - f_{x, r_A,\gamma}(y)f_{y, r_B,\gamma}(x), 
\]
where the \textit{smooth indicator functions} with center $x$ for $r>0$ and $\gamma>0$ is defined as 
\[
f_{x, r,\gamma}(y) = 
\begin{cases}
	\exp\left(-\frac{1}{\left(\|x-y\|^\gamma - r^\gamma\right)^2} + \frac{1}{r^{2\gamma^2}}\right) & \text{if } \|x-y\| < r\\
	0 & \text{otherwise}
\end{cases},
\]
$r_A = \min d(x_l, x_s), x_l, x_s\in A$, $r_B = \min d(y_l, y_s), y_l, y_s\in B$, and $\gamma$ is a shape parameter. With this, a kernel for datasets $C$ and $D$ in $2^{\mathcal{X}}$ can be defined as 
\[
K(C, D) = \sum_{x\in C} \sum_{y\in D} K_{A, B}(x, y).
\]
For $C = A$ and $D = B$, $\mu_{K_{A,B}}(A\cap B) = K(A, B)$ can be interpreted as a measure for $A\cap B$ by counting the common points. With $\mu_{K_{A,B}}(A\cup B) = N = \#(A\cup B)$, it follows that  $\mu_{K_{A,B}}(A\triangle B) = N - \mu_{K_{A,B}}(A\cap B)$. In general, $\mu_{K_{C,D}}(C\cap D) = K(C, D)$ can be interpreted as a measure for $C\cap D$ by counting the common points using $\stackrel{A(F_1)}{=}$ and $\stackrel{B(F_2)}{=}$ as equality operators, so taking the distance between $C$ and $D$ is conditioned to a resolution level ($r_A$ and $r_B$) determined by $A$ and $B$.
Therefore, the kernel distance between datasets is defined as 
\[
d_K(C, D) = 1 - \frac{K(C, D)}{N}, 
\]
where $N = \#(C\cup D)$. This is a semimetric. For $C=A$ and $D=B$ and the sizes for both sets increasing, it holds $\mu_{K_{A,B}}(A\cap B) \stackrel{n_1, n_2\to\infty}{\to} \mu(A\cap B)$ and $\mu_{K_{A,B}}(A\cup B) \stackrel{n_1, n_2\to\infty}{\to} \mu(A\cup B)$, so the limit of the kernel distance is the Jaccard distance for datasets, that is $1-\frac{\mu(A \cap B)}{\mu(A\cup B)}$. This divergence can also be interpreted as a kernel-based method.

\subsection{Graph-based methods}
In the following, methods using similarity graphs are presented. First, graph-based methods based on different similarity graphs are discussed. Then, methods based on the important case of the nearest neighbor graph are shown.
\subsubsection{General graph-based methods}
Graph-based methods to compare distributions are especially popular in testing. \textcite{arias-castro_consistency_2016} present a general framework of graph-based tests: recall that the pooled sample is defined as \[\{Z_1,\dots,Z_N\} = \{X_1,\dots,X_{n_1}, Y_1,\dots, Y_{n_2}\}.\] Let $\mathcal{G}$ be a directed graph with this pooled sample as the node set and write $Z_i\to Z_j$ if there is an edge from $Z_i$ to $Z_j$ in $\mathcal{G}$. Reject $H_0$ for small values of 
\[	T_{\mathcal{G}}(Z) = \#\{i\le n_1, j > n_1: Z_i\to Z_j\} + \#\{i\le n_1, j > n_1: Z_j\to Z_i\},
\]
that is the number of neighbors in the graph from different samples. Many of the methods presented below fall within this framework. For the $K$-nearest neighbor graph as $\mathcal{G}$ under the assumption of distinct values in the pooled sample, the test by \textcite{schilling_multivariate_1986} is given which is a special case of the general approach of \textcite{friedman_nonparametric_1973}. The minimum spanning tree starting with the complete graph weighted by Euclidean distances on the other hand results in the multivariate runs test of \textcite{friedman_multivariate_1979}. A minimum distance matching gives the test by \textcite{rosenbaum_exact_2005}.

\textcite{mukhopadhyay_nonparametric_2020} also try to generalize different graph-based tests into a single framework. They note that the tests by  \textcite{weiss_two-sample_1960}, \textcite{friedman_multivariate_1979}, \textcite{chen_new_2017}, \textcite{chen_weighted_2018}, \textcite{rosenbaum_exact_2005}, and \textcite{biswas_distribution-free_2014} have the following steps in common: 
\begin{enumerate}
	\item Construct a weighted undirected graph $\mathcal{G}$ based on pairwise Euclidean distances on the pooled sample.
	\item Compute a subgraph $\mathcal{G}^*$ that contains a certain optimal subset of edges (e.g.\ shortest Hamiltonian path).
	\item Compute cross-match statistics by counting the number of edges between samples from two different populations.
\end{enumerate}
All of the tests mentioned will be described in more detail below.

\paragraph{Tests based on minimal spanning trees (Friedman-Rafsky test)}\phantomsection\label{friedman_multivariate_1979}
One of the first and best-known graph-based tests is the multivariate runs test by \textcite{friedman_multivariate_1979}. It generalizes the Wald-Wolfowitz runs test to the multivariate domain based on a minimal spanning tree of pooled sample points. \textcite{henze_multivariate_1999} proved later that the test is asymptotically distribution-free and universally consistent. \textcite{chen_ensemble_2013} on the other hand observe that power decreases with the imbalance of sample sizes in their simulations. \textcite{chen_weighted_2018} investigate this problem in more detail. \textcite{biswas_nonparametric_2014} highlight that the test is rotation invariant and invariant under location change and homogeneous scale transformation and that it can be used even when the dimension of data is larger than the sample size, but they also derive sufficient conditions for failure for $p \to \infty$. \textcite{biswas_distribution-free_2014} give sufficient conditions for failure where power converges to 0 as $p\to\infty$ and criticize that the test is not distribution-free for finite samples. \textcite{sarkar_graph-based_2020} again show situations where power is very low and formal conditions under which power decreases to 0 for increasing $p$. \textcite{chen_new_2017} observe that in practice (simulations), the test has low or even no power for scale alternatives when the dimension is moderate to high unless the sample size is ``astronomical'' due to the curse of dimensionality.\\
\textcite{friedman_multivariate_1979} propose a second test in their paper that is a generalization of the KS test to the multivariate domain based on a minimal spanning tree of pooled sample points. The test needs a reasonable distance measure between points. An approximation of the null distribution is used for testing. \textcite{friedman_multivariate_1979} themselves show that the test has either no power for scale-only or no power for location-only alternatives. In addition to that, \textcite{chen_graph-based_2013} demonstrate that the test does not work well on categorical data due to ties. \\
Both tests of \textcite{friedman_multivariate_1979} are implemented in the \texttt{R} package \texttt{GSAR} \parencite{GSAR} and \texttt{gTests} \parencite{gTests}. Note that in the \texttt{GSAR} implementation the test statistic is standardized by empirical mean and standard deviation instead of the theoretical values under $H_0$ as in the original definition.

\paragraph{Tests based on optimal non-bipartite matching (Rosenbaum's cross-match test)}\phantomsection\label{rosenbaum_exact_2005}
Another well-known test is the cross-match test by \textcite{rosenbaum_exact_2005}. Here, an optimal non-bipartite matching is formed in the pooled sample based on the inter-point distances. The number of pairs containing one observation from the first distribution and one from the second is considered as a test statistic. Consistency and the asymptotic distribution of the test statistic are shown under the assumption of discrete distributions with finite support. The computational cost for finding an optimal non-bipartite matching of $N$ subjects is $\mathcal{O}(N^3)$. In case of an odd pooled sample size $N$, one observation needs to be discarded. The test is not applicable for partially ordered responses. \textcite{chen_graph-based_2013} note that the test does not work well on categorical data due to ties and show via simulation that power decreases with the imbalance of sample sizes. In general, different distance measures can be used for the optimal non-bipartite matching. \textcite{biswas_nonparametric_2014} note that the test can be used even when the dimension of data is larger than the sample size if the Euclidean distance is used. Due to the high computational cost, a greedy algorithm that reduces the cost to $\mathcal{O}(N^2)$ might be employed, but \textcite{huang_efficient_2017} point out that solving with the greedy heuristic does not guarantee finding the optimum. Furthermore, \textcite{sarkar_graph-based_2020} show situations where power is very low. 
On the other hand, \textcite{deb_multivariate_2021} state that the test by \textcite{rosenbaum_exact_2005} is one of two tests for the multivariate two-sample problem that is exactly distribution-free, computationally feasible, and consistent against all alternatives. 
Consistency against all fixed alternatives is shown by \textcite{arias-castro_consistency_2016}. For the consistency of the general cross-match statistic, they need the assumptions that densities w.r.t. Lebesgue measure of both distributions exist, that the sample sizes are comparable in the sense that their ratio converges to a fixed constant in $(0,1)$, the assumption of bounded out- and in-degree in the graph as well as that the out-degree is essentially constant and long edges are essentially absent and that the dimension is constant. They note that the graph-based setting exhibits a typical curse of dimensionality although literature is silent on that topic. The required conditions cover the minimum spanning tree used by \textcite{friedman_multivariate_1979} \parencite{henze_multivariate_1999}, nearest-neighbor graphs \parencite{schilling_multivariate_1986} and general matchings (shown here). Additionally, \textcite{arias-castro_consistency_2016} show that the null distribution as studied by \textcite{rosenbaum_exact_2005} and \textcite{heller_sensitivity_2010} is available in closed form and coincides with the permutation distribution. The test is implemented in the \texttt{R} package \texttt{crossmatch} \parencite{crossmatch}.

\paragraph{Extensions of Friedman-Rafsky and Rosenbaum tests}\phantomsection\label{chen_graph-based_2013}
\textcite{chen_graph-based_2013} propose a graph-based test for categorical data with a large number of categories and a sparsely populated contingency table that extends the Fried\-man-Rafsky and Rosenbaum tests to categorical data. They assume that a distance matrix is given on the set of categories and that there are not many ties in these distances. In the presence of ties, the resulting graphs are not unique anymore and the number of possible graphs grows fast with the number of ties. Their tests work by either averaging over all optimal graphs for a certain graph type (e.g.\ MST) or by taking the union of all optimal graphs. An analytic form and an asymptotic form for both types of tests are proposed. The analytic form of their first proposed test requires the enumeration of all MSTs on categories which might not be computationally feasible but can be bypassed by assuming that instead of the distance matrix, the similarity is directly represented by a graph with the categories as nodes. For the asymptotic normality of the test statistics, assumptions on cell counts and graph structure are required and the number of categories has to go to infinity. The computational cost of the test is $\mathcal{O}(K^2)$ with $K$ number of categories for the Rosenbaum version resp. $\mathcal{O}(M)$ with $M$ number of minimum spanning trees on categories for the Friedman-Rafsky version or $\mathcal{O}(K^2)$ for the bypassed version. The tests are implemented in the R package \texttt{gTests} \parencite{gTests}.

\paragraph{Tests based on the shortest Hamilton path}\phantomsection\label{biswas_distribution-free_2014}
\textcite{biswas_distribution-free_2014} present a multivariate generalization of two-sample run tests based on the shortest Hamilton path using Euclidean distances that is applicable to high-dimensional data and small sample sizes. It is also invariant under location change, rotation, and homogeneous scale transformations and distribution-free in finite-sample situations. Finding the shortest Hamilton path is NP-complete for complete graphs so instead a heuristic search algorithm is used that does not always yield the optimum. Under several assumptions, consistency for $p \to \infty$ is shown for the test, but \textcite{sarkar_graph-based_2020} show situations where power is very low. They also note that the computational complexity is still $\mathcal{O}(N^2\log N)$ with the heuristic method based on Kruskal's algorithm. \textcite{deb_multivariate_2021} therefore conclude that the test is extremely expensive to compute and possibly not applicable even for moderate sample sizes.

\paragraph{Tests based on orthogonal perfect matchings, minimum spanning tree or nearest neighbors}\phantomsection\label{petrie_graph-theoretic_2016}
\textcite{petrie_graph-theoretic_2016} presents tests based on new graphs using orthogonal perfect matchings as well as on minimum spanning trees or nearest neighbors. The construction for the new graph type works by first finding the optimal perfect matching on the data, then finding the optimal perfect matching without the edges from the first matching, and so on until the $K$th matching is reached. The graph is then given as the union of these matchings. $K\approxeq 0.15N$ is suggested as a heuristic choice. The test is intended for continuous data. It can be used to compare multiple samples. An asymptotic normal test is proposed that is claimed to have good HDLSS performance. \textcite{mukherjee_distribution-free_2022} on the other hand observe that it tends to have low power as the dimension and/ or number of samples to be compared increase and point out that its mathematical properties have not been investigated. The test for Euclidean data and using the optimal non-bipartite matching as a graph is implemented in the \texttt{R} package \texttt{multicross} \parencite{multicross}.

\paragraph{Test based on similarity graphs}\phantomsection\label{chen_new_2017}
\textcite{chen_new_2017} propose a new test based on a similarity graph constructed over the pooled sample that has higher power for differences in location as well as in scale in contrast to former tests that often only have high power for one of those. Consistency is only shown for continuous distributions that differ on a set of positive measures. Additionally, certain conditions for the similarity graph are required that are fulfilled by a $K$-MST with $K = \mathcal{O}(1)$. The new test statistic is given as the quadratic form of the vector of the numbers of edges connecting observations within the same sample for both samples centered with its expectation under the permutation null distribution and the inverse covariance matrix of these numbers under the permutation null distribution. The test statistic cannot be determined if all nodes in the chosen graph have the same degree or if the graph is star-shaped since in these cases the covariance matrix is singular. \textcite{chen_new_2017} recommend not to use the test if the graph is very close to one of these cases since then the inversion of the covariance matrix is already ill-conditioned. The test is implemented in the package \texttt{gTests}\parencite{gTests}

\paragraph{Extension of edge-count tests for imbalanced data}\phantomsection\label{chen_weighted_2018}
\textcite{chen_weighted_2018} aim to improve edge-count tests for unequal sample sizes by weighting. They show consistency only for continuous distributions and if the graph is the MST based on Euclidean distance. The test is also only more powerful than former edge-count tests under locational alternatives, not under general alternatives. The test is implemented in the \texttt{R} package \texttt{gTests}\parencite{gTests}. \textcite{pan_ball_2018} observes that the efficiency of the test is limited by the choice of the number of neighbors. 

\paragraph{Extension of edge-count tests for categorical data}\phantomsection\label{zhang_graph-based_2019}
	\textcite{zhang_graph-based_2019} extend the generalized edge-count test of \textcite{chen_new_2017} and the weighted edge-count test of \textcite{chen_weighted_2018} for categorical data following the approach of \textcite{chen_graph-based_2013}, i.e.\ using either the union of all optimal graphs or averaging over the edge-counts of all optimal graphs. Additionally, they propose a new group of graph-based tests for categorical data, the \textit{extended max-type edge-count tests}. These consist of two components. First, the weighted edge-count statistic of the extension of the test of \textcite{chen_weighted_2018} standardized by its mean and variance under $H_0$ and multiplied with a factor $\kappa$ is considered. Second, the absolute difference of the edge counts of points within the first sample that are connected by an edge and points in the second sample that are connected by an edge, again standardized by its mean and variance under $H_0$, is taken into account. The test statistic of the extended max-type edge-count test is then given by the maximum of these two. Again, two versions are proposed. The first version is based on the union of all optimal graphs and the second version is based on averaging over all optimal graphs. The resulting tests are claimed to be effective for both location and scale alternatives. Without prior knowledge about the difference between the distributions, a choice of $\kappa\in\{1.31, 1.14, 1\}$ is recommended based on a small simulation study. Asymptotic null distributions are derived for all proposed statistics under several conditions on the similarity graph and the class distributions. All asymptotic as well as permutation versions of the tests are implemented in the \texttt{R} package \texttt{gTests}\parencite{gTests}. For numerical data, the newly proposed max-type test is also implemented in this package.

\paragraph{Extensions of nearest neighbor, Rosenbaum, and Friedman-Rafsky test}\phantomsection\label{sarkar_graph-based_2020}
\textcite{sarkar_graph-based_2020} modify graph-based tests to overcome weak performance caused by distance concentration to achieve higher power for high-dimensional data with low sample sizes (HDLSS). Under certain assumptions including uniformly bounded fourth moments, the order of correlations between inter-point distances and the convergence of mean distances and traces of covariance matrices for $p\to\infty$, consistency is shown under $p \to \infty$ for fixed sample size for the modified tests of \textcite{biswas_distribution-free_2014} and \textcite{rosenbaum_exact_2005} and under additional assumptions on sample size also for the nearest neighbor test of \textcite{henze_multivariate_1988} and \textcite{schilling_multivariate_1986} as well as the MST-run test of \textcite{friedman_multivariate_1979}. Depending on the choice of distance, the computation of distances between two points has a cost of $\mathcal{O}(pN)$ compared to $\mathcal{O}(p)$ for Euclidean distance or other distances.

\paragraph{Power of tests}
\textcite{bhattacharya_asymptotic_2020} presents results regarding the limiting distribution under general alternatives as well as power and consistency for Friedman-Rafsky and nearest neighbor tests. The results are based on several different assumptions.

\paragraph{Test based on orthonormal polynomials}\phantomsection\label{mukhopadhyay_nonparametric_2020}
A new class of distribution-free tests for the high-dimensional $k$-sample problem based on new nonparametric tools and connections with spectral graph theory is proposed by \textcite{mukhopadhyay_nonparametric_2020}. 
Their motivation is that classical multivariate rank-based $k$-sample tests like \textcite{puri_nonparametric_1971} or \textcite{oja_multivariate_2004} are not applicable for $p>N$. They demand several desirable properties of tests:
\begin{enumerate}
	\item should be robust and not unduly influenced by outliers (current methods even perform poorly for datasets that are contaminated by only a small percentage of outliers)
	\item should allow testing beyond location-scale alternatives
	\item valid for a combination of discrete and continuous covariates
	\item should provide insight into why the hypothesis was rejected
	\item should work for $k$-sample problem (all former tests only work for two-sample).
\end{enumerate}
For their test, a nonparametrically designed set of orthogonal functions (LP polynomials) is obtained by orthonormalizing a set of functions constructed as orthonormal polynomials of mid-distribution transforms. These are used for the construction of a polynomial kernel of degree 2 that encodes the similarity between two $p$-dimensional data points in the LP-transformed domain. 
The values of the kernel Gram matrix are then used as weights on a graph with the pooled sample as vertices. The idea is to cluster points for the graph into $k$ groups that have higher connectivity and compare how closely related this clustering is to the true memberships to the $k$ distributions. Then testing for homogeneity becomes a problem of testing independence which can be accomplished by determining whether all of the LP comeans are zero. The test statistic has an asymptotic $\chi^2_{(k-1)^2}$ distribution. It has to be explicitly chosen for which moments to test for equality. Implicitly this requires the corresponding moments of the distributions to exist. The test is implemented in the \texttt{LPKsample} package \parencite{LPKsample} in \texttt{R}.

\paragraph{Extension of Rosenbaum test for $k$-sample problem}\phantomsection\label{mukherjee_distribution-free_2022}
\textcite{mukherjee_distribution-free_2022} propose a generalization of the test by \textcite{rosenbaum_exact_2005} to the $k$-sample problem that is exactly distribution-free. It can be applied if inter-point distances are well-defined. If distributions have densities w.r.t. Lebesgue measure on $\mathbb{R}^p$, the test is universally consistent. As for the original test by \textcite{rosenbaum_exact_2005}, the pooled sample size $N$ has to be even, or one observation has to be discarded. For the test, the optimal non-bipartite matching on the pooled sample is calculated. Then a matrix of cross-match counts is constructed whose entries are given by the number of matches with one observation coming from one sample and the other from another sample for each pair of samples. The test statistic is given as the Mahalanobis distance of the observed cross-counts under the null hypothesis. The test was implemented in the \texttt{R} package \texttt{multicross} \parencite{multicross}. 

\paragraph{Tests for the $k$-sample problem for high-dimensional and non-Eucli\-dean data}\phantomsection\label{song_new_2022}
	\textcite{song_new_2022} propose three new tests for the $k$-sample problem, especially for high-dimensional and non-Euclidean data. Their main idea is to use not only the between-sample edges of a similarity graph on the pooled sample, i.e.\ the edges connecting points from different samples, but also the within-sample edges, i.e.\ the edges connecting points from the same sample, to use as much information as possible. Let $R^W$ denote the vector containing the numbers of within-sample edges for each of the $k$ samples and $R^B$ denote the vector containing the numbers of between-sample edges for all $k(k-1)$ pairs of different samples. Then, the first test statistic is given by 
	\begin{align*}
		S &= S^W + S^B\\
		S^W &= \left(R^W - \E(R^W)\right)^T \Sigma_W^{-1} \left(R^W - \E(R^W)\right)\\
		S^B &= \left(R^B - \E(R^B)\right)^T \Sigma_B^{-1} \left(R^B - \E(R^B)\right),
	\end{align*}
	where $\E$ and $\Sigma$ denote the expectation and covariance matrix under the permutation null hypothesis. The second test statistic is based on the vector $R^A$ of all linearly independent numbers of edges between and within samples, i.e.\ all numbers of edges between all pairs of samples including the pairs of a sample with itself except for the pair of the sample $(k-1)$ with the $k$th sample. The test statistic is then defined as 
	\[
	S^A = \left(R^A - \E(R^A)\right)^T \Sigma_A^{-1} \left(R^A - \E(R^A)\right), 
	\]
	where again $\E$ and $\Sigma$ denote the expectation and covariance matrix under the permutation null hypothesis. While $\Sigma_W$ is shown to be always invertible, no such proof exists for $\Sigma_B$ and $\Sigma_A$. Therefore, \textcite{song_new_2022} suggest checking the invertibility numerically before applying the test and using a generalized inverse if necessary. Formulas for the expectations and covariance matrices under the permutation null are given in Theorem 2.1 of \textcite{song_new_2022}. Moreover, \textcite{song_new_2022} show that under some assumptions on the similarity graph that are fulfilled by a $K$-MST with $K=\mathcal{O}(1)$, $S^W \to \chi^2_k$, $S^B \to \chi^2_b$, $S^A \to \chi^2_a$ asymptotically, where $b = \text{rank}(\Sigma_B)$ and $a = \text{rank}(\Sigma_A)$. The asymptotic distribution of $S$ is more complicated and hard to compute in practice, therefore it is suggested to use a fast test instead. This fast test combines tests using $S^W$ and $S^B$ and takes the Bonferroni-adjusted $p$-value of both these tests. Alternatively, a permutation test can be performed. Consistency of the asymptotic tests based on $S$ and $S^A$ against all alternatives is shown in the multivariate setting for the $K$-MST and under the condition that $\Sigma_A$ is invertible. The tests are implemented in the \texttt{R} package \texttt{gTestsMulti} \parencite{song_gtestsmulti_2022}.  

\paragraph{Graph-based tests based on denser graphs}\phantomsection\label{zhu_limiting_2024}
	\textcite{zhu_limiting_2024} present a review of graph-based tests as well as new theoretical results based on less strict assumptions. 
	Their motivation is that the assumptions made on the graphs e.g.\ by \textcite{friedman_multivariate_1979}, \textcite{chen_new_2017}, and \textcite{chen_weighted_2018} to show asymptotic distributions of the test statistics are quite strict and often not fulfilled in practice, especially for denser graphs. 
	However, empirically an improved performance of tests based on denser graphs, e.g.\ the 5-MST instead of the MST, was observed. 
	Therefore, constructing tests based on denser graphs is promising for improving the power. 
	\textcite{zhu_limiting_2024} derive less strict sufficient conditions under which the asymptotic null distributions of the original, weighted, generalized, and max-type edge count statistic hold.
	These allow for using much denser graphs than the conditions derived before.
	Additionally, simulations on the newly derived assumptions are presented.

\subsubsection{Methods based on nearest neighbors}

An important subgroup of graph-based tests are nearest-neighbor type tests. \textcite{chen_graph-based_2013} claim that they do not work well for categorical data in general. 

\paragraph{Weiss test based on spheres}\phantomsection\label{weiss_two-sample_1960}
One of the first approaches for the multivariate two-sample problem goes back to \textcite{weiss_two-sample_1960}. The procedure presented there needs the assumption that for both distributions, piecewise continuous and bounded densities exist. The test statistic is given by the proportion of points from the first sample where no point of the second sample is contained in the sphere around the respective point from the first sample with radius $1/2$ of the distance to its nearest neighbor from the first sample. This yields a multivariate analog of the Wald-Wolfowitz run test. The test is not distribution-free, but the calculation of its critical value is possible under assumptions on densities under $H_0$. The test statistic is invariant under translations and rotations of space or under linear stretching of each of the axes by the same factor. Disadvantages of the procedure are that the test statistic lacks symmetry (roles of the first and second sample not interchangeable) and as pointed out by \textcite{henze_multivariate_1988} the test lacks proof of consistency.

\paragraph{Nearest neighbor test of Friedman and Steppel}\phantomsection\label{friedman_nonparametric_1973}
The idea of nearest-neighbor type tests dates back to \textcite{friedman_nonparametric_1973} and was motivated by assessing the influence of different features on a target variable by splitting the feature dataset according to values of the target and comparing if the subsets differ in their distributions. It is assumed that the distributions have existing density functions. The $K$ nearest neighbors in the pooled sample that originate from the first sample are counted and the distribution of these frequencies is compared to that expected under the null hypothesis by a permutation procedure. One way to do this is to compare the frequency distribution of $K$ nearest neighbors that originate from the first sample in the first sample to that in the second sample. Under the null, these are expected to be equal. \textcite{friedman_nonparametric_1973} suggest performing a test based on a $t$-statistic that compares the mean frequencies from both samples or alternatively use another test for comparison of univariate distributions. This test may be asymptotically optimal, but in finite samples is relatively insensitive to differences in scale between the two multivariate samples. An alternative to fix this is to compare frequency distributions (or the distribution of summed frequencies from both samples) directly with that expected under the null via a $\chi^2$ goodness-of-fit test. The null distribution is in general hard to derive but can be approximated by binomial distribution $\text{Bin}(K, n_1/N)$. The choice of $K$ and of the metric to determine the nearest neighbors is left open. For consistency, $K$ should be a function of the total sample size $N$ that goes to $\infty$ for $N\to\infty$ such that $K(N)/N\to 0$ for $N\to\infty$. More important than the choice of $K$ is the choice of the metric, the features to use, and their scaling. \textcite{friedman_nonparametric_1973} recommend scaling the data by the inverse covariance matrix if no prior knowledge is given. More features only improve the power of the test if they contain information concerning the hypothesis under test, otherwise adding features decreases power. \textcite{friedman_nonparametric_1973} give no general recommendation for the choice of metric (Minkowski $q$-metrics for $q = 1, 2, \infty$ are considered) and list different algorithms for the nearest neighbor calculation giving $\mathcal{O}(2^pN\log_2N)$, $\mathcal{O}(p[Kp\Gamma(p/2)/2]^{1/p}N^{2-1/p})$ or for brute force $\mathcal{O}(pN^2)$ cost for the computation.

\paragraph{Nearest neighbor test of Schilling and Henze}\phantomsection\label{schilling_multivariate_1986}
\textcite{schilling_multivariate_1986} developed a two-sample test based on nearest neighbor type coincidences and \textcite{henze_multivariate_1988} proved its asymptotic properties. Their test is probably the best-known nearest-neighbor test. It is based on the ideas of \textcite{friedman_nonparametric_1973} and \textcite{rogers_convergence_1978}. The test statistic is the proportion of all $K$ nearest neighbor comparisons based on the (Euclidean) distance in which a point and its neighbor belong to the same sample. A scaled version of this test statistic is shown to be asymptotically normal, which motivates an asymptotic test. The method is in general only applicable for continuous distributions since then the probability of ties in the calculation of the nearest neighbors is zero. \textcite{schilling_multivariate_1986} originally proposed to use randomization of the ranks if distances are tied. The test is shown to be consistent against all alternatives. Weighted versions of the test statistic are given by \textcite{schilling_multivariate_1986}. The generalization to the multisample problem is presented by \textcite{henze_multivariate_1988}. The determination of the number $K$ of nearest neighbors to consider is left open. \textcite{biswas_distribution-free_2014} state that the test is not exactly distribution-free and \textcite{sarkar_graph-based_2020} show situations where its power is very low and formal conditions under which power decreases to zero for increasing $p$. \textcite{chen_ensemble_2013} show via simulation that the test's power decreases with the imbalance of the sample sizes. \textcite{aslan_new_2005} reported that according to a private communication with Henze, there is no recipe for how to choose the number of neighbors. \textcite{biswas_nonparametric_2014} mention that the test statistic is rotation invariant and invariant under location change and under homogeneous scale transformation and can be used even when the dimension of the data is larger than the sample size. They also derive sufficient conditions when power decreases to zero for $p \to \infty$.\\
\textcite{henze_almost_1992} derive sufficient conditions for almost sure convergence of a class of sequences of symmetric test statistics for the $k$-sample problem that includes, e.g.\ the test statistics of \textcite{schilling_multivariate_1986} and \textcite{henze_multivariate_1988}. They assume absolutely continuous Lebesgue densities.

\phantomsection\label{barakat_multivariate_1996}
\textcite{barakat_multivariate_1996} further generalize Schilling's nearest neighbor test to circumvent choosing the number of nearest neighbors. Their test statistic is the sum of edge counts for all values of $K$ for the $K$-nearest neighbor graph. Alternatively, it can be seen as a term relying on the sample sizes and a quantity that can be interpreted as follows. First, randomly an observation $x$ is selected from the pooled sample, then one observation $x_1$ is randomly selected from the sample to which the first selected observation belongs, and one observation $x_2$ from the other sample. There are $n_1 n_2 (N-2)$ such choices, i.e.\ sets of such three observations. Then the number of cases for which the first selected observation $x$ is closer to the observation $x_1$ from the same sample than to the observation $x_2$ from the other sample is calculated, and a correction term depending on the sample sizes is added. The resulting test is equivalent to a sum of Wilcoxon rank sums. It requires samples in the Euclidean space $\mathbb{R}^p$ and it is assumed that there are no ties in ranking w.r.t. to nearness.

\paragraph{Nearest neighbor test for categorical data}\phantomsection\label{nettleton_testing_2001}
\textcite{nettleton_testing_2001} propose a test for the two or $k$-sample problem with categorical components. A function that gives the distance between any two data vectors is defined and the number of edges in a nearest-neighbor graph that connect observations from different samples is counted. The test works by adding up values of distance functions over dimensions and calculating the number of edges that link data points from different groups based on a nearest-neighbor graph of the pooled sample. The $p$-value of the test can be determined both by permutation testing or by an asymptotic test. The distance function that maps $\{0,\dots,K -1\}^2$ with $K$ denoting the number of classes to $\mathbb{R}$ is chosen depending on the application (e.g.\ Hamming distance for binary data) and there are no clear general recommendations for this choice. According to \textcite{nettleton_testing_2001}, the procedure needs a ``few minutes using a personal computer'' to calculate an estimate of the conditional $p$-value. It might therefore not be feasible if a large number of tests needs to be performed. 

\paragraph{Nearest neighbor test for continuous data (Hall and Tajvidi test)}\phantomsection\label{hall_permutation_2002}
\textcite{hall_permutation_2002} propose a permutation test based on ranking the pairwise distances between data points. Their test is only applicable for continuous distributions with identical support. The distance measure should be symmetric and does not have to satisfy the triangle inequality. Similar to nearest neighbor tests, the number of $j$ nearest neighbors in the pooled sample that belong to the same sample as the point under consideration are determined for both samples. The test statistic then is a weighted sum of powers of the absolute deviations of these numbers from their expectations under $H_0$ over all sample points and all possible values of $j$. The choice of the power and the weight functions is left open. Weight functions of the form $w_i(j) = 1$, $w_i(j) = j$, and $w_i(j) = n_i + 1 - j$ are proposed. For theoretic results, a weight function is required that converges to a non-degenerate function when viewed as a function of $j/n_2$. The test can distinguish between local alternatives that are distant $n_2^{-1/2}$ from the null hypothesis if the distance is chosen as the Euclidean distance and both distributions have continuous densities. According to \textcite{biswas_distribution-free_2014}, the test is not distribution-free. \textcite{biswas_nonparametric_2014} mention that the test statistic is rotation invariant and can be used even when the dimension of data is larger than the sample size. \textcite{montero-manso_two-sample_2019} claim that the test is valid for infinite dimensional Euclidean spaces since it can take any dissimilarity function as distance. 

\paragraph{Extension of Schilling-Henze test for unbalanced sample sizes}\phantomsection\label{chen_ensemble_2013}
\textcite{chen_ensemble_2013} use a test based on the nearest neighbor method of \textcite{schilling_multivariate_1986} and subsampling to improve the unsatisfactory performance of two-sample tests when sample sizes are unbalanced. The finite sample distribution of the resulting test statistic is unknown but asymptotic and permutation approaches are presented. Consistency is shown when the ratio of sample sizes either goes to a finite limit or tends to infinity. For this, it is assumed that the distributions are absolutely continuous with respect to the Lebesgue measure and that there are no ties for the identification of nearest neighbors. The size of the subsample from the larger of the two samples needs to be chosen to calculate the test statistic. Based on simulations, a subsample size equal to the size of the smaller sample is recommended.

\paragraph{Nearest neighbor test for high dimension low sample size setting}\phantomsection\label{mondal_high_2015}
\textcite{mondal_high_2015} propose a new multivariate two-sample test based on nearest neighbor type coincidences suitable also for the high dimension low sample size (HDLSS) regime that has higher power than the test of \textcite{schilling_multivariate_1986} and \textcite{henze_multivariate_1988} in certain situations. The test statistic modifies the one of \textcite{schilling_multivariate_1986} and \textcite{henze_multivariate_1988} by subtracting the expected values under $H_0$ from both proportions of nearest neighbors from the same sample and taking either the absolute value of this difference or squaring it. Under similar conditions as in other papers for the HDLSS regime, consistency for fixed sample size and $p \to \infty$  is shown for both variants. Also, conditions are derived under which the tests are not consistent for increasing dimensionality. The tests are shown to be asymptotically distribution-free for $N\to \infty$, but a permutation procedure is used in practice where $H_0$ is rejected for large values of the test statistics. Moreover, consistency for fixed $p$ and $N\to\infty$ is shown for distributions with continuous densities. The choice of the number of nearest neighbors to consider is left open. \textcite{mondal_high_2015} consider $K =3$ neighbors in all examples and applications.

\subsection{Methods based on inter-point distances}
Many methods are based on analyzing the distributions of inter-point distances in and between the samples. A theoretical justification for methods based on inter-point comparisons based on a univariate function (e.g.\ a distance) is given by \textcite{maa_reducing_1996}. They show that equality of distributions of in-sample comparisons (i.e.\ $\|X - X^{\prime}\|$ and $\|Y - Y^{\prime}\|$) together with equality of distributions of between-sample comparisons (i.e.\ $\|X - Y\|$) between points is equivalent to the equality of distributions of the samples. This holds in general for discrete distributions. For the continuous case, some restrictions on the density function are needed, including the existence of expectations and a second condition that is for example fulfilled if one of the densities is bounded or continuous.

The advantages of using tests based on inter-point distances according to \textcite{montero-manso_two-sample_2019} are that
\begin{itemize}
	\item it reduces the dimension of the problem,
	\item the use is not limited to dealing with continuous data,
	\item tests can be conducted whenever distances are available even though the original observations are not accessible,
	\item the versatility to choose a proper distance facilitates the introduction of prior domain knowledge,
	\item it is intuitively expected that employing a suitable distance should increase the test power.
\end{itemize}

\subsubsection{Energy Statistic}\phantomsection\label{zech_new_2003}
The most popular statistic based on inter-point distances is the so-called energy statistic. It was proposed by \textcite{zech_new_2003} and by \textcite{aslan_statistical_2005}, where the concept of statistical energy of statistical distributions similar to electric charge distributions is introduced, which was later on also proposed by \textcite{szekely_testing_2004}. Independent from that, \textcite{baringhaus_new_2004} introduced a test based on the difference of the sum of all Euclidean distances between random vectors belonging to different samples and $1/2$ of both sums of distances between random vectors belonging to the same sample, which they call the Cramér test. Its test statistic is equal to the energy statistic. The Cramér test is not distribution-free and needs the assumption that expectations of both distributions exist. It is shown to be consistent against any fixed alternative $F_1\ne F_2$ with finite expectations. Convergence of a Bootstrap version of the test is shown as well. The test is invariant w.r.t. orthogonal linear transformations. It is implemented in the \texttt{R} package \texttt{cramer} \parencite{cramer}. According to \textcite{sarkar_high-dimensional_2018}, the Cramér test needs the two distributions to differ in their locations or average variances to perform well in the HDLSS setup. \textcite{biswas_nonparametric_2014} note that the test is rotation invariant and invariant under location changes and homogeneous scale transformations, it can be used even when the dimension of data is larger than the sample size, and under conditions similar to those made in \textcite{biswas_nonparametric_2014}, a similar consistency result can be shown for the Cramér test as for the test introduced by \textcite{biswas_nonparametric_2014}. On the other hand, they demonstrate situations in which the test fails to detect differences in distributions.

A comprehensive review of the literature on the energy statistic and its applications is given in \textcite{szekely_energy_2017}. We focus on the results for the two-sample situation here, although applications of the energy statistic also include for example one-sample goodness-of-fit tests, clustering, or testing for independence \parencite{szekely_energy_2017}. We extend the presentation of the two-sample energy statistic by \textcite{szekely_energy_2017} using the references therein as well as additional literature.

In the following, we present the energy statistic and its application in two-sample testing according to \textcite{szekely_testing_2004}. \textcite{aslan_statistical_2005} give a slightly more general form of the statistic since they leave the choice of the distance function between points open (for discussion of properties, see below) while \textcite{szekely_testing_2004} define the statistic using the Euclidean distance. They propose a distribution-free test for the equality of two or more multivariate distributions. The approximate permutation test uses Euclidean distances between elements of the samples. Its computational complexity is independent of the dimension and the number of datasets. The test is motivated by the lack of distribution-free extensions of approaches for the two-sample problem based on comparing EDFs (e.g.\ Kolmogorov-Smirnov and Cramér-von-Mises test) to the multivariate case as well as the lack of extensions of tests for the multivariate problem relying on ML to the general $k$-sample problem due to the distributional assumptions.\\
The test statistic of \textcite{szekely_testing_2004} relies on the $e$-distance between finite sets: The $e$-distance $e(\mathcal{X}, \mathcal{Y})$ between disjoint nonempty subsets $\mathcal{X} = \{X_1,\dots, X_{n_1}\}$ and $\mathcal{Y} = \{Y_1,\dots, Y_{n_2}\}$ of $\mathbb{R}^p$ is defined as
\begin{align*}
	e(\mathcal{X}, \mathcal{Y}) = &\frac{2}{n_1 n_2}\sum_{i = 1}^{n_1}\sum_{j = 1}^{n_2} \|X_i - Y_j\|_2  \\
	&- \frac{1}{n_1^2} \sum_{i = 1}^{n_1}\sum_{j = 1}^{n_1} \|X_i - X_j\|_2 - \frac{1}{n_2^2} \sum_{i = 1}^{n_2}\sum_{j = 1}^{n_2} \|Y_i - Y_j\|_2,
\end{align*}
where $\|\cdot\|_2$ is the Euclidean norm. Its population equivalent is given by 
\[
\mathcal{E}(X, Y) = 2\E[\|X - Y\|_2] - \E[\|X- X^{\prime}\|_2] - \E[\|Y-Y^{\prime}\|_2],
\]
where $X^{\prime}$ and $ Y^{\prime}$ denote independent copies of $X$ and $Y$, respectively.\\
Given $\mathcal{X}_1,\dots, \mathcal{X}_k$, $k\ge2$ independent random samples of random vectors in $\mathbb{R}^p$ with sizes $n_1,\dots,n_k$, let $N = \sum_{i = 1}^k n_i$. Denote the $e$-distance of each pair of samples $(\mathcal{X}_i, \mathcal{X}_j)$, $i\ne j$, by $\mathcal{E}_{n_i, n_j} (\mathcal{X}_i, \mathcal{X}_j) = e(\mathcal{X}_i, \mathcal{X}_j)$. Then the $k$-sample test statistic is given by the sum of the $e$-distances for all $k(k - 1)/2$ pairs of samples: 
\[
T_\text{Energy} = \sum_{1\le i<j\le k} \frac{n_i n_j}{n_i + n_j} \mathcal{E}_{n_i, n_j} (\mathcal{X}_i, \mathcal{X}_j) = \sum_{1\le i<j\le k} \frac{n_i n_j}{n_i + n_j} e(\mathcal{X}_i, \mathcal{X}_j).
\]
Large values of the test statistic lead to rejection of the null hypothesis. To obtain the (approximate) null distribution of the test statistic a permutation test is performed by drawing $B$ Bootstrap samples of the pooled sample and partitioning each Bootstrap sample into sets of the same sizes as $n_i, i = 1,\dots,k$. For each Bootstrap sample, the test statistic for these new sets is calculated and the null hypothesis is rejected if the observed value of the test statistic is larger than the $(1-\alpha)$-quantile of the empirical distribution of test statistics from the Bootstrap samples. The test is implemented in the \texttt{R} \parencite{R_4_1_2} package \texttt{energy} \parencite{energy}.\\
According to \textcite{szekely_energy_2017}, the energy distance is invariant w.r.t.\ distance-preserving transformations (e.g.\ translation, reflection, angle-preserving rotation of coordinate axes) of data, i.e.\ rigid motion invariant. Moreover, it is scale invariant. It can be seen as a weighted $L^2$ distance between characteristic functions and the specific weight function is the only solution for such a weighted $L^2$ distance between characteristic functions so that the distance is rotation and scale invariant (under some technical assumptions). For equal sample sizes, the sample energy distance is the square of a metric on the sample space.

In \textcite{szekely_energy_2013} a discussion and illustration of the theory and application of energy statistics are given. A generalization of the energy statistic is given for which a continuous, monotonic decreasing function of the Euclidean distance between points needs to be chosen. \textcite{szekely_energy_2013} choose $-\log$ such that the test is scale invariant. Moreover, they recommend standardizing all variables with the mean and standard deviation of the pooled sample to avoid a single variable dominating the value of the test statistic.

Another generalization of the energy statistic is given by taking each distance to the power of $\alpha$, $\alpha\in(0,2]$. For $0<\alpha<2$, it still holds that the statistic is nonnegative with equality to zero if and only if both distributions are equal. The latter property does not hold in the case of $\alpha = 2$ \parencite{szekely_energy_2017}. When using a different metric than the Euclidean metric, non-negativity of the resulting energy statistic is equivalent to the condition that the metric space in which the random variables take their values has negative type while the property that the statistic is equal to zero if and only if the distributions are equal is equivalent to the condition that that metric space has strong negative type. This holds e.g.\ for Euclidean spaces and separable Hilbert spaces \parencite{szekely_energy_2017}. A metric is said to be of negative type if there exists a mapping $f : \mathcal{X} \to L^2$ such that $d(x, y) = \|f(x) - f (y)\|_2^2$ for every $x, y \in \mathcal{X}$.

\textcite{li_asymptotic_2018} derives the asymptotic null distribution of the energy statistic and shows under some assumptions that the test is more powerful for location than for scale differences.

\textcite{chakraborty_new_2021} show that energy distance based on the usual Euclidean distance cannot completely characterize the homogeneity of two high-dimensional distributions, but only detects equality of means and the traces of covariance matrices in the high-dimensional setup. They criticize the energy distance based on Euclidean distances and define a new test with complexity linear in the dimension of the data that is capable of detecting homogeneity between the low-dimensional marginal distributions in the high-dimensional setup.
They generalize the energy statistic by replacing the Euclidean distances with a newly defined semimetric 
	\[
	K(x, y) := \sqrt{\rho_1(x_{(1)}, y_{(1)}) + \dots + \rho_m(x_{(m)}, y_{(m)})}, 
	\]
	where $\rho_i$ are metrics or semimetrics on $\mathbb{R}^{p_i}, i = 1,\dots,m,$ and the vectors $x$ and $y$ are partitioned into $m$ groups as $x = (x_{(1)},\dots,x_{(m)})$, where $x_{(i)}\in\mathbb{R}^{p_i}$ and $\sum_{i=1}^m p_i = p$ (analogously for $y$). \textcite{chakraborty_new_2021} focus on the case where each $\rho_i$ is a metric of strong negative type on $\mathbb{R}^{p_i}, i = 1,\dots,m$. In that case $K(x, y)$ is a metric of strong negative type on $\mathbb{R}^p$. They define a $t$-test based on their newly proposed metric.
They need several moment assumptions in their analysis and several other assumptions. The new test is shown to be able to detect a wider range of alternatives than the energy statistic but cannot detect differences beyond the equality of the low-dimensional marginal distributions with non-trivial power. No resampling-based inference is needed for their test, but a homogeneity metric as well as a grouping of samples is needed. 

\textcite{rizzo_disco_2010} show that the energy test can be seen as the treatment sum of squares in an ANOVA interpretation of the $k$-sample problem. They use a different measure of dispersion for univariate or multivariate responses based on all pairwise distances between-sample elements for ANOVA. With this, they derive their so-called \textit{distance components (DISCO) decomposition} for powers of distances in $(0,2]$ that gives a partition of the total dispersion in the samples into components analogous to the variance components in ANOVA.  The resulting distance components determine a test for the general hypothesis of equal distributions. For each index in $(0,2)$ this determines a nonparametric test for the multi-sample problem that is statistically consistent against general alternatives. For an index equal to two, it equals the usual ANOVA F-test. Their test statistic is somewhat similar to a generalization of the energy statistic where each of the differences is taken to the power $\alpha$ (given that $\E(\|X\|^\alpha) < \infty,\, \E(\|Y\|^\alpha) < \infty$). 
The new test is performed via permutation testing. Its asymptotic null distribution is a quadratic form (constants not given). The test is consistent against all alternatives with finite second moments. The choice of the index $\alpha$ is difficult. In general, the computational costs for calculating Gini means, in terms of which the test statistic can be formulated, is $\mathcal{O}(N^2)$, for $\alpha= 1$ it can be linearized and computation time reduces to $\mathcal{O}(N\log N)$. The simplest and most natural choice for $\alpha$ is one, for heavy-tailed distributions one may want to apply a small $\alpha$. The test is implemented by permutation Bootstrap in the \texttt{R} package \texttt{energy} \parencite{energy}.

\textcite{huang_efficient_2017} propose a \textit{Randomly Projected Energy Statistics test} based on random projections and energy statistics to lower the computational costs from $\mathcal{O}(N^2$) to $\mathcal{O}(mN \log N)$, with $m$ denoting the number of random projections. For practical use, the number of random projections needs to be determined. \textcite{huang_efficient_2017} derive the asymptotic distribution of usual energy statistics and of the randomly projected one under conditions on expectations and variances and show that the modified version has nearly the same asymptotic efficiency as the usual energy statistic.

\textcite{deb_multivariate_2021} present a rank version of energy statistic. Using the theory of measure transportation (optimal transport) a general framework for distribution-free, nonparametric tests based on multivariate ranks is provided. According to the authors, their test is nonparametric, exactly distribution-free, computationally feasible, and consistent against all alternatives under absolute continuity of the distributions. For consistency, no moment conditions are necessary, which enables the usage of heavy-tailed distributions. The test statistic is invariant under scaling and addition of a vector ($a + b Z, b\in \mathbb{R}, a\in \mathbb{R}^p$). The worst-case complexity for rank assignment is $\mathcal{O}(N^3)$. The calculation given ranks takes $\mathcal{O}(n_1 n_2 p)$. The resulting test is exactly equivalent to the Cramér-von Mises test for $p = 1$. An extension to the $k$-sample setting is possible. \texttt{R} code for the test is available on GitHub (\url{https://github.com/NabarunD/MultiDistFree.git}). The method can also be seen as a rank-based method.

\textcite{al-labadi_bayesian_2022} propose an extension of the energy test to a Bayesian test for the $k$-sample problem, based on belief ratios. Their test is shown to be consistent. For the test, a prior has to be specified. \textcite{al-labadi_bayesian_2022} choose a Dirichlet prior, but the choice of its parameters is not clear. Recommendations based on simulation are given. Additionally, a parameter in the belief ratio needs to be chosen. No implementation of the test is given, but a pseudocode algorithm is presented.

\subsubsection{Other methods based on inter-point distances}
\paragraph{Rigid motion invariant test}\phantomsection\label{baringhaus_rigid_2010}
\textcite{baringhaus_rigid_2010} define rigid motion (length and angle preserving transformation) invariant tests based on inter-point distances between samples and inter-point distances within each sample. The test is based on the Cramér test by \textcite{baringhaus_new_2004} which is equivalent to the energy test and the test by \textcite{szabo_variable_2002, szabo_multivariate_2003}. Therefore it requires distributions with finite expectations. The Cramér test itself is rigid motion invariant (rigid motion $Qx + a$ where $Q$ is an orthogonal matrix and $a$ is a vector). The new test statistic generalizes the test statistic analogous to the Cramér test statistic by using a continuous function $\phi$ such that $\phi(\|x-y\|^2)$ is a negative definite kernel. The following conditions on the function $\phi$ are needed for consistency against all fixed alternatives. The resulting test statistic is nonnegative and zero if and only if $H_0$ is true, and one assumes w.l.o.g. that $\phi(0)=0$ and $\phi$ is nonnegative. These assumptions are e.g.\ fulfilled for all distributions with finite support if and only if  $\phi(\|x-y\|^2)$ is negative definite, which is equivalent to $\phi$ having a completely monotone derivative on (0,$\infty$). For the existence of the test statistic, moment assumptions on distributions are needed that make sure that integrals over $\phi(\|X\|^2)$  exist. Different examples for functions are given, including as special cases the Cramér test, the test by \textcite{bahr_ein_1996}, and the test by \textcite{szabo_variable_2002}. The test is not (asymptotically) distribution-free, but its asymptotic distribution can be approximated using a Bootstrap approach. It is shown to be consistent. 
Since the null distribution of the test statistic and also the asymptotic null distribution depend on the common underlying distribution, the critical value needs to be approximated by Monte Carlo samples from the empirical distribution of the pooled sample or by bootstrapping. Efficiencies are examined under certain alternatives. \textcite{baringhaus_rigid_2010} give recommendations for the choice of the function $\phi$ based on simulations. Overall they recommend $\phi(z) = \log(1 + z)$ for general alternatives and for the Cramér test for location alternatives. An extension to the $k$-sample problem is possible. \textcite{tsukada_high_2019} give geometric interpretations of the tests. The tests are implemented with the recommended choices of $\phi$ for general use, for location alternatives, for scale alternatives, and $\phi$ corresponding to the \textcite{bahr_ein_1996} test in the R package \texttt{cramer} \parencite{cramer}.

\paragraph{Triangle test}\phantomsection\label{liu_triangle_2011}
\textcite{liu_triangle_2011} define a triangle test. First, one point from one of the samples and two points from the other sample are randomly selected. Then, it is examined how often the distance between the two observations from the same distribution is the largest, the middle, or the smallest in the triangle formed by these three observations. The test is asymptotically distribution-free under the null hypothesis of equal, but unknown continuous distribution functions, and it is well-defined when the number of variables $p$ is larger than the number of observations $N$. Its computational complexity is independent of $p$. According to \textcite{biswas_distribution-free_2014}, it is not distribution-free in finite samples. \textcite{biswas_nonparametric_2014} note that the test is rotation invariant.

\paragraph{Test for high dimension low sample size setting}\phantomsection\label{biswas_nonparametric_2014}
\textcite{biswas_nonparametric_2014} propose a test based on inter-point distances for high dimension, low sample size (HDLSS) setups, which is directly motivated by results of \textcite{maa_reducing_1996}. The test is invariant under location change, rotation, and homogeneous scale transformations and can be used even if the dimension is much larger than the sample size. \textcite{biswas_nonparametric_2014} derive results for increasing dimension and fixed sample sizes under assumptions (similar to those of \textcite{hall_geometric_2005}) about increasing information with increasing dimensions, uniformly bounded fourth moments, weak dependence among component variables, and related to sample size. Under these assumptions, \textcite{biswas_nonparametric_2014} show consistency of their test for $p \to \infty$. If finite second moments of both distributions exist, additionally under $H_0$ the asymptotic distribution is shown to be a weighted chi-square distribution (asymptotically distribution free), and consistency for $N \to \infty$ and $n_1/n_2 \to \text{const}$ is shown. \textcite{sarkar_high-dimensional_2018} show via simulations that the test has limitations in the HDLSS setup. The two distributions must differ in their locations or average variances to perform well in the HDLSS setup. \textcite{tsukada_high_2019} gives a geometric interpretation of the test.

\paragraph{Extensions of the Cramér test and the test for the HDLSS setting}\phantomsection\label{sarkar_high-dimensional_2018}
\textcite{sarkar_high-dimensional_2018} aim to improve the tests based on mean inter-point distances that use the Euclidean distance (Cramér Test by \textcite{baringhaus_new_2004} and the test by \textcite{biswas_nonparametric_2014}), by using a new class of distance functions instead. Block variants of the new tests are given, but the choice of the block size is left open. They show consistency under certain assumptions on the distributions and on the sample size for $p \to \infty$ for modified tests without and with blocking.

\paragraph{Cramér-von Mises test on inter-point distances}\phantomsection\label{montero-manso_two-sample_2019}
\textcite{montero-manso_two-sample_2019} develop a Cramér-von Mises test on inter-point distances motivated by \textcite{maa_reducing_1996} that compares the whole distribution of inter-point distances instead of individual moments. Therefore, the univariate distributions of the pairwise distances within and between samples are compared using a Cramér-von Mises-type statistic. The resulting test statistic is called \textit{distribution of distances (DD)} statistic. The test is applicable to a broad range of both continuous and discrete distributions since only the mild regularity conditions of \textcite{maa_reducing_1996} are needed. Still, the theoretical results are only derived for continuous distributions. The asymptotic power of the test as $p$ goes to infinity is not studied. For computing distances, a symmetric, real-valued, nonnegative function is required that fulfills mild regularity conditions but does not have to fulfill the triangle inequality. This function $d$ needs to fulfill the condition $d(x, y) = 0$ iff $x = y$ and $d(ax + b, ay + b) = |a|d(x, y)$. For consistency of the test statistic also a bounded support of the distance is required. The test becomes asymptotically distribution-free under the null hypothesis and its critical value is obtained via a permutation approach. The computational cost is $\mathcal{O}(N^2 \log(N))$.

\paragraph{Modification of rigid motion invariant and HDLSS tests}\phantomsection\label{tsukada_high_2019}
\textcite{tsukada_high_2019} propose new criteria based on the tests by \textcite{baringhaus_rigid_2010} and by \textcite{biswas_nonparametric_2014}. The first test statistic is the length of the difference between the vector $\hat{\mu}$ consisting of estimated means of $\|X - Y\|, \|X - X^{\prime}\|$ and $\|Y - Y^{\prime}\|$ and the vector that projects $\hat{\mu}$ onto the line with direction vector $(1, 1, 1)^T$ via the origin. The \textcite{biswas_nonparametric_2014} test is the squared sum of $(2, -1, -1)^T \hat{\mu}$ and $(0, 1, -1)^T \hat{\mu}$. The second new test statistic uses a weighted sum of non-squared terms. Under the moment assumptions of \textcite{hall_geometric_2005} (bounded fourth moments, $\rho$-mixing condition), results for fixed sample size and increasing dimension are derived. Under additional assumptions on the trace of the covariance matrices and the difference of the mean vectors and under the assumption of equal sample sizes that are not too small, consistency for $p\to\infty$ is shown for the second test. The asymptotic null distribution is derived for the second test under the assumption of finite second moments and for $N\to \infty$. The test is asymptotically distribution-free. Under the same assumptions, consistency for the second test is shown. In simulations, the power of the second test is stable for high-dimensional data and large samples. On the other hand, \textcite{tsukada_high_2019} state that they ``expected the power of the [first proposed] test to be comparable to that of the [\textcite{biswas_nonparametric_2014}] test, but its performance was disappointing''. Therefore they recommend the second test if there is no information that the two population covariance matrices are nearly identical, and they recommend the \textcite{baringhaus_rigid_2010} test if it is known that the two population covariance matrices are equal.

\subsection{Methods based on kernel (mean) embeddings}\label{sec:kernel}

The general idea of kernel mean embeddings is to extend feature maps $\phi$ as used by other kernel methods (e.g.\ in the context of kernel support vector machines) to the space of probability distributions by representing each distribution $F$ as a mean function 
\begin{equation}
	\phi(F) = \mu_{F} := \int_\mathcal{X} K(x, \cdot) \dif F(x) = \E_{F}(K(X, \cdot)), \label{def.emb}
\end{equation}
where $K:\mathcal{X}\times\mathcal{X}\to\mathbb{R}$ is a symmetric and positive definite kernel function. A reproducing kernel Hilbert space (RKHS) $\mathcal{H}$ of functions on the domain $\mathcal{X}$ with kernel~$K$ is a Hilbert space of functions $f: \mathcal{X}\to\mathbb{R}$ with dot product $\langle\cdot,\cdot\rangle$ that satisfies the reproducing property 
\begin{align*}
	\langle f(\cdot), K(x,\cdot)\rangle &= f(x)
	\;\Rightarrow\; \langle K(x,\cdot), K(x^{\prime},\cdot)\rangle = K(x,x^{\prime}),
\end{align*}
such that the linear map from a function to its value at $x$ can be viewed as an inner product. \\
In the following, we always assume that the integral (\ref{def.emb}) exists. Then the kernel mean embedding as given above is essentially a transformation of the distribution $F$ to an element in the reproducing kernel Hilbert space (RKHS) $\mathcal{H}$ corresponding to the kernel~$K$ \parencite{muandet_kernel_2017}. For characteristic kernels, the kernel mean representation captures all information about the distribution $F$, i.e.\ the map $F\mapsto\mu_F$ is injective, which implies $\|\mu_{F_1} - \mu_{F_2}\|_{\mathcal{H}} = 0 \Leftrightarrow F_1 = F_2$ \parencite{fukumizu_dimensionality_2004, sriperumbudur_injective_2008, sriperumbudur_hilbert_2010}. Therefore the kernel mean embeddings can be used for comparing distributions. Conditions that ensure the characteristic property are given in \textcite{sriperumbudur_hilbert_2010} (e.g.\ by showing that integrally strictly positive definite kernels are characteristic) and in \textcite{sriperumbudur_universality_2011}. For more details on kernel mean embeddings and their applications refer to the comprehensive review of \textcite{muandet_kernel_2017} and the papers cited therein. Here, we only give a brief overview of the main aspects regarding the problem of comparing two distributions.

\subsubsection{Maximum Mean Discrepancy}\label{sec.mmd}
The following section is largely based on the main points from Section 3.5 in \textcite{muandet_kernel_2017}, supplemented by additional findings from the sources cited therein as well as more recent findings.\\
Building on the ideas given above, a kernel mean embedding can be used to define a metric for probability distributions, the so-called \textit{Maximum Mean Discrepancy (MMD) }
\begin{equation}
	\text{MMD}(\mathcal{H}, F_1, F_2) = \|\mu_{F_1} - \mu_{F_2}\|_{\mathcal{H}}.
\end{equation}
It was proposed in the context of two-sample testing by \textcite{gretton_kernel_2006} but enjoys increasing popularity in different applications like data integration \parencite{borgwardt_integrating_2006}, generative adversial networks \parencite{li_mmd_2017, sutherland_2017_generative, binkowski_demystifying_2021}, testing for independence \parencite{gretton_kernel_2012}, and goodness-of-fit testing \textcite{jitkrittum_informative_2018}.\\
The MMD can equivalently be expressed as 
\[
\text{MMD}(\mathcal{H}, F_1, F_1) = \sup_{f\in\mathcal{F}} \left(\int f(x)\dif F_1(x) - \int f(x)\dif F_2(x)\right),
\] 
with $\mathcal{F}$ the unit ball in a universal RKHS $\mathcal{H}$, and therefore belongs to the class of integral probability measures \parencite[][cf. Section \ref{sec.ipm}]{muller_integral_1997}. MMD is bounded by the Wasserstein distance (\ref{eq:wasserstein}) and up to a constant also by the total variation distance (\ref{eq:tot.var}) \parencite[Theorem 2.1]{sriperumbudur_hilbert_2010}, so if two distributions are close w.r.t. one of those distances, they are also close according to MMD. The MMD can also be defined on other function spaces $\mathcal{F}$, which leads to a generalization of some further metrics like the Kolmogorov-Smirnov statistic or the Earth Mover's distances \parencite{gretton_kernel_2012}.\\\\
Another connection between MMD and other methods presented is that for translation invariant kernels, MMD can be written as 
\[
\text{MMD}(\mathcal{H}, F_1, F_2) = \int_{\mathbb{R}^p} |\phi_{F_1}(\omega) - \phi_{F_2}(\omega)|^2\dif\Lambda(\omega),
\]
where $\Lambda$ is the spectral measure appearing in Bochner's theorem and $\phi_{F_1}$, $\phi_{F_1}$ are the characteristic functions of $F_1$ and $F_2$ \parencite[Corollary 4]{sriperumbudur_hilbert_2010}. So for translation invariant kernels it can be interpreted as the $L^2(\Lambda)$ distance between the characteristic functions. See Section \ref{sec:char.fun} for more methods based on comparisons of characteristic functions. \\
Another representation of MMD in terms of the associated kernel function that is useful for estimation is
\[
\text{MMD}^2(\mathcal{H}, F_1, F_2) = \E_{X,X^{\prime}}\left[K(X, X^{\prime})\right] - 2 \E_{X,Y}\left[K(X, Y)\right] + \E_{Y,Y^{\prime}}\left[K(Y, Y^{\prime})\right]
\]
where $X, X^{\prime}\sim F_1$ and $Y,Y^{\prime}\sim F_2$ are independent copies.     
$\text{MMD}^2(\mathcal{H}, F_1, F_2)$ can be estimated by the $U$-statistic
\begin{align}\label{mmd.u}
	\widehat{\text{MMD}}^2(\mathcal{H}, X, Y)_{U} &=  \frac{1}{n_1(n_1-1)}\sum_{i=1}^{n_1}\sum_{\substack{j=1 \\ j\neq i}}^{n_1} K\left(x_i, x_j\right) \\\nonumber 
	& \quad\;+ \frac{1}{n_2(n_2-1)}\sum_{i=1}^{n_2}\sum_{\substack{j=1 \\ j\neq i}}^{n_2} K\left(y_i, y_j\right)\nonumber
	- \frac{2}{n_1 n_2}\sum_{i=1}^{n_1}\sum_{\substack{j=1 \\ j\neq i}}^{n_2} K\left(x_i, y_j\right)\\\nonumber
	&= \frac{1}{n_1(n_1-1)}\sum_{i=1}^{n_1}\sum_{\substack{j=1 \\ j\neq i}}^{n_1} h_K\left((x_i, y_i), (x_j, y_j)\right) \text{, if } n_1 = n_2, \end{align}
where $h_K((x, y), (x^\prime, y^\prime)) = K(x, x^\prime) - K(x,y^\prime) - K(y, x^\prime) + K(y, y^\prime)$  \parencite{smola_hilbert_2007}. This estimator is unbiased \parencite{gretton_kernel_2006}. \textcite{sutherland_unbiased_2019} presents an unbiased estimator of the variance of the squared MMD estimator and the difference of two correlated squared MMD estimators at essentially no additional computational cost.\\
Based on the above reformulations, MMD could also be seen as an IPM, a method based on comparing characteristic functions, or a method based on inter-point distances.\\
Under $H_0: F_1 = F_2$, if $n_1 = n_2$ and $\E\left[h_K^2\right]<\infty$, it holds
\[
n_1\widehat{\text{MMD}}^2(\mathcal{H}, X, Y)_{U} \stackrel{D}{\to}\sum_{l = 1}^{\infty} \lambda_l[g_l^2 - 2]
\] 
with $g_l\sim N(0, 2)$ i.i.d., $\lambda_i$ solutions to $\int_\mathcal{X} \tilde{K}(x, x^\prime)\psi_i(x)\dif p(x) = \lambda_i\psi_i(x^\prime)$, and centered RKHS kernel \[\tilde{K}(X_i, X_j) = K(X_i, X_j) - \E_X(K(X_i, X)) -\E_X(K(X, X_j)) + \E_{X, X\prime}(K(X, X^\prime)).\] Given a finite sample approximation of the $(1 - \alpha)$-quantile of the null distribution of $n\widehat{\text{MMD}}^2(X, Y)_{U}$, this can be used for testing $H_0$ against $H_1: F_1 \ne F_2$. The quantiles can be approximated by bootstrapping or by fitting Pearson curves using the first four moments \parencite{smola_hilbert_2007}, or through a Gamma approximation of moments and by approximating the eigenvalues in the above expression by their empirical counterparts, which can be obtained from the Gram matrix \parencite{gretton_fast_2009}. The methods of \textcite{gretton_fast_2009} give a consistent estimate of the null distribution computed from the eigenspectrum of the Gram matrix on the pooled sample. They might therefore be preferable since Bootstrap is computationally costly and the Pearson curve fitting method has no consistency or accuracy guarantees. According to \textcite{song_generalized_2021} the MMD Bootstrap test performs poorly in experiments if (only) variance differs between high-dimensional distributions.\\
Alternatively, an asymptotic test based on the following asymptotic distribution shown by \textcite{muandet_kernel_2017} based on the work of \textcite{gretton_kernel_2012} can be performed: 
\[
\sqrt{n_1}\left[\widehat{\text{MMD}}^2(\mathcal{H}, X, Y)_{U} - \text{MMD}^2(\mathcal{H}, F_1, F_2)\right]\stackrel{D}{\to}N(0, \sigma^2_{XY}),
\]
where again $n_1 = n_2$ and $\E\left[h_K^2\right]<\infty$ is assumed and in addition it is assumed that $H_1: F_1 \ne F_2$ holds and $\sigma^2_{XY}$ is defined as
\begin{align*}
	\sigma^2_{XY} & = 4\left(\E_{(X,Y)}\left[\E_{(X^{\prime}, Y^{\prime})}\left(h_K\left((X, Y),(X^{\prime}, Y^{\prime})\right)\right)^2\right]\right. \\
	& \qquad - \left.\left[\E_{(X,Y), (X^{\prime}, Y^{\prime})}\left(h_K\left((X, Y),(X^{\prime}, Y^{\prime})\right)\right)\right]^2\right).
\end{align*}
The convergence rate of $1/\sqrt{n_1}$ of the statistic to its population value is independent of $p$ \parencite{sriperumbudur_empirical_2012}, but \textcite{muandet_kernel_2017} warn that the dimension may show up in a constant term which can make the upper bound arbitrarily large for high-dimensional data. \textcite{danafar_testing_2014} additionally note that the distribution of MMD degenerates under the null hypothesis and its estimator also degenerates under the null and has no consistency or accuracy guarantee.\\

\paragraph{Linear-time statistics}
The cost for computing $\widehat{\text{MMD}}^2(\mathcal{H}, X, Y)_{U}$ is $\mathcal{O}\left(N^2\right)$ \parencite{gretton_kernel_2012}. To circumvent the quadratic cost, \textcite{gretton_kernel_2012} propose an unbiased linear-time statistic 
\begin{align}
	\widehat{\text{MMD}}^2(\mathcal{H}, X, Y)_{U, l} &=  \frac{1}{n_1}\sum_{i=1}^{\lfloor n_1 /2\rfloor} h_K\left((x_{2i-1}, y_{2i-1}), (x_{2i}, y_{2i})\right) \text{, if } n_1 = n_2,\label{mmd.u.l}
\end{align}
for which the same convergence to a normal distribution can be shown with the only difference that its variance is only half as large as that for the quadratic-time statistic. \\
\phantomsection\label{zaremba_b-test_2013} Another way to speed up the calculation is given by \textcite{zaremba_b-test_2013}. The motivation behind their modification is manifold. The test statistic is degenerate under the null hypothesis, and its asymptotic distribution takes the form of an infinite weighted sum of independent $\chi^2$ variables. Further, the methods for estimating the null distribution in a consistent way (Bootstrap or method by \textcite{gretton_fast_2009}) are computationally demanding with costs of $\mathcal{O}(N^2)$ with a large constant or $\mathcal{O}(N^3)$ with a smaller constant, and Pearson curve fitting has no consistency guarantees. To solve these problems, they define a family of block tests for MMD. The choice of block size means a trade-off between power and computation time. To obtain an asymptotic Gaussian null distribution, the size of blocks $B$ needs to be chosen such that $n_1/B \to \infty$ for $n_1=n_2 \to \infty$. The assumptions made are the same as for quadratic-time MMD. Additional conditions for second moments are required for convergence of the test statistic. Due to the asymptotic Gaussian distribution, the critical values for testing are easy to compute. A choice for the size of blocks $B$ is needed to perform the test, and only a heuristic choice of $\lfloor\sqrt{n_1}\rfloor$ is proposed by \textcite{zaremba_b-test_2013}. Moreover, like with the normal MMD test, the kernel needs to be chosen. \\
\phantomsection\label{zhao_fastmmd_2015}\textcite{zhao_fastmmd_2015} instead use the connection between MMD and characteristic functions to define an efficient test called fastMMD test. The idea is to equivalently transform MMD with shift-invariant kernels into amplitude expectation of a linear combination of sinusoid components based on Bochner's theorem and the Fourier transform \parencite{rahimi_random_2007}. For this, they make use of sampling of Fourier transforms. By that, the complexity is reduced from $\mathcal{O}(N^2p)$ to $\mathcal{O}(LNp)$, where $L$ is the number of basis functions for approximating kernels, which determines the approximation accuracy. Spherically invariant kernels allow for further acceleration to $\mathcal{O}(LN\log p)$ by using the Fastfood technique \parencite{le_fastfood_2013}. \textcite{zhao_comparing_2021} show convergence of their estimates.\\
Two important modifications to linear-time tests are given by \phantomsection\label{chwialkowski_fast_2015}\textcite{jitkrittum_interpretable_2016}. They define two semimetrics on probability distributions using the sum of differences in expectations of analytic functions evaluated at either spatial or frequency locations. The goal is to choose features such that the distinguishability of distributions is maximized. Therefore, the lower bound on test power for tests using the features is optimized. This leads to two different linear-time tests.\\
The tests are based on analytic representations of probability distributions presented by \textcite{chwialkowski_fast_2015}. The difference is that there the features are chosen at random, while here the lower bound for the test power is derived, which can be used to optimize the choice of the features.\\
The first test is the \textit{Mean Embedding (ME)} test, which evaluates the difference of mean embeddings at locations chosen to maximize the test power lower bound (spatial features). The second test, the \textit{Smooth Characteristic Function (SCF)} test, uses the difference of two smoothed empirical characteristic functions, evaluated at points in the frequency domain, which are chosen such that the same criterion is maximized (frequency features). The optimization of the mean embedding kernel/ frequency smoothing function is performed on held-out data. The ME and SCF test are defined in \textcite{chwialkowski_fast_2015} as follows: for the test samples, $X = \{X_1, \dots, X_n\}$ and $Y = \{Y_1, \dots, Y_n\}$ i.i.d. according to $F_1$ and $F_2$ are given. Both tests evaluate the hypotheses $H_0: F_1 = F_2$ versus $H_1: F_1 \ne F_2$. \\
The test statistic for the ME test is given by
\begin{align*}
	T_{\text{ME}} &= n \bar{Z}_n^T S_n^{-1} \bar{Z}_n, \text{ with}\\
	\bar{Z}_n &= \frac{1}{n} \sum_{i=1}^{n} Z_i,\\
	S_n &= \frac{1}{n-1} \sum_{i = 1}^{n} (Z_i - \bar{Z}_n)(Z_i - \bar{Z}_n)^T, \\
	Z_i &= \{K(X_i, V_j) - K(Y_i, V_j)\}_{j = 1}^J\in\mathbb{R}^J.
\end{align*}
The test statistic depends on the positive definite kernel $K: \mathcal{X}\times\mathcal{X}\to\mathbb{R}$, $\mathcal{X}\subseteq \mathbb{R}^p$ and the set of $J$ test locations $\mathcal{V} = \{V_1, \dots, V_J\}\subseteq\mathbb{R}^p$. It is asymptotically chi-square distributed \[T_{\text{ME}}\stackrel{H_0, asymp}{\sim}\chi^2_J\] and can be seen as a form of Hotelling's $T^2$ statistic. $T_{\text{ME}}$ is a semimetric, since it can be seen as squared normalized $L^2(\mathcal{X}, V_J)$ distance of the mean embeddings of the empirical measures $F_{1,n} = \frac{1}{n} \sum_{i = 1}^n \delta_{x_i}$ and $F_{2,n} = \frac{1}{n} \sum_{i = 1}^n \delta_{Y_i}$ where $V_J = \frac{1}{J}\sum_{i = 1}^J \delta_{V_i}$, and $\delta_x$ is the Dirac measure concentrated at $x$. \\
The SFC test statistic $T_{\text{SCF}}$ is defined in the same way as $T_{\text{ME}}$, but uses a modified $Z$: 
\begin{align*}
	Z_i =& \{\hat{l}(X_i)\sin(X_i^TV_j) - \hat{l}(Y_i)\sin(Y_i^TV_j), \\
	& \quad\hat{l}(X_i) \cos(X_i^TV_j) - \hat{l}(Y_i)\cos(Y_i^TV_j)\}_{j = 1}^J\in\mathbb{R}^{2J}, \\
	\hat{l}(x) =& \int_{\mathbb{R}^p} \exp(-i u^Tx) l(u) \dif u \quad(\text{Fourier transform of } l(x))
\end{align*}
Here $l:\mathbb{R}^p\to\mathbb{R}$ is an analytic translation-invariant kernel (i.e.\ $l(x-y)$ defines a positive definite kernel for $x$ and $y$). The locations $\mathcal{V}=\{V_1,\dots,V_J\}\subset\mathbb{R}^p$ are in the frequency domain. The test statistic is again asymptotically $\chi^2$ distributed 
\[
T_{\text{SCF}}\stackrel{H_0, asymp}{\sim}\chi^2_{2J}
\]
and can be interpreted as a normalized version of the $L^2(\mathcal{X}, V_J)$ distance of the empirical smooth characteristic functions $\phi_{F_1}(v)$ and $\phi_{F_2}(v)$, where $\phi_F(v) = \int_{\mathbb{R}^p} \varphi_F(w)l(v - w) \dif w$ with $\varphi_F(w) = \E_{X\sim F}\left[\exp\left(i w^TX\right)\right]$ is the characteristic function of $F$. Therefore, it could also be classified as a method based on comparing characteristic functions.
\textcite{jitkrittum_interpretable_2016} denote the degrees of freedom of the $\chi^2$ distribution for both tests as $J^\prime$. They use a modification of the test statistic with regularization parameter $\gamma_n$:
\[
T_{\text{ME/SCF}} = n \bar{Z}_n^T (S_n + \gamma_n I)^{-1} \bar{Z}_n,
\]
to obtain a higher stability of the matrix inversion. The asymptotic distribution under the null hypothesis stays the same as long as $\gamma_n\to 0$ for $n\to\infty$.  Simulations on high-dimensional text and image data show that the tests are comparable to the state-of-the-art quadratic-time MMD test of \textcite{gretton_optimal_2012}, but in contrast to the MMD tests return human-interpretable features explaining the test results. The test statistics depend on the set of test locations $\mathcal{V}$ and the kernel parameter $\sigma$. \textcite{jitkrittum_interpretable_2016} propose to set $\theta = \{\mathcal{V}, \sigma\} = \arg\max_{\theta} \lambda_n = \arg\max_{\theta} \mu^T\Sigma^{-1}\mu$, where $\lambda_n = n \mu^T\Sigma^{-1}\mu$ with $\mu= \E_{F_1,F_2}(Z_1)$, and $\Sigma = \E_{F_1,F_2}\left[\left(Z_1 - \mu\right)\left(Z_1 - \mu\right)^T\right]$ is the population counterpart of $T_{ME/SCF}$. Since a dependency between $\theta$ and the data used for testing would affect the null distribution, it is proposed to split the dataset in half and first use one half $\mathcal{D}^{tr}$ of $\mathcal{D} = (\mathcal{D}_1, \mathcal{D}_2)$ for optimizing $\theta$ via gradient ascent on $T_{\text{ME/SCF}}^{tr}$ (in theory one should maximize $\lambda_n$ but $\mu$ and $\Sigma$ are unknown) and then perform the actual test using the test statistic $T_{\text{ME/SCF}}^{te}$ on the other half $\mathcal{D}^{te}$ of the dataset. Convergence of the test statistic to $\lambda_n$ is guaranteed for $n\to\infty$ over all kernels in a family of uniformly bounded kernels (e.g.\ Gaussian kernel class) and all test locations in an appropriate class. \textcite{jitkrittum_interpretable_2016} use the isotropic Gaussian kernel class $\mathcal{K}_g = \{K_{\sigma}: (x,y) \to \exp(-(2\sigma^2)^{-1}\|x-y\|^2_2)\arrowvert\sigma>0\}$, where $\sigma$ is constrained to be in a compact set and $\mathbb{V} = \{\mathcal{V}\arrowvert$ any two locations are at least $ \varepsilon$ distance apart, and all test locations have their norms bounded by $\zeta\}$, where $\mathcal{V}$ is the set of test locations as defined above. The authors conduct experiments to compare their proposed ME and SCF tests with the versions from \textcite{chwialkowski_fast_2015} (ME and SCF with $\sigma$ optimized by grid search and random test locations) as well as with the quadratic-time and linear-time version of the MMD test \parencite{gretton_kernel_2012} and the standard two-sample Hotelling's $T^2$ test. The newly proposed SCF test outperforms ME in terms of power and also as the (quadratic-time) MMD test, while the linear-time MMD test performs worst. The quadratic-time MMD test becomes computationally infeasible for $p\in[5, 1500]$ and $n = 10000$. The observed type I error rate is too high for Hotelling's $T^2$ in high dimensions since an accurate estimation of the covariance matrix gets more difficult. The performance of the linear-time MMD test drops quickly with increasing dimension $p$, while the ME and SCF test with optimization show the slowest decrease in power with increasing dimension. On real data (text data/ image data) sometimes the ME test performs best and sometimes the ME and SCF test both perform well. Additionally, the learned location can be interpreted (e.g.\ by counting how often a specific word or pixel is chosen as a test location and looking at those that are chosen more often). The number of test locations $J$ has to be chosen manually. 

\paragraph{Other modifications to MMD}\phantomsection\label{danafar_testing_2014}
There are several other modifications that do not aim at reducing the computational cost but focus on other aspects.\\
\textcite{danafar_testing_2014} present a regularized Maximum Mean Discrepancy test for the comparison of multiple distributions. The regularizer is set provably optimal for maximal power such that there is no need for tuning by the user. The presented test is consistent under conditions on second moments. It has higher asymptotic power and higher power in small samples than the MMD and kernel Fisher discriminant analysis (KFDA) tests (\textcite{eric_testing_2007}, see below), but still a computational cost of $\mathcal{O}(N^2)$. Experiments show higher relative efficiency, compared to MMD and KFDA.

\phantomsection\label{cheng_two-sample_2020}\textcite{cheng_two-sample_2020} propose a new kernel-based MMD statistic that can be made more powerful to distinguish certain alternatives when distributions are locally low-dimensional. The idea is to incorporate local covariance matrices and to construct an anisotropic kernel. The test's consistency is proven under mild assumptions on the kernel, as long as $\|f_1 - f_2\|\sqrt{n}\to\infty$. A finite-sample lower bound of the testing power is derived under the assumption that the distributions are continuous, compactly supported, and have densities w.r.t. the Lebesgue measure, and that $1 < p \ll \min(n_1, n_2)$. A set of reference points or a reference distribution and a covariance field, respectively, are required to conduct the test. Under the same assumptions as for consistency, \textcite{cheng_two-sample_2020} show that convergence of the power to 1 is at least as fast as $\mathcal{O}(N^{-1})$. The cost for computing one empirical estimate of the test statistic is $\mathcal{O}(N N_R)$, where $N_R$ is the number of reference points.

\phantomsection\label{kirchler_two-sample_2020}\textcite{kirchler_two-sample_2020} propose a two-sample testing procedure based on a learned deep neural network representation. Instead of the kernel function that gives a feature representation, deep learning is used to obtain a suitable data representation. \textcite{kirchler_two-sample_2020} aim to overcome the problem that the MMD test depends critically on the choice of the kernel function and therefore ``might fail for complex, structured data such as sequences and images, and other data where deep learning excels''. At the same time, they want to improve the classifier two-sample test of \textcite{lopez-paz_revisiting_2017} (see Section \ref{sec.c2st}) that needs a train/test split of the data. The new test instead first maps the data onto a hidden layer of a deep neural network that was trained on an independent, auxiliary dataset. This transformed data is then compared using the MMD test statistic of \textcite{gretton_kernel_2012} or a variant of it, or alternatively using the kernel FDA test \parencite{harchaoui_testing_2008} (see below). The corresponding procedures are called \textit{Deep Maximum Mean Discrepancy (DMMD)} test and\textit{ Deep Fisher Discriminant Analysis (DFDA)} test, respectively. For the class of deep ReLU networks with a $\tanh$ activation function in the final layer, an asymptotic test based on an asymptotic normal or $\chi^2$ distribution of the DMMD and DFDA test statistic is presented. For this, the covariance matrix of the learned feature map must exist and for DFDA it additionally must be invertible. Consistency of the tests can be shown under several assumptions on the neural network and its training and the assumption that the transfer task on which the deep neural network is fitted is not too far from the original task. There are no explicit directions on how to choose the transfer task since the theoretically optimal choice depends on the true distributions and the Bayes rate for the transfer task. So, if there is enough data, splitting is the safe way that guarantees the similarity of transfer and original task. 

\phantomsection\label{song_generalized_2021}The new test of \textcite{song_generalized_2021} makes use of common patterns in moderate and high dimensions. It is aimed at solving the curse of dimensionality for kernel two-sample tests. It takes into account the variance-covariance matrix of the first two terms in (\ref{mmd.u}). The test is implemented in the \texttt{R} package \texttt{kerTests} \parencite{kerTests}. There are two corner cases in which the test statistic is not well-defined. In general, two conditions on the kernel and data are made that are usually fulfilled if there is no major outlier in the data and if one uses a Gaussian kernel with the median heuristic as described below. Under these, an asymptotic normal distribution for the test statistic is shown.

\paragraph{Choice of kernel function and parameters}
All methods described so far depend on a kernel function. The choice of this kernel function is nontrivial. Although there are many proposals on how to choose it, the optimal choice remains an open problem \parencite{muandet_kernel_2017}.\\
In general, as stated at the beginning, characteristic kernels are preferred since they ensure that the MMD is zero if and only if the two distributions coincide. Details on conditions for kernels being characteristic are given in \textcite{sriperumbudur_kernel_2009}, \textcite{sriperumbudur_universality_2011} and \textcite{simon-gabriel_kernel_2018}. Concrete examples are listed in Table 3.1 of \textcite{muandet_kernel_2017}. Still, the class of characteristic kernels is large and leaves some room for decision.\\
Probably the most popular class of characteristic kernels are radial basis function (RBF) kernels. But even within this class, there are different proposals on how to choose the RBF kernel parameter. A heuristic for the choice of the kernel size for the RBF kernel is to set its parameter $\sigma$ to the median distance between points in the pooled sample. The empirical MMD is zero both for a kernel size of zero and an infinitely large kernel size \parencite{gretton_kernel_2006}.\\
A simulation study conducted by \textcite{gretton_kernel_2006} shows that for low sample sizes, the threshold based on Pearson curves performs better in terms of type I error, while for high sample sizes, the Bootstrap threshold is preferred due to the lower computational cost. In a simulation study, all in all, the method outperforms competing methods ($t$-test, Friedman-Rafsky Kolmogorov-Smirnov generalization \parencite{friedman_multivariate_1979}, Biau-Györfi test \parencite{biau_asymptotic_2005}, Hall-Tajvidi test \parencite{hall_permutation_2002}, or is at least close to the best-performing method \parencite{gretton_kernel_2006}.\\
Later, \textcite{gretton_optimal_2012} propose to choose the kernel such that the test power is maximized for a given significance level. Therefore a kernel is selected from a particular family $\mathcal{K}$ of kernels. This family is defined as 
\[
\mathcal{K} = \left\{K: K = \sum_{u = 1}^{d} \beta_u K_u, \sum_{u = 1}^d \beta_u = D, \beta_u \ge 0\;\forall u\in \{1,\dots,d\}\right\}
\]
with a constant $D > 0$ and $\{K_u\}_{u = 1}^d$ a set of positive definite functions $K_u:\mathcal{X}\times\mathcal{X}\to\mathbb{R}$ which are assumed to be bounded, i.e.\ $|K_u|\le C\,\forall u\in\{1,\dots,d\}$. Then each kernel $K\in\mathcal{K}$ corresponds to exactly one RKHS $\mathcal{H}_K$ and the test statistic becomes 
\[
\widehat{\text{MMD}}^2(\mathcal{H}_K, F_1, F_2)_{U,l} = \sum_{u = 1}^{d} \beta_u\eta_u(F_1, F_2) = \E(\beta^T h)  = \beta^T \eta,
\] 
where
\[
\eta_u = \E_{XX^\prime YY^\prime} [h_{K_u}((X, Y), (X^\prime, Y^\prime))]
\]
and $h = (h_{K_1}, \dots, h_{K_d})^T$, $\beta = (\beta_1,\dots,\beta_d)^T$, and $\eta = (\eta_1,\dots,\eta_d)^T\in\mathbb{R}^d$. The authors here make use of the asymptotically unbiased linear-time estimate of \textcite{gretton_kernel_2012} given in (\ref{mmd.u.l}). To maximize the Hodges and Lehmann asymptotic relative efficiency (i.e.\ the power at a given significance level $\alpha$) for the test based on the asymptotic normal distribution of the linear-time statistic, the following quadratic optimization program needs to be solved:
\[
\min\left\{\beta^T\left(\hat{Q} + \lambda_m I\right)\beta: \beta^T\hat{\eta} = 1, \beta \succeq 0\right\},
\]
if $\hat{\eta}$ has at least one positive entry. $\hat{Q}$ is a linear-time empirical estimate of the covariance matrix $\mathbb{C}\text{ov}(h)$ and \[\hat{\eta} = (\widehat{\text{MMD}}^2(\mathcal{H}_{K_1}, F_1, F_2)_{U,l}, \dots, \widehat{\text{MMD}}^2(\mathcal{H}_{K_d}, F_1, F_2)_{U,l}).\] The optimization is performed on a training set of $m$ points $(X_i, Y_i),\, i = 1, \dots,m$, and this training set and the data points used for testing are disjoint. In particular, all estimates needed for the optimization are calculated from this training data. If no entry of $\hat{\eta}$ is positive, a single base kernel $K_u$ with the largest $\hat{\eta}_u/\hat{\sigma}_{K_u,\lambda}$ is arbitrarily selected since it is unlikely that the test statistic computed on the test data will exceed the always positive threshold. 
\[
\hat{\sigma}_{K,\lambda} = \sqrt{\beta^T\left(\hat{Q} + \lambda_m I \right)\beta} = \sqrt{\hat{\sigma}_K^2 + \lambda_m\|\beta\|^2_2}
\] 
is a regularized standard deviation estimate.\\
\textcite{gretton_optimal_2012} conducted simulations that show that their strategy for choosing an optimal kernel yields better results than other strategies, such as the aforementioned heuristic of setting the kernel size to the median distance between points in the aggregate sample or the strategy of maximizing the MMD test statistic proposed by \textcite{sriperumbudur_kernel_2009}. Their method is only outperformed by choosing the kernel with the highest ratio $\hat{\eta}_u/\hat{\sigma}_{K_u,\lambda}$ if a single best kernel exists. Otherwise, if a linear combination of kernels is needed, that strategy fails and the proposed optimal choice performs better in terms of power.\\
Another proposal for choosing the kernel is made by \textcite{liu_learning_2020} where deep kernels are used. The proposed kernel has the form 
\[
K_{\omega}(x, y) = [(1-\varepsilon)\kappa(\phi_{\omega}(x), \phi_{\omega}(x)) + \varepsilon] q(x, y),
\]
where $\phi_{\omega}$ is a deep neural network with parameters $\omega$ that extracts features, and $\kappa$ is a simple kernel (e.g.\ Gaussian with lengthscale $\sigma_{\phi}$) on those features. $q$ is a simple characteristic kernel on the input space and $0<\varepsilon<1$. This allows for an extremely flexible choice of kernels that can learn complex behavior. The parameters $\omega$ are selected by maximizing the ratio of the MMD to its variance, which asymptotically maximizes the power of the test. This is done in a similar train-test manner as in \textcite{gretton_optimal_2012}, but here the proportion of the data assigned to the train set is optimized as well. Also, an improved estimator of the variance of the MMD estimator as proposed by \textcite{sutherland_unbiased_2019} is used. This approach can be understood as a generalization of the evaluation of the accuracy of classifiers proposed by \textcite{lopez-paz_revisiting_2017}, but instead of cross-entropy, the test power is maximized. \\
The learning of the deep kernel is performed using minibatches of size $m$ if the dataset is large. For each minibatch, the cost is $\mathcal{O}(mE + m^2C)$ with the term $mE$ typically dominating for moderate $m$. Here, $E$ denotes the cost of computing an embedding $\phi_{\omega}$ and $C$ the cost of computing the deep kernel. Testing is performed as a permutation test as proposed by \textcite{sutherland_2017_generative}, instead of using the asymptotic distribution like proposed before, i.e.\ approximating the null distribution by drawing $n_{\text{perm}}$ new samples $X^{\prime}$ and $Y^{\prime}$ from the pooled sample and calculating the test statistic on these samples. The permutation approach takes $\mathcal{O}(NE+N^2C+N^2n_{\text{perm}})$ time.\\
It is shown theoretically that for reasonably large $n$ and if the optimization process succeeds, the found kernel generalizes nearly optimally instead of just overfitting to the training data. Furthermore, the resulting test is compared to the one proposed by \textcite{gretton_optimal_2012} and the SCF and ME tests \parencite{chwialkowski_fast_2015, jitkrittum_interpretable_2016} as well as the classifier two-sample tests of \textcite{lopez-paz_revisiting_2017} and \textcite{cheng_classification_2022} in terms of type I error and power on several synthetic and real-world datasets. \textcite{liu_learning_2020} find that all tests keep the nominal type I levels and that the deep kernel MMD test generally has the highest power across a range of settings.\\
The MMD test along with different choices for kernels and many other kernel-based methods is implemented in the \texttt{R} package \texttt{kernlab} \parencite{kernlabR, kernlab}.

\subsubsection{Other kernel-based methods}
\paragraph{Kernel Fisher discriminant analysis test}\phantomsection\label{eric_testing_2007}
\textcite{eric_testing_2007} propose test statistics based on kernel Fisher discriminant analysis (kernel FDA). It is assumed that the kernel function is bounded for all probability measures $\Prob$ and that the RKHS associated with the kernel is dense in $L^2(\Prob)$. Additionally, assumptions are made on eigenvalues of covariance matrices of both distributions. Both assumptions on the kernel are needed in the proof of consistency for the test, but only the first of each assumption is needed to show asymptotic normality. The resulting asymptotic normal distribution is independent of the kernel and an additional regularization parameter that must be chosen.

\paragraph{Tests based on symmetric kernels}\phantomsection\label{fromont_kernels_2012}
\textcite{fromont_kernels_2012} present testing procedures based on a general symmetric kernel. Critical values of the tests are chosen by a wild Bootstrap or permutation Bootstrap approach. An aggregation method enables overcoming the difficulty of choosing a kernel and/or kernel parameters. It is demonstrated that the aggregated tests may be optimal in a classical statistical sense and non-asymptotic properties are shown for the aggregated tests. Therefore, the assumption is made that densities exist with respect to some non-atomic $\sigma$-finite measure and are square-integrable. A kernel needs to be chosen, but suggestions for this choice are given. An alternative test based on the conditional distribution of the test statistic given the sample is shown to be an exact level $\alpha$ test.

\paragraph{Findings on kernel and distance-based tests I}
Two important findings regarding kernel- and distance-based tests are given in \textcite{sejdinovic_equivalence_2013} and \textcite{ramdas_decreasing_2015}.

\textcite{sejdinovic_equivalence_2013} establish a relationship between the energy test and the MMD test by showing that the energy statistic can be seen as a special case of MMD for a certain kernel function. For that, they give a generalized form of the energy statistic by replacing the Euclidean norm with other norms. They also determine the class of distributions for which these tests are consistent against all alternatives. In simulations, they show that the energy test is inferior regarding power. They make the same assumptions as \textcite{szekely_testing_2004} for the introduction and analysis of the energy statistic.

\textcite{ramdas_decreasing_2015} show that tests based on kernel embeddings or based on distances between pairs of points are not well-behaved for high-dimensional data, in contrast to general belief. Instead, they show that the power decreases at least polynomially in dimension for fair alternatives.

\paragraph{ME and SCF test based on $L^1$ distance}\phantomsection\label{scetbon_comparing_2019}
\textcite{scetbon_comparing_2019} present a test using the $L^1$ instead of the $L^2$ distance between kernel-based distribution representatives and define new ME and SCF test statistics based on that. They show that a sequence of Borel measures converges weakly towards a measure if and only if the $L^q, q\ge1$, distance of their mean embeddings to the mean embedding of that measure converges to zero, i.e.\ that the $L^q, q\ge1$, distance metrizes weak convergence. They also show that their $L^1$ version rejects the null hypothesis better than the $L^2$ version under $H_1$ with high probability. The new tests are shown to be consistent for $N \to \infty$ and $n_1/ N \to \text{const}$, and the asymptotic distribution of the test statistic is shown to be a Nakagami distribution.

\paragraph{Kernel-based quadratic distance}\phantomsection\label{chen_kernel_2020}
\textcite{chen_kernel_2020} introduce a generalization of the MMD statistic to kernel-based quadratic distance. They give a review of two-sample tests that includes some of the tests presented here in less detail and also a review of the MMD literature that is not as detailed as that from \textcite{muandet_kernel_2017}.
The test from \textcite{chen_kernel_2020} is based on the \textit{kernel-based quadratic distance} introduced by \textcite{lindsay_quadratic_2008} 
\[
d_K(F_1, F_2) = \int\int K(s, t)\dif(F_1 - F_2)(s)\dif(F_1 - F_2)(t),
\]
with a nonnegative definite kernel $K$. An estimator is presented that relies on an appropriately centered kernel. Its limiting distributions under the null and the alternative and the exact variance under the null can be derived. However, those cannot be used for the construction of a critical value since the null distribution is an infinite sum that depends on eigenvalues of the centered kernel. Optimal tuning parameters can be chosen based on the ideas of \textcite{lindsay_kernels_2014} for the one-sample test. Moreover, an extension to the $k$-sample problem is presented. The practical calculation of the test statistic under the assumption that the common distribution under $H_0$ belongs to a family of parametric distributions as well as the concrete form of the test statistic under a normal assumption are shown. Alternatively, a nonparametric calculation is possible by using the mixing distribution of $F_1$ and $F_2$, or its empirical counterpart, as the centering distribution for the kernel. Corresponding critical values for these two versions can be calculated by a parametric or nonparametric Bootstrap or in both cases with a permutation procedure.

\paragraph{Findings on kernel and distance-based tests II}
\textcite{zhu_interpoint_2021} present situations in which the energy and MMD permutation tests are inconsistent. They show that the class of two-sample tests based on inter-point distances (generalized energy statistics) including MMD with Gaussian or Laplacian kernels and the energy statistic as well as the generalized energy statistic using the $L^1$ instead of the Euclidean distance are inconsistent when two high-dimensional distributions correspond to the same marginal distributions but differ in other aspects. Additionally, they derive the limiting distribution of a test statistic based on inter-point distances under low and medium sample sizes for increasing dimensions. They also show that under HDLSS and HDMSS, the energy statistic and MMD test are consistent if the sum of component-wise means or variances are not too small. On the other hand, if the sum of component-wise mean and variance differences are both of order $o(\sqrt{p} / \sqrt{n_1 n_2})$, then these tests suffer from a substantial power loss under HDLSS and have trivial power under HDMSS. Under HDLSS, they have trivial power if additionally, the sum over squared covariance differences is $o(p)$. The $L^1$-norm-based test also experiences a power drop under HDLSS and has trivial power under HDMSS if the marginal univariate distributions are the same. Under HDLSS, it has trivial power when the distributions have the same bivariate marginal distributions. For the analysis, \textcite{zhu_interpoint_2021} make assumptions on the existence of means and variances and additional moment and weak dependence assumptions on the components of $X$ and $Y$. \textcite{zhu_interpoint_2021} argue that in low dimensions the $L^1$-norm is not suitable since an $L^1$ distance of zero does not imply $F_1 = F_2$.

\paragraph{Bayesian kernel test}\phantomsection\label{zhang_bayesian_2022}
\textcite{zhang_bayesian_2022} define a Bayesian kernel paired two-sample test based on modeling the difference between kernel mean embeddings in the RKHS. Their test is based on the framework of \textcite{flaxman_bayesian_2016} and automatically selects kernel parameters relevant to the problem. The use of a kernel allows for the use of the test beyond Euclidean spaces. In contrast to most other methods, they do not need the assumption that samples are independent of each other, but only the assumptions on the kernel as for the MMD. The test is conditional on the choice of the family of kernels. \textcite{zhang_bayesian_2022} focus on Gaussian RBF kernels in their analysis. The test statistic is based on the Bayes factor. \textcite{zhang_bayesian_2022} propose to model the witness function with a Gaussian process prior under the alternative model and to use a Gaussian noise model for the empirical witness vector given the bandwidth parameter. They derive the posterior distribution of the bandwidth parameter if it is unknown with a Gamma(2,2) prior under both null and alternative models and marginalize over it so that this parameter no longer has to be selected.

\paragraph{Kernel Measure of Multi-Sample Dissimilarity (KMD)}\phantomsection\label{huang_kernel_2022}
\textcite{huang_kernel_2022} define a nonparametric kernel measure of multi-sample dissimilarity (KMD). Denote the dataset membership of each point in the pooled sample $\{Z_1,\dots, Z_N\}$ by $\{\Delta_1,\dots, \Delta_N\}$. If $\frac{n_i}{N}\to \pi_i\in(0,1)$ for $N\to\infty$ such that $\sum_i \pi_i = 1$ then $\{(\Delta_i, Z_i)\}_{i=1}^N$ can approximately be seen as an i.i.d.\ sample from $(\tilde{\Delta}, \tilde{Z})$ with distribution $\mu$ specified by $\Prob(\tilde{\Delta} = i) = \pi_i, i = 1,\dots, M$ and $\tilde{Z}|\tilde{\Delta} = i \sim F_i$. Let $(\tilde{Z}_1, \tilde{\Delta}_1), (\tilde{Z}_2, \tilde{\Delta}_2) $ i.i.d.\ samples from $\mu$ and $(\tilde{Z}, \tilde{\Delta}), (\tilde{Z}, \tilde{\Delta}^{\prime}) \sim \mu$ with $\tilde{\Delta}, \tilde{\Delta}^{\prime}$ conditionally independent given $\tilde{Z}$. Denote by $K$ a kernel function over the space $\{1,\dots,k\}$, e.g.\ the discrete kernel $K(x, y) := \mathbbm{1}(x = y)$. Then, the \textit{kernel measure of multi-sample dissimilarity} (KMD) is defined as
\[
\eta(F_1,\dots,F_k) := \frac{\E\left[K(\tilde{\Delta}, \tilde{\Delta}^{\prime})\right] - \E\left[K(\tilde{\Delta}_1, \tilde{\Delta}_2)\right]}{\E\left[K(\tilde{\Delta}, \tilde{\Delta})\right] - \E\left[K(\tilde{\Delta}_1, \tilde{\Delta}_2)\right]}.
\]
It has a lower bound of 0 that is attained if and only if the $k$ distributions coincide and an upper bound of 1 that is attained if and only if all distributions are mutually singular. Monotonicity of $\eta$ for location and scale alternatives for $k=2$, $\mathcal{X} = \mathbb{R}^p, p\ge1$ and log-concave distributions is shown such that values of KMD  in $(0,1)$ can be interpreted reasonably. Moreover, it is a member of the multi-distribution $f$-divergence as defined by \textcite{garcia-garcia_divergences_2012} (see Section \ref{sec:f.div}) and therefore fulfills all properties of the $f$-divergences. An estimator of $\eta$ can be defined as follows. Given the pooled sample $Z_1, \dots, Z_N$ and the corresponding sample memberships $\Delta_1,\dots, \Delta_N$ let $\mathcal{G}$ be a geometric graph on $\mathcal{X}$ such that an edge between two points $Z_i$ and $Z_j$ in the pooled sample implies that $Z_i$ and $Z_j$ are close, e.g.\ the $K$-nearest neighbor graph with $K\ge 1$ or the MST. Denote by $(Z_i,Z_j)\in\mathcal{E}(\mathcal{G})$ that there is an edge in $\mathcal{G}$ connecting $Z_i$ and $Z_j$. Moreover, let $o_i$ be the out-degree of $Z_i$ in $\mathcal{G}$. Then, an estimator for $\eta$ is defined as 
\[
\hat{\eta} := \frac{\frac{1}{N} \sum_{i=1}^N \frac{1}{o_i} \sum_{j:(Z_i,Z_j)\in\mathcal{E}(\mathcal{G})} K(\Delta_i, \Delta_j) - \frac{1}{N(N-1)} \sum_{i\ne j} K(\Delta_i, \Delta_j)}{\frac{1}{N}\sum_{i=1}^N K(\Delta_i, \Delta_i) - \frac{1}{N(N-1)} \sum_{i\ne j} K(\Delta_i, \Delta_j)}.
\]
This estimator is consistent and asymptotically normally distributed under assumptions similar to those of \textcite{deb_multivariate_2021} on the geometric graph and for characteristic kernel functions. The $k$-sample test based on KMD is shown to be consistent against all alternatives where at least two distributions are unequal and \textcite{huang_kernel_2022} provide a complete characterization of the asymptotic power and detection threshold of the test for $\mathcal{X} = \mathbb{R}^p$ and assuming that $P_i$ has a density w.r.t.\ the Lebesgue measure. Under $H_0$ the permutation and unconditional distribution of the estimator of KMD are both asymptotically normal and if $\mathcal{X}$ is a Euclidean space and the common distribution under $H_0$ has a Lebesgue density and under assumptions on the graph, the asymptotic null distribution is distribution-free. The test can be seen as a generalization of the two-sample statistic of the $K$-nearest neighbor test of \textcite{schilling_multivariate_1986} and \textcite{henze_multivariate_1988} or \textcite{petrie_graph-theoretic_2016}. It could therefore also be assigned to the class of graph-based tests. It is implemented in the R package \texttt{KMD} \parencite{KMD}.  
For the $K$-nearest neighbor graph (with $K\ge1$ fixed) the calculation of $\hat{\eta}$ has computational complexity $\mathcal{O}(KN\log N)$. The use of the nearest neighbor graph rather than MST is recommended because of flexibility and computational convenience. Moreover, they recommend to use $K =1$ for $K$-NN graph for estimation of $\eta$. For testing, larger values of $K$ are recommended.

\subsection{Methods based on binary classification}
\paragraph{Classifier tests of Friedman}\phantomsection\label{friedman_multivariate_2004}
The idea of measuring divergence between two distributions via separation and the misclassification error can be traced back as far as the 50's and 60's \parencite{rao_advanced_1952, ali_general_1966}. Later, \textcite{friedman_multivariate_2004} brings up the idea of using a binary classifier to distinguish between distributions generating the two datasets. For that, a binary classifier is trained on the pooled dataset $\mathcal{D} = \{(X_i, 1)\}_{i = 1}^{n_1} \cup \{(Y_i, -1)\}_{i = 1}^{n_2} =: \{(Z_i, L_i)\}_{i = 1}^N$. This binary classifier provides scores $s_i$ for the confidence that sample $i$ belongs to the first dataset ($L_i = 1$). The scores for the first and the second dataset can be seen as random samples from respective probability distributions with densities $f_+$ and $f_-$. Thus, a univariate two-sample test for equality of these densities, i.e.\ $H_0: f_+(s) = f_-(s)$, e.g.\ chi-squared, Kolmogorov-Smirnov, Mann-Whitney or $t$-test, can be used to compare the distributions. To perform such a test, there are two options. First, the data are split into a training and test set and only the training set is used to train the classifier while the test set is used to perform the test, making use of the known null distribution of the respective test statistic. Second, all observations are used for training the classifier as well as for testing. Then the null distribution of the test statistic from the univariate test is not valid. Instead, a permutation test is performed by randomly permuting the labels and calculating the test statistic values for the classifiers trained on the permuted data. The empirical $(1-\alpha)$-quantile of these statistics can then be used as the critical value. The power of the test highly depends on the classifier but is likely not very sensitive to the choice of the univariate test statistic. The sensitivity in the choice of the classifier can be exploited to obtain a higher power by choosing a classifier that fits the differences of distributions that are of particular interest. Moreover, depending on the classifier, the differences in the distributions can further be examined after a rejection of the null hypothesis, e.g.\ for decision trees.

\paragraph{Classifier two-sample tests (C2ST)}\phantomsection\label{sec.c2st}
The general idea of \textcite{lopez-paz_revisiting_2017} is to use a binary classifier for classifying to which of two datasets a sample belongs (here labeled by 0 and 1). If the datasets are generated from the same distribution, the accuracy should be close to chance level, otherwise, the classifier should be able to distinguish between the two distributions and hence the accuracy should be higher than chance level. A \textit{Classifier Two-Sample Test (C2ST)} based on these considerations learns a representation of the data on the fly, and its test statistic is in interpretable units. Moreover, the predictive uncertainty allows interpreting where the distributions differ.\\
For the definition of the test statistic, w.l.o.g.\ assume that $n_1 = n_2$ and that two samples are given over the same sample space. The C2ST then consists of five steps: 
\begin{enumerate}
	\item Construct the dataset 
	\[
	\mathcal{D} = \{(X_i, 0)\}_{i = 1}^{n_1} \cup \{(Y_i, 1)\}_{i = 1}^{n_2} =: \{(Z_i, L_i)\}_{i = 1}^N
	\]
	consisting of the samples from both datasets labeled with their membership to the two datasets.
	\item Shuffle $\mathcal{D}$ at random and split it into a disjoint training and test set $\mathcal{D}^{\text{tr}}$ and $\mathcal{D}^{\text{te}}$ with $n_{\text{te}} = |\mathcal{D}^{\text{te}}|$.
	\item Train a binary classifier $f:\mathcal{X}\to[0,1]$ on $\mathcal{D}^{\text{tr}}$ such that $f(z_i)$ is an estimate of the conditional probability distribution $p(L_i = 1\arrowvert Z_i)$.
	\item Calculate the C2ST statistic on $\mathcal{D}^{te}$ 
	\[
	\hat{T}_{\text{C2ST}} = \frac{1}{n_{\text{te}}} \sum_{(Z_i, L_i)\in \mathcal{D}^{\text{te}}} \mathbb{I}\left[\mathbb{I}\left(f(Z_i) > \frac{1}{2}\right) = L_i\right],
	\]
	which is the accuracy of the test set. $\mathbb{I}$ denotes the indicator function. The accuracy should be close to chance level if $F_1 = F_2$ and should be greater than chance level for $F_1 \ne F_2$, since then the classifier should identify distributional differences between the two samples.
	\item Calculate a $p$-value using the null distribution of the C2ST statistic, which is approximately $N(\frac{1}{2}, \frac{1}{4n_{\text{te}}})$.
\end{enumerate}
Maximizing the power of a C2ST is a trade-off between a large training set, to optimize the classifier, and a large test set $n_{te}$, to better evaluate the performance of the classifier. \\
The test statistic is interpretable as the percentage of samples that are correctly classified. Furthermore, the values $f(z_i)$ along with the true labels $l_i$ explain which samples were correctly or wrongly classified and with how much confidence. This provides information on where the two distributions differ. Using the classification-based approach also inherits the interpretability of the classifier to explain which features are most important for distinguishing between the two distributions.\\
In a simulation study, \textcite{lopez-paz_revisiting_2017} compare C2ST using a neural network and C2ST using a $K$-NN classifier against the Wilcoxon-Mann-Whitney test, KS test, and Kuiper test for one-dimensional data, and additionally the MMD test, ME test and SCF test for one-dimensional as well as multi-dimensional data. They repeat the experiments from \textcite{jitkrittum_interpretable_2016}. In all cases, C2ST shows a good performance. They observe that C2ST is better or nearly as good as SCF and MMD in the multi-dimensional case and nearly as good as the Kuiper and the ME test in the one-dimensional case.\\
\textcite{cai_two-sample_2020} argue that disadvantages of the C2ST are that the use of train/test data for estimating the prediction accuracy makes the test less efficient in data utilization and can slow down the computation. They show that a more powerful test can be derived by not using the prediction accuracy directly (see below). The test is implemented in the R package \texttt{Ecume} \parencite{Ecume}.

\paragraph{Regression based test}\phantomsection\label{kim_global_2019}
\textcite{kim_global_2019} derive a test that is intended for high-di\-men\-sio\-nal and complex data. A regression approach is used so the test can efficiently handle different types of data structures depending on the chosen regression model. Local differences can be identified with statistical confidence. The test gives a general framework for both global and local two-sample problems and for high-dimensional and non-Euclidean data. It is assumed that the densities of both distributions exist. The idea of the test is similar to that in other approaches based on binary classification. The equivalent null hypothesis based on regression for a binary outcome that determines the membership of data points is that the regression function does not depend on the features.
The test statistic measures the empirical distance between the regression function $\Prob(Y = 1| X = x)$ and the class probability $\Prob(Y = 1)$ which both take values in $(0,1)$. The power of the test can be related to the mean integrated squared error (MISE) of the chosen regression estimator. The null distribution of the test statistic is unknown and depends on the regression model and the distribution of the data. Therefore, a permutation test is performed. \textcite{kim_global_2019} use Fisher's LDA as the regression method and show optimality under the assumption of normal distributions with equal covariance matrices. In general, a train/ test split is required for the method. \textcite{kim_global_2019} assume that the MISE is smaller than a positive constant times an $o(1)$ term and that the permutation critical value is uniformly bounded by this term up to some constant factor with high probability. Then, the procedure yields a level $\alpha$ test, and for sufficiently large $N$ and for sufficiently large differences between the distributions, the type II error of the test is bounded. \textcite{kim_global_2019} use a linear smoother as the regression method (e.g.\ kNN regression, kernel regression, or local polynomial regression) for theoretical analysis. The convergence rates can be used for calculations on test errors. Note that the authors call their test regression-based, but model $\Prob(Y = 1| X = x)$ like in many of the other classification approaches.

\paragraph{Test based on the logit function of a classifier}\phantomsection\label{cheng_classification_2022}
\textcite{cheng_classification_2022} follow a slightly different approach for using a binary classifier network to distinguish between data from two different distributions. They train a classifier network and use the difference between both datasets of the provided logit function as the test statistic. An advantage of using networks is that the algorithm scales to large samples. Also, the use of networks is motivated by generalizing discriminative networks used in generative adversarial networks (GANs) from the goodness-of-fit problem to two-sample problems.\\
For the calculation of the test statistic, it is assumed w.l.o.g.\ that $N = n_1 + n_2$ is an even integer. Then, the test is performed via the following steps:
\begin{enumerate}
	\item Split the dataset $\mathcal{D}$ constructed as in \textcite{lopez-paz_revisiting_2017} into two halves used as training and test set with $n_1^{\text{te}}$ and $n_2^{\text{te}}$ denoting the number of samples from dataset one and two, respectively, in the test set. 
	\item Training: Train a binary classification neural network on the training set using softmax loss. This gives estimated class probabilities
	\begin{align*}
		\Prob(l = 0|z) &= \frac{\exp(u_{\theta}(z))}{\exp(u_{\theta}(z)) + \exp(v_{\theta}(z))}, \\
		\Prob(l = 1|z) &= \frac{\exp(v_{\theta}(z))}{\exp(u_{\theta}(z)) + \exp(v_{\theta}(z))}
	\end{align*}
	with $u_{\theta}(z)$ and $v_{\theta}(z)$ activations in the last hidden layer of the network and $\theta$ the network parametrization. The \textit{logit} is then defined as 
	\[
	f_{\theta} = u_{\theta} - v_{\theta}
	\]
	\item Testing: The test statistic is computed as 
	\[
	\hat{T}_{CC} = \frac{1}{n_1^{\text{te}}} \sum_{x\in X^{1, \text{te}}} f_{\theta}(x) - \frac{1}{n_2^{\text{te}}} \sum_{y\in X^{2, \text{te}}} f_{\theta}(y)
	\]
	with $f_{\theta}$ parametrized by a trained neural network and $X^{1, \text{te}}$ and $X^{2, \text{te}}$ denoting the subsets of the test set corresponding to the first and the second dataset. The critical value $\tau$ is calculated by a permutation test where the labels on the test set are randomly permuted $m_{\text{perm}}$ times and the test statistic is recomputed each time using the permuted labels. $\tau$ is set to the empirical $(1-\alpha)$-quantile of these test statistics.
\end{enumerate}
The test statistic can be viewed as estimating the symmetric KL divergence $\text{KL}(F_1, F_2) + \text{KL}(F_2, F1)$ (see Section \ref{sec:f.div}).\\
Under the assumption that the training is terminated after a fixed number of epochs, the overall complexity of the test is $\mathcal{O}(N)$. Under certain assumptions regarding the neural network and the densities of $F_1$ and $F_2$, the test is asymptotically consistent. Moreover, a reduction of the needed network complexity for densities on or near low-dimensional manifolds in ambient space is shown.\\
In a simulation, the test is compared to the one proposed by \textcite{lopez-paz_revisiting_2017} and to different kernel choices for the MMD test, where the kernel bandwidth is chosen as the median of the pairwise distances among all samples, as proposed in \textcite{gretton_kernel_2012}. \textcite{cheng_classification_2022} observe better performance of their test than for the C2ST and in certain settings (especially high dimensional data) also than for the MMD tests.

\paragraph{Test based on classification tree}\phantomsection\label{yu_two-sample_2007}
\textcite{yu_two-sample_2007} describe a two-sample test motivated by candidate gene association studies from the perspective of supervised machine learning. The estimated prediction error of a classification tree is used as a test statistic. A simulation study shows that the nominal type I error holds, but the power is sensitive to the chosen estimator for prediction error. The $.632+$ estimator results in the best overall performance. One advantage of the use of classification trees is that it enables the use of missing data since a tree can handle them via the use of surrogate variables.

\paragraph{Direction-projection-permutation (DiProPerm) test}\phantomsection\label{wei_direction-projection-permutation_2016}
\textcite{wei_direction-projection-permutation_2016} concentrate on the HDLSS setting and propose the so-called \textit{direction-projection-permutation (DiProPerm)} test as a tool to assess whether a binary linear classifier detects statistically significant differences between high-dimensional distributions. The main idea is to work directly with the one-dimensional projections induced by the binary linear classifier. According to \textcite{wei_direction-projection-permutation_2016}, consistency is a nontrivial property in the HDLSS asymptotic regime, but certain variations of DiProPerm are consistent. In HDLSS settings, for ease of interpretability linear classifiers are preferable to more complicated ones like random forests. The test statistic is a univariate two-sample statistic applied to the projection onto the normal vector of a separating hyperplane. A permutation test is performed. In general, the choice of the classifier is open, but \textcite{wei_direction-projection-permutation_2016} recommend using the distance weighted discrimination (DWD) classifier \parencite{marron_distance-weighted_2007}. Also, different test statistics can be chosen (e.g.\ difference in means, $t$-test statistic, AUC). The theoretical analysis is performed only for the centroid projection direction and on the mean difference (MD) statistic and the $t$-statistic because these have simple closed-form expressions. Similar assumptions are needed for HDLSS asymptotic theory as in \textcite{biswas_nonparametric_2014}. Under these assumptions, the test is only shown to be consistent for the alternative of unequal means. The proof of consistency is performed only under certain alternatives (equal means, different covariance matrices) and only for centroid-$t$-statistic, while the test based on centroid-MD is inconsistent in this setting. \textcite{montero-manso_two-sample_2019} mention that the test is not distribution-free. The test is implemented in the R package \texttt{diproperm} \parencite{diproperm}.

\paragraph{Classification probability test}\phantomsection\label{cai_two-sample_2020}
\textcite{cai_two-sample_2020} present a test, called the \textit{Classification Probability Test (CPT)}, based on estimates of classification probabilities from a classifier trained on the samples. It can be applied whenever there is an appropriate classifier to consistently estimate the classification probabilities. In contrast to other classification-based tests, this test is not based on classification accuracy. Instead of testing $H_0: F_1 = F_2$ directly, the idea is to equivalently test for hypotheses on the joint distribution of the data points and their dataset labels. For this, the odds ratio (OR) of probabilities that the label of a given feature point is one is used as a proxy for the likelihood ratio (LR) since $\text{LR} = \text{OR} \cdot \text{const}$ in this case. Since the test is an approximation of the LR test, asymptotically there should be no loss of information in contrast to the classification accuracy test proposed by \textcite{kim_classification_2021}. For the test, it is assumed that a consistent estimator of the classification probability is given. According to \textcite{cai_two-sample_2020}, more research is needed on sufficient conditions for that. In addition, the assumption is made that the density functions of both distributions exist. A permutation test is performed. The test statistic estimates the KL divergence whenever the law of large numbers holds. An advantage of the test is that it does not need any density or density ratio estimation but only class probability estimates that can be obtained efficiently by different classification algorithms. The test performance generally depends on the underlying distribution and the classifier. Under the condition of uniform consistency for the estimation of class probabilities, the test is asymptotically most powerful. This uniform consistency condition is strong and artificial. Therefore, a second test is proposed based on more heuristic arguments that the two-sample test is equivalent to determining if the mapping of observations to class probabilities is a constant function. For this test, the variance of the estimated class probabilities is considered as a test statistic and again a permutation test is performed. In both cases, a classifier has to be chosen. \textcite{cai_two-sample_2020} propose to choose it by $K$-fold cross-validation which is computationally intensive. \textcite{cai_two-sample_2013} do not mention it, but probably some sort of training set is needed to train the classifier in the first step.

\paragraph{Testing for deviation of classification accuracy from chance}\phantomsection\label{kim_classification_2021}
\textcite{kim_classification_2021} analyze a general test based on checking if the accuracy of a classifier is significantly different from chance and compare it with Hotellings $T^2$ test. If the true error remains by at least $\varepsilon>0$ better than chance as $p, N \to \infty$, then the permutation test is consistent. It is also computationally efficient. The permutation test offers exact control of the type I error rate and is consistent if the number of permutations is greater than $(1 - \alpha) / \alpha$. A test based on a Gaussian approximation is also shown to be consistent. It is simple but has no finite sample guarantees. \textcite{kim_classification_2021} focus their analysis on tests for Gaussian or elliptical distributions. For performing the test, a train/ test split is required. 

\paragraph{Test based on random forests}\phantomsection\label{hediger_use_2021}
\textcite{hediger_use_2021} provide a two-sample test based on the classification error of random forests that is applicable for any distribution. It requires almost no tuning, but for an asymptotic version of the test, both train and test set are required. Alternatively, an out-of-bag (OOB) based permutation test can be performed. OOB statistics can be used to increase the sample efficiency compared to the test based on a holdout sample. The variable importance measures of the random forest provide insights into sources of distributional differences. The test is implemented in the R-package \texttt{hypoRF} \parencite{hypoRF}.

\paragraph{Critique on accuracy based test}
\textcite{rosenblatt_better-than-chance_2021} criticize tests that analyze whether the estimated accuracy of a classifier is significantly better than chance level. Such tests can be underpowered compared to a ``bona fide statistical test'' and are also computationally more demanding. They examine candidate causes for low power, including the discrete nature of the accuracy test statistic, the types of signals that accuracy tests are designed to detect, the inefficient use of data, and a suboptimal regularization. For the analysis, they assume that the number of samples is in the order of the dimension or smaller. They demonstrate that in the high-dimensional regime accuracy tests never have more power than two-sample location or goodness-of-fit (GOF) tests. Problems with accuracy tests are that data splitting reduces the effective sample size, required regularization for testing seems to differ from that for predicting, and discretization makes the permutation tests conservative. The last point can not be captured in theoretical analyses as it decreases with sample size. Therefore, they recommend choosing a two-sample location or GOF test over an accuracy test and using appropriate regularization. For the use of accuracy tests, they recommend using larger test sets, regularization, and resampling with replacement. The results are fully based on a simulation study. No theoretical results are provided. 

\subsection{Distance and similarity measures for datasets}  

\paragraph{Distance and similarity based on metafeatures}\phantomsection\label{feurer_initializing_2015}
\textcite{feurer_initializing_2015} define a distance measure between datasets. They intend to use it for speeding up Sequential Model-based Bayesian Optimization (SMBO) for hyperparameter tuning by using configurations that performed well on similar datasets for initialization (meta-learning). Under the assumption that each dataset $\mathcal{D}^{(i)}$ can be described by a set of $K$ metafeatures $\boldmath{m}^{i} = (m_1^i,\dots,m_K^i)$ they propose two distance measures. The first one uses the $q$-norm of the difference between metafeatures of the datasets 
\[
D_{Fq}(\mathcal{D}^{(i)}, \mathcal{D}^{(j)}) = \|\boldmath{m}^i - \boldmath{m}^j\|_q.
\]
The second one measures similarity w.r.t.\ performance of different hyperparameter settings by using the negative Spearman correlation between ranked results of a fixed set of $n$ hyperparameter settings $\theta_l, l =1,\dots,n$, on both datasets 
\[
D_{Fc}(\mathcal{D}^{(i)}, \mathcal{D}^{(j)}) = 1 - \mathbb{C}\text{or}([g^{\mathcal{D}^{(i)}}(\theta_1),\dots,g^{\mathcal{D}^{(i)}}(\theta_n)], [g^{\mathcal{D}^{(j)}}(\theta_1),\dots,g^{\mathcal{D}^{(j)}}(\theta_n)]),
\]
with $g^{\mathcal{D}^{(i)}}$ denoting the target function. In the context of finding the most similar of the datasets for which the hyperparameters have already been tuned to a new dataset for which the tuning has not yet been performed, the distance from the old datasets to the new one cannot be calculated, since the $g^{\mathcal{D}^{(i)}}(\theta_l)$ are not known for this new dataset. Instead the distances are estimated using regression to learn a function mapping from pairs of metafeatures $\left(\boldmath{m}^i, \boldmath{m}^j\right)$ to $D_{Fc}(\mathcal{D}^{(i)}, \mathcal{D}^{(j)})$ based on the metafeatures and pairwise distances of the old datasets.
\textcite{feurer_initializing_2015} suggest 46 metafeatures found in the literature. These metafeatures can be categorized into 
\begin{itemize}
	\item simple metafeatures (describe basic dataset structure, e.g.\ number of features),
	\item PCA metafeatures,
	\item information-theoretic metafeatures (measure entropy),
	\item statistical metafeatures (use descriptive statistics to characterize dataset, e.g.\ kurtosis or dispersion of label distribution),
	\item landmarking metafeatures (are based on running several fast machine learning algorithms that can capture different properties of the dataset, e.g.\ linear separability).
\end{itemize}

\paragraph{Gromov-Hausdorff distance of metric measure spaces}\phantomsection\label{memoli_distances_2017}
\textcite{memoli_distances_2017} defines a distance between datasets via the Gromov-Hausdorff metric between metric measure spaces. The idea is to represent data as a metric space endowed with a probability measure \textit{(metric measure space)}
and then determine the distance between these metric measure spaces. Given two metric measure spaces $(\mathcal{X}, d_{\mathcal{X}}, \mu_{\mathcal{X}})$ and $(\mathcal{Y}, d_{\mathcal{Y}}, \mu_{\mathcal{Y}})$ corresponding to the two datasets, denote by $\mathcal{U}(\mu_{\mathcal{X}}, \mu_{\mathcal{Y}})$ the collection of all couplings between $\mu_{\mathcal{X}}$ and $\mu_{\mathcal{Y}}$, i.e.\ of all measures $\mu$ over $\mathcal{X}\times\mathcal{Y}$ such that the push-forward of $\mu$ (i.e.\ the measure $\mu \circ f^1 (A) = \mu(f^-1(A))$ for some measurable function $f$) for the first canonical projection $\pi_1$ is equal to $\mu_{\mathcal{X}}$, $\mu\circ\pi_1^{-1} = \mu_{\mathcal{X}}$, and analogously $\mu\circ\pi_2^{-1} = \mu_{\mathcal{Y}}$. Then, the \textit{Gromov- distance of order q} \parencite{memoli_gromovwasserstein_2011} is defined as 
\begin{align*}
	d_{\text{GW}, p}(\mathcal{X}, \mathcal{Y}) := \frac{1}{2} \inf_{\mu\in\mathcal{U}(\mu_{\mathcal{X}}, \mu_{\mathcal{Y}})} \left(\int\int \right. & |d_{\mathcal{X}}(x, x^{\prime}) - d_{\mathcal{Y}}(y, y^{\prime})|^q\\
	& \left.\mu(\dif x\times \dif y) \mu(\dif x^{\prime}\times \dif y^{\prime})\right)^{1/q}.
\end{align*}
This means that the function $(x, y, x^{\prime}, y^{\prime}) \mapsto |d_{\mathcal{X}}(x, x^{\prime}) - d_{\mathcal{Y}}(y, y^{\prime})|^q$ is integrated over the measure $\mu\otimes\mu$ for any $\mu\in\mathcal{U}(\mu_{\mathcal{X}}, \mu_{\mathcal{Y}})$ and the infimum with respect to $\mu$ is determined \parencite{memoli_distances_2017}.
For $q\ge1$ this defines a proper distance on the collection of isomorphism classes of metric measure spaces \parencite{memoli_gromovwasserstein_2011, memoli_distances_2017}. The calculation in practice remains unclear. Also the choice of the metrics $d_{\mathcal{X}}$ and $d_{\mathcal{X}}$ might be nontrivial in practice. 

\paragraph{Similarity based on method ranking}\phantomsection\label{leite_selecting_2012}
\textcite{leite_selecting_2012} work on meta-learning in situations where it is not possible to evaluate and compare all combinations of learning algorithms and their possible parameter settings. For that, they develop a new technique called \emph{active testing} that intelligently selects the most promising competitor for the next round of cross-validation based on prior duels between algorithms on similar datasets. Therefore, they characterize datasets based on the pairwise performance differences between algorithms. Their idea is that if the same algorithms win, tie or lose in comparisons, then the datasets are expected to be similar at least in terms of effects on learning performance. They propose four ways to calculate dataset similarity. The first measure, called AT0, is not of interest since it assumes the same similarity for all pairs of datasets and is only used as a baseline. The second one, AT1, works as AT0 at the beginning before any tests on the new data were performed. Then, in each of the next iterations of cross-validation (CV) on the new data, the similarity is estimated based on the most recent CV test as follows. All datasets for which the new current best algorithm is better than the old one are assigned a similarity value of 1, and all other datasets have a similarity value of 0. An alternative is to set the similarity to the difference of relative landmarks (performance gain of the new best compared to the old best) for all datasets for which the new current best algorithm is better than the old one and then normalize these values to the range between 0 and 1. The third measure, ATW, works like AT1 but uses all CV tests carried out on the new dataset and calculates the Laplace-corrected ratio of results in which the datasets had the same results. The last measure, called ATx, works similarly to ATW but it is required that all pairwise comparisons yield the same outcome. In that case, the similarity is set to one, and otherwise to zero. \textcite{leite_selecting_2012} present experiments to compare the different approaches. The results show that ATW and AT1 provide good performance using a small number of CV tests. Nonetheless, they believe that the results could be improved by using classical information-theoretic measures and/ or sampling landmarks for measuring the dataset similarity.\\
In \textcite{leite_exploiting_2021}, an improved version is presented that outperforms the previous active testing data similarity measures. For this, the performance gain of each algorithm on each dataset compared to the current best algorithm is estimated as the ratio of the performances of these algorithms divided by the ratio of the runtimes required for training the learners to the power of a parameter $q$. The authors recommend $q = 1/32$. The performance gain is estimated as this quantity minus one, if the resulting value is positive, and zero otherwise. The similarity of datasets is measured via the (weighted or unweighted) correlation of these estimated performance gains of all algorithms on the respective datasets.

\paragraph{Deep Dataset Dissimilarity Measures}\phantomsection\label{calderon_ramirez_dataset_2022} 
\textcite{calderon_ramirez_dataset_2022} define another set of dataset dissimilarity measure, called the \textit{deep dataset dissimliarity measures (DeDiMs)}. Their motivation is to asses a distribution mismatch between labeled and unlabelled data in semi-supervised deep learning (SSDL) and therefore to quantify the difference between datasets. In total, four distances are defined: two Minkowski-based distance measures and two nonparametric density-based dataset divergence measures. The general steps presented for calculation, given two datasets $\mathcal{D}^{a}$ and $\mathcal{D}^{b}$, are as follows:
\begin{enumerate}
	\item Draw a random subsample of $\mathcal{D}^{a}$ and $\mathcal{D}^{b}$ of size $\tau$ and denote these subsamples as $\mathcal{D}^{a, \tau}, \mathcal{D}^{b, \tau}$.
	\item Transform the observation $x_{i\bullet}\in\mathbb{R}^p$ of dataset $i \in\{a,b\}$ using a feature extractor $g$ to obtain the feature vector $h_i = g(x_{i\bullet})\in\mathbb{R}^{p^{\prime}}$. This yields the feature sets $H^{a, \tau}, H^{b, \tau}$.
\end{enumerate}
For calculating the Minkowski-based distance sets, afterward, the following steps are performed: 
\begin{enumerate}
	\item Calculate $\hat{d}_i = \min_k \|h_i - h_k\|_q$ for $q = 1$ (Manhatten distance) or for $q = 2$ (Euclidean distance) for each of $\mathcal{C}$ samples $h_i$ of $H^{a, \tau}$, where $h_k$ is the closest feature vector from $H^{b, \tau}$. This yields a list of distances $d_{\ell_q}(\mathcal{D}^{a}, \mathcal{D}^{b}, \tau, \mathcal{C}) = \left\{\hat{d}_1,\dots,\hat{d}_{\mathcal{C}}\right\}$.
	\item Calculate a reference list of distances for the same samples of the dataset $\mathcal{D}^{a}$ to itself (intra-dataset distance)  $d_{\ell_q}(\mathcal{D}^{a}, \mathcal{D}^{a}, \tau, \mathcal{C}) = \left\{\check{d}_1,\dots,\check{d}_{\mathcal{C}}\right\}$.
	\item Calculate the absolute differences between reference and inter-dataset distances $d_c = |\hat{d}_c - \check{d}_c|$ as well as their average reference subtracted distance $\bar{d}$ and the $p$-value of a Wilcoxon test on these differences.
\end{enumerate}
This approach can be seen as a method based on inter-point distances.
For the calculation of the density-based distances, the following steps are performed instead: 
\begin{enumerate}
	\item Compute the normalized histogram for each dimension $r = 1, \dots, p^{\prime}$ in the feature space to approximate the density function $f_{r,a}$ based on $H^{a, \tau}$ and $f_{r,b}$ based on $H^{b, \tau}$.
	\item Compute the sum of the dissimilarities between the density functions $f_{r, a}$ and $f_{r, b}$ for the Jensen-Shannon divergence ($d_{JS}$) or the cosine distance ($d_{C}$): $\hat{d}_i = \sum_{r= 1}^{p^{\prime}} \delta_g(f_{r, a}, f_{r,b}),$ $g = JS, C$ for all $\mathcal{C}$ samples (assumption: variables are statistically independent).
	\item Compute the intra-dataset distances $\check{d}_1,\dots,\check{d}_{\mathcal{C}}$.
	\item Calculate the absolute differences between reference and inter-dataset distances $d_c = |\hat{d}_c - \check{d}_c|$ as well as their average reference subtracted distance $\bar{d}$ and the $p$-value of a Wilcoxon test on these differences.
\end{enumerate}
This approach can be seen as a method based on comparing density functions or as a divergence.
The dissimilarity measures do not fulfill the conditions of a metric or pseudo-metric since the distance of a dataset to itself is in general not exactly zero and symmetry properties are not fulfilled. The distances are evaluated in a simulation study with regard to their ability to detect a  distribution mismatch and to increase SSDL performance. Both goals are achieved.

\paragraph{Distance based on optimal transport} \phantomsection\label{sec.ot}
\textcite{alvarez-melis_geometric_2020} define a distance between datasets relying on optimal transport. They motivate the need for such distances by stating that methods to combine, adapt, and transfer knowledge across datasets need a notion of distance between datasets while ``the notion of distance between datasets is an elusive one, and quantifying it efficiently and in a principled manner remains largely an open problem''. They criticize that current methods to quantify the distance of two datasets are often heuristic, and highly dependent on tuning and on the architecture of a certain task. Also, many of the other proposals do not take the target variable into account. Therefore, \textcite{alvarez-melis_geometric_2020} propose a new distance between datasets that is model-agnostic, does not involve training, can compare datasets even if their label sets are disjoint, and has a theoretical footing. Their empirical results also show a good correlation with how hard a transfer-learning task is.\\
The definition of their distance heavily relies on the optimal transport (OT) problem. Therefore, we define this first in the following. Consider a complete, separable metric space $\mathcal{X}$ and a probability measures $\alpha, \beta\in\mathcal{P}(\mathcal{X})$. The optimal transport according to \textcite{kantorovitch_translocation_1958} is defined as 
\[
\text{OT}(\alpha, \beta) := \min_{\pi\in\Pi(\alpha, \beta)} \int_{\mathcal{X}\times\mathcal{X}} c(x, y) \dif \pi(x, y), \label{eq:OT}
\]
where $c:\mathcal{X}\times\mathcal{X} \to \mathbb{R}^+$ is a cost function, the so-called ground cost, and 
\[
\Pi(\alpha, \beta) := \{\pi_{1,2}\in\mathcal{P}(\mathcal{X}\times\mathcal{X})\arrowvert \pi_1 = \alpha, \pi_2 = \beta\}
\]
is the set of joint distributions over the product space $\mathcal{X}\times\mathcal{X}$ with marginal distributions $\alpha$ and $\beta$. If $\mathcal{X}$ is provided with a metric $d_{\mathcal{X}}$, it is natural to use this as ground cost. In the special cases of $c(x, y) = d_{\mathcal{X}}(x, y)^q$ with $q\ge1$, the term 
\begin{equation} 
	W_q(\alpha, \beta) := \text{OT}(\alpha, \beta)^{1/q}\label{eq:wasserstein}
\end{equation} 
is the $q$-Wasserstein distance, for $q = 1$ also called Earth Mover's Distance. Finite samples as usually given in practice implicitly define discrete measures for which the pairwise cost can be represented as a cost matrix. The OT then becomes a linear program. Solving this is often difficult due to its cubic complexity. The entropy-regularized problem
\[
\text{OT}_{\varepsilon}(\alpha, \beta) := \min_{\pi\in\Pi(\alpha, \beta)} \int_{\mathcal{X}\times\mathcal{X}} c(x, y) \dif \pi(x, y) + \varepsilon H(\pi \alpha\otimes\beta),
\] 
where $H(\pi \alpha\otimes\beta) = \int \log(\dif\pi / \dif\alpha\, \dif\beta) \dif \pi$ is the relative entropy and $\varepsilon$ gives a time vs. accuracy trade-off, is more efficiently to solve. Based on this, the \textit{Sinkhorn divergence} \parencite{genevay_learning_2018}
\[
\text{SD}_{\varepsilon}(\alpha, \beta) = \text{OT}_{\varepsilon}(\alpha, \beta) - \frac{1}{2} \text{OT}_{\varepsilon}(\alpha, \alpha) - \frac{1}{2} \text{OT}_{\varepsilon}(\beta, \beta)
\]
can be calculated.\\
\textcite{alvarez-melis_geometric_2020} define a dataset $\mathcal{D}$ as a set of feature-label pairs $z:= (x, y)\in\mathcal{X}\times\mathcal{Y} =: \mathcal{Z}$ over a feature space $\mathcal{X}$ and a label set $\mathcal{Y}$. They focus on classification and therefore assume $\mathcal{Y}$ to be finite. Moreover, for simplicity, it is assumed that two datasets $\mathcal{D}_1$ and $\mathcal{D}_2$ are given whose feature spaces have the same dimensionality. It is not required as an assumption, but \textcite{alvarez-melis_geometric_2020} find it useful to think of samples in datasets as being drawn from joint distributions $F_1(x,y)$ and $F_2(x,y)$. \\
To define the distance without relying on external models or parameters, a metric on $\mathcal{Z}$ is needed. Given metrics on $\mathcal{X}$ and $\mathcal{Y}$ one could define $d_{\mathcal{Z}}(z, z^\prime) = (d_{\mathcal{X}}(x, x^{\prime})^q + d_{\mathcal{Y}}(y, y^{\prime})^q)^{1/q}$ for $q\ge 1$, but $d_{\mathcal{Y}}$ is  rarely readily available. Since information about the occurrence of $y$ in relation to feature vectors $x$ is given, instead the metric in $\mathcal{X}$ can be used to compare labels. Let 
\[
N_{\mathcal{D}}(y) := \{x\in\mathcal{X}\arrowvert(x,y)\in\mathcal{D}\}
\]
be the set of feature vectors with label $y$ and $n_y = |N_{\mathcal{D}}(y)|$ its cardinality. The labels are to be represented by their distribution over the feature space $y\mapsto\alpha_y(X):=\Prob(X| Y = y)$. The set $N_{\mathcal{D}}(y)$ can be understood as a finite sample of that. That given, choosing a distance between labels is equal to choosing a divergence between the associated distributions. \textcite{alvarez-melis_geometric_2020} propose OT as an ideal choice since it yields a true metric, it is computable from finite samples, and it is able to deal with sparsely supported distributions. $d_{\mathcal{X}}^q$ can be used as the optimal transport cost which results in the $q$-Wasserstein distance $W_q^q(\alpha_y, \alpha_{y^{\prime}})$ (see (\ref{eq:wasserstein})) between labels. With this, the distance between feature-label pairs can be defined as 
\[
d_{\mathcal{Z}}(z, z^\prime) := (d_{\mathcal{X}}(x, x^{\prime})^q + W_q^q(\alpha_y, \alpha_{y^{\prime}}))^{1/q}.
\]
This distance can be used in optimal transport to finally define a distance between measures (i.e.\ datasets):
\[
d_{\text{OT}}(\mathcal{D}_1, \mathcal{D}_2) = \min_{\pi\in\Pi(\alpha, \beta)} \int_{\mathcal{Z}\times\mathcal{Z}} d_{\mathcal{Z}}(z, z^\prime)^q \dif \pi(z, z^\prime)
\]
This defines a true metric on $\mathcal{P}(\mathcal{Z})$ which \textcite{alvarez-melis_geometric_2020} call the \textit{Optimal Transport Dataset Distance (OTDD)}. \\
There are different approaches to represent the distributions $\alpha_y$, depending on the size of the dataset. 
In the first approach, the samples in $N_{\mathcal{D}}(y)$ can be treated as support points of a uniform empirical measure so that $\alpha_y= \sum_{x\in N_{\mathcal{D}}(y)} \frac{1}{n_y} \delta_x$. When applying this, in every evaluation of $d_{\mathcal{Z}}(z, z^\prime)$ an OT problem needs to be solved which yields a total worst-case $\mathcal{O}(N^5\log N)$ complexity and makes this approach only feasible for small to medium-sized datasets. For these, e.g.\ when $p\gg N$, in simulations for $N\lesssim 5000$, it might even be faster than a proposed second approach. For this second approach, each $\alpha_y$ is modeled as a Gaussian $N(\hat{\mu}_y, \hat{\Sigma}_y)$ with $\hat{\mu}_y$ the sample mean and $\hat{\Sigma}_y$  the covariance of $N_{\mathcal{D}}(y)$. Then, the 2-Wasserstein distance has an analytic form, known as Bures-Wasserstein distance. The distance defined using this approach is denoted as $d_{\text{OT}-\mathcal{N}}$ or Bures-OTDD. It might be the only feasible approach for $n\gg p$ very large.\\
It holds $d_{\text{OT}-\mathcal{N}}(\mathcal{D}_1, \mathcal{D}_2)\le d_{\text{OT}}(\mathcal{D}_1, \mathcal{D}_2)\le d_{\text{UB}}(\mathcal{D}_1, \mathcal{D}_2)$ for any two datasets, where $d_{\text{UB}}$ is a distribution-agnostic OT upper bound defined by the OT distance using a certain cost function. For datasets of sizes $n_1$ and $n_2$ with $k_1$ and $k_2$ classes, dimension $p$ and maximum class size $m$, both distances cause costs of $\mathcal{O}(n_1n_2\log(\max\{n_1, n_2\})\tau^{-3})$ for solving the outer OT problem $\tau$-approximately, and the worst-case complexity for computing label-to-label pairwise distances is $\mathcal{O}(n_1 n_2(p + m^3\log m + p m^2))$ for $d_{\text{OT}}$ and $\mathcal{O}(n_1 n_1 p + k_1 k_2 p^3 + p^2 m(k_1 + k_2))$ for $d_{\text{OT}-\mathcal{N}}$. Under more assumptions and simplifications, additional speed-ups are possible. To speed up the calculations it is also possible to use the Sinkhorn divergence with approximate OT solution for the inner OT problem.\\
\textcite{alvarez-melis_geometric_2020} suggest assessing how realistic assumptions such as the use of Gaussian distributions or the choice of the entropy regularization parameters are before using their method, in order to avoid an unreliable distance estimation.
The OTDD can alternatively be seen as an inter-point distance-based method.

\subsection{Comparison based on summary statistics}

\paragraph{DataSpheres}\phantomsection\label{johnson_comparing_1998}
\textcite{johnson_comparing_1998} aim to develop a fast, inexpensive me\-thod for massive high-dimensional datasets that does not rely on any distributional assumptions. The idea is to generate a so-called \textit{DataSphere} (map of the dataset) which is a summary of the data, and compare these DataSpheres. The DataSphere can be generated in two passes over the data and can also be further aggregated. It partitions data into sections and represents each section through a set of summaries, which \textcite{johnson_comparing_1998} call \textit{profiles}. Then, tests for these profiles can be used to determine which datasets changed and where. For these tests, a set of weaker hypotheses that only need the profile information is used instead of testing if the joint distribution of the variables is the same for the two datasets.\\
For the construction of the DataSpheres, the following assumptions are made for the dataset $\mathcal{D}^T = (X_{1\bullet},\dots,X_{n\bullet})$: each $X_i\in\mathbb{R}^{p + 1}$ with $p = v + c$ consists of $p + 1$ attributes of which $c$ are categorical, $v$ are value attributes and one attribute is the dataset membership with value 1 or 2. Let $S_j = \{X_{i\bullet}:$ dataset membership attribute has value $j\},\, j\in\{1, 2\}$ and let $C$ be a particular value of the categorical variables in $D$. The \textit{subpopulation} $\mathcal{D}[C_j]$ is defined as the tuples in $\mathcal{D}$ that have value $C$ in their categorical features and value $j$ in their dataset membership attribute. $V[C_j]$ is defined as the projection of $\mathcal{D}[C_i]$ to the value attributes.
Now, for each value of $C$ that is present in $D$, it is examined if the distribution of $V[C_1]$ differs significantly from the distribution of $V[C_2]$. Therefore, $\mathcal{D}$ is partitioned into $K$ \textit{layers} $\{\mathcal{D}_\ell\}_{\ell= 1}^K$ that are more homogeneous than the entire dataset. This is achieved by defining each layer as a set of data points that are within the same (Mahalanobis) distance range from a center of the data cloud (defined as the vector of trimmed means). The cutoffs for the ranges are defined using a fast approximative quantiling algorithm, so each layer contains the same number of data points. Additionally, directional information is included through the use of \textit{pyramids}: a $d$ dimensional set can be partitioned into $2d$ pyramids $P_{\ell\pm}, i=1,\dots,d$ 
\begin{align*}
	P_{\ell+} &= \{X_{i\bullet}: |\tilde{x}_{i\ell}| > |\tilde{x}_{ij}|, \tilde{x}_{i\ell} > 0, j = 1,\dots,d, j\ne \ell\} \\
	P_{\ell-} &= \{X_{i\bullet}: |\tilde{x}_{i\ell}| > |\tilde{x}_{ij}|, \tilde{x}_{i\ell} < 0, j = 1,\dots,d, j\ne \ell\}
\end{align*}
with $\tilde{x}$ the normalized vectors. The tops of all pyramids meet at the center of the data cloud. A \textit{section} $S(\mathcal{D}_\ell, P_{\ell\pm}, C)$ is now defined as the data points with categorical attributes $C$ such that the value attributes lie in layer $\mathcal{D}_\ell$ and pyramid $P_{\ell\pm}$. Sections are summarized through sets of statistics, called profiles $P(\mathcal{D}_\ell, P_{\ell\pm}, C)$. For a dataset comparison the number of data points, the vector of means of value attributes, and the covariance matrix are used as statistics in the profile. A collection of profiles is called \textit{data map} of a dataset. A data map can be seen as a representation of the dataset.\\
The authors propose the use of two different tests. The first test is the \textit{multinomial test for proportions}. It compares the proportion of points falling into each section within a subpopulation. The second test is the \textit{Mahalanobis} $D^2$ test (same as Hotelling's test), which is used to establish the closeness of the multivariate means of each layer within each subpopulation for the two datasets. Both tests are described in detail by \textcite{rao1973linear}. Two different tests are used since for passing the tests it is sufficient but not necessary that the joint distribution in the two datasets is the same.

\paragraph{Constrained minimum (CM) distance}\phantomsection\label{tatti_distances_2007}
\textcite{tatti_distances_2007} defines a distance of two datasets that is based on summary statistics but also takes into account their correlation. The so-called Constrained Minimum (\textit{CM}) Distance can be computed in cubic time. \textcite{tatti_distances_2007} lists several properties that a distance of datasets should fulfill: First of all, it should be a metric since metric theory is a well-known area and metrics have many theoretical and practical advantages. It also should take the statistical nature of the datasets into account, e.g.\ the distance should approach zero for an increasing number of data points when both datasets are generated from the same distribution. Finally, it should be quick to evaluate since data may be high dimensional. Motivated by these requirements the CM distance is defined. \\
For this, first, define a \textit{feature function} $S:\mathcal{X}\to\mathbb{R}^m$ that maps points from the sample space $\mathcal{X}$ to a real vector. The \textit{frequency} $\theta\in\mathbb{R}^m$ of $S$ with respect to dataset $\mathcal{D}$ is the average of the values of $S$ 
\[
\theta= \frac{1}{N}\sum_{i = 1}^{N} S(X_{i\bullet}).
\]
Let $\mathcal{P}$ be the set of all distributions on $\mathcal{X}$. Then, a distribution $F\in\mathcal{P}$ satisfies the frequency~$\theta$ if $\E_{F}(S) = \theta$. 
Assume that the points in $\mathcal{X}$ can be enumerated as $\mathcal{X}= \{1, 2,\dots,|\mathcal{X}|\}$. Then, each distribution $F\in\mathcal{P}$ can be represented by a vector $u\in\mathbb{R}^{|\mathcal{X}|}$ with elements $u_i = f(i)$.
Define a \textit{constrained space}
\[
\mathcal{C}(S, \theta) = \left\{u\in\mathbb{R}^{|\mathcal{X}|} \left| \sum_{i\in\mathcal{X}} S(i) u_i = \theta, \sum_{i\in\mathcal{X}} u_i = 1\right. \right\}
\]
of distributions satisfying $\theta$. Then, interpreting the distributions as geometrical objects,  $\mathcal{C}(S, \theta)$ is an affine space since the constraints defining it are vector products. This implies that the constrained spaces for two different frequencies $\theta_1$ and $\theta_2$ are parallel. The distance between two parallel affine spaces can be measured by the shortest segment going from a point in the first space to a point in the second space, and this segment can be found by taking the points from both spaces that have the shortest norm. Motivated by this, the \textit{Constrained Minimum (CM) Distance} is defined as follows. Given two datasets $\mathcal{D}_1$ and $\mathcal{D}_2$ pick a vector from each constrained space having the shortest norm 
\[
u_i = \arg\min_{u\in C(S, S(\mathcal{D}_i))} \|u\|_2, i = 1, 2,
\]
and define the CM distance between the datasets as 
\[
D_{\text{CM}}(\mathcal{D}_1, \mathcal{D}_2|S) = \sqrt{|\mathcal{X}|}\|u_1 - u_2\|_2.
\]
The vectors $u_1$ or $u_2$ may have negative elements, thus the CM distance is not a distance between two distributions but rather a distance based on the frequencies of a given feature function motivated by the geometrical interpretation of the distribution sets. For calculation purposes, the CM distance can be rewritten as 
\[
D_{\text{CM}}(\mathcal{D}_1, \mathcal{D}_2|S)^2 = (\theta_1 - \theta_2)^T\mathbb{C}\text{ov}^{-1}(S)(\theta_1 - \theta_2),
\]
with
\[
\mathbb{C}\text{ov}(S) = \frac{1}{|\mathcal{X}|}\sum_{\omega\in\mathcal{X}} S(\omega)S(\omega)^T - \left(\frac{1}{|\mathcal{X}|}\sum_{\omega\in\mathcal{X}}S(\omega)\right)\left(\frac{1}{|\mathcal{X}|}\sum_{\omega\in\mathcal{X}}S(\omega)\right)^T.
\]
The CM distance fulfills the following properties: $D_{\text{CM}}(\mathcal{D}_1, \mathcal{D}_2|S)$ is a pseudo metric.	If $\mathcal{D}_1$ and $\mathcal{D}_2$ have the same number of items and $\mathcal{D}_1$, $\mathcal{D}_2$, and $\mathcal{D}_3$ are datasets with the same features, then $D_{\text{CM}}(\mathcal{D}_1\cup\mathcal{D}_3, \mathcal{D}_2\cup\mathcal{D}_3|S) = (1 - \varepsilon) D_{\text{CM}}(\mathcal{D}_1, \mathcal{D}_2|S)$ with $\varepsilon = \frac{|\mathcal{D}_3|}{|\mathcal{D}_1| + |\mathcal{D}_3|}$. This means that adding external data to the original datasets makes the distance smaller. Furthermore, adding extra features cannot decrease the distance. Also, for $T(\omega) = AS(\omega) + b$ with an invertible $N\times N$ matrix $A$ and a vector $b\in \mathbb{R}^N$, it holds that $D_{\text{CM}}(\mathcal{D}_1, \mathcal{D}_2|T) = D_{\text{CM}}(\mathcal{D}_1, \mathcal{D}_2|S)$.\\
Proposals for the choice of a feature function $S$ are means of features or means and pairwise correlations or frequent itemsets.\\
For binary data and $S$ chosen as the conjunction function, i.e.\ $S$ is one if all components of an observation are one, and zero otherwise, or as the parity function, i.e. $S$ is one if an odd number of components of an observation are one, and zero otherwise, the CM distance reduces to a more simple form. In these cases, it can be calculated as 
\[
	D_{\text{CM}}(\mathcal{D}_1, \mathcal{D}_2|S) = 2\|\theta_1 - \theta_2\|_2.
\]
Note that the factor $\sqrt{2}$ instead of 2 that is stated in the original publication of \textcite{tatti_distances_2007} in formula (4) and in Example 3 is not correct as can be seen from the proof of Lemma 8 from which these formulas follow. From $\E(S^2) = \E(S) = 0.5$, it follows that $\mathbb{V}\text{ar}(S) = \E(S^2) - \E(S)^2 = 0.25$ and therefore $\mathbb{C}\text{ov}(S) = 0.25I$ instead of $0.5I$ as claimed in Lemma 8, where $I$ denotes the identity matrix.

\subsection{Different testing approaches}\label{sec:testing}
\paragraph{General Bootstrap test}\phantomsection\label{romano_bootstrap_1989}
\textcite{romano_bootstrap_1989} studies the asymptotic behavior of some nonparametric tests and shows that under fairly general conditions Bootstrap and randomization tests are equivalent (i.e.\ the difference in critical functions evaluated at the observed data tends to 0 in probability). The results hold for general applications and the $k$-sample problem is only one application among others. A very general test statistic for $k$-sample problems is presented. Its exact form is not specified. The test of \textcite{bickel_distribution_1969} is a special case for $p$-dimensional data and $k=2$, the KS test is a special case for $p = 1$ and $k = 2$. \textcite{romano_bootstrap_1989} shows consistency for Bootstrap and permutation tests under some assumptions on the weights for the test statistic and on the distributions of the data.\\

\paragraph{Weighted Bootstrap test}\phantomsection\label{burke_multivariate_2000}
\textcite{burke_multivariate_2000} designs a test using a weighted Bootstrap method based on independent random variables instead of sampling from the uniform distribution. Additionally, uniform confidence bands for the distribution function of multivariate data are constructed. Asymptotically consistent multivariate versions of the KS test and the Cramér-von Mises test are proposed.\\

\paragraph{Test based on projections I}\phantomsection\label{ping_bootstrap_2000}
\textcite{ping_bootstrap_2000} considers the two- and $k$-sample problem. Projection pursuit-type statistics are used to overcome the sparseness of data points in high-dimensional space. The limiting distributions of the test statistics are not tractable and depend on the underlying distribution. Therefore, the properties of a Bootstrap approximation are examined. An approximation for statistics based on a number theoretic method is used for computational reasons. This number-theoretic method chooses directions for projections from the unit sphere. The presented tests are projection versions of the KS-test, CvM-test, and Anderson test. For the theoretical results, only continuous distributions are considered. Consistency is discussed implicitly by proving that the test statistics tend to infinity with probability one as $n_i\to\infty$.
\\

\paragraph{Test based on empirical Bayes factors}\phantomsection\label{chen_bayesian_2014}
In \textcite{chen_bayesian_2014}, empirical Bayes factors constructed from independent Polya tree priors are proposed as a test statistic for the two-sample problem. From this, $p$-values can be obtained by permuting the group membership indicator. The test was proposed to test whether data distributions are the same across several subpopulations. Initially, it was designed for univariate distributions only but an extension to multivariate distributions is also provided. Both versions are applicable to the $k$-sample problem. The goal of \textcite{chen_bayesian_2014} is to design a test that performs almost as well as the $t$-test for approximately normal data, but substantially better for non-normal data. Their test statistic is the ratio of marginal densities under $H_1$ and $H_0$. The permutation test rejects $H_0$ for large values of the test statistic. In the limiting case, the test corresponds to the likelihood ratio test based on normal data. For approximately normal data, it behaves similarly to a $t$-test but pronounced data-driven deviations from normality are also taken into account. \textcite{chen_bayesian_2014} are able to give the exact closed-form expression for the marginal density due to the conjugate property of the Polya tree. However, this prior is only suitable for continuous data. \textcite{chen_bayesian_2014} center the Polya tree at the normal distribution since they assume that ``many datasets are approximately normal, and therefore centering at normal can improve power compared to other nonparametric models that assume nothing''. 

The test of \textcite{chen_bayesian_2014} extends the former approaches of \textcite{holmes_two-sample_2015} and \textcite{ma_coupling_2011} to the $k$-sample problem and to censored data. According to simulations, their new test has higher power. For testing, several parameters are chosen via heuristics. The computational cost is $\mathcal{O}(pN^2)$ in the multivariate case. According to \textcite{chen_bayesian_2014}, in their examples computing permutation $p$-values took less than 5 minutes in each case, using \texttt{R} on an ``old Windows-based laptop''. For Bayes factors based on an infinite Polya tree, posterior consistency can be shown.

\paragraph{Projections obtained by maximization of a smooth test statistic}\phantomsection\label{zhou_two-sample_2017}
\textcite{zhou_two-sample_2017} propose a test that modifies Neyman's smooth test and extends it to the multivariate case based on projection pursue. They use a Bootstrap method to compute the critical value. Similar to \textcite{ghosh_distribution-free_2016}, they apply the idea that $H_0$ is equivalent to $H_0: u^TX=_d u^TY\,\forall u\in \mathcal{S}^{p-1}$ with $\mathcal{S}^{p-1}$ denoting the unit sphere in $\mathbb{R}^p$. They assume that the two sample sizes are comparable ($c_0n_1\le n_2\le n_1, c_0\in(0,1]$) and that $n_2\le n_1$. For the projections in the directions of each $u$ vector, \textcite{zhou_two-sample_2017} use multiple vectors $u\in \mathcal{S}^{p-1}$ and calculate a univariate smooth-type test statistic which is the supremum norm of a vector of means of several orthonormal functions applied to values of the distribution function of $u$ evaluated at the cross product of $u$ with the observations of the second dataset. The choice of the orthonormal functions remains unclear. The final test statistic is the (scaled) maximum of test statistics for different $u$ vectors. $H_0$ is rejected for large values. The limiting distribution of the test statistic may not exist, therefore a Gaussian process approximation of the test statistic and its estimator are given. Multiplier Bootstrap is proposed for testing. For the analysis of the test, \textcite{zhou_two-sample_2017} make the assumption of absolute continuous distribution functions and the assumption that the $d$ orthonormal functions from $[0,1]\to \mathbb{R}$ are twice differentiable, with $d\le n_1$. For $n_2\to\infty$, additional assumptions on the maximum over the supremum norm of each function and its first and second derivative are made. The assumptions are fulfilled for normalized Legendre polynomials with $d = o((n_2/\log n_2)^{1/9})$ and for a trigonometric series with  $d = o((n_2/\log n_2)^{1/4})$. Then the difference between the $\alpha$ level and the type I error of smooth test tends to zero for $n_2\to\infty$. Moreover, power against local alternatives tends to~1 for $n_2,d\to\infty$, for normalized Legendre polynomials with $d = o(n_2^{1/9})$, and for trigonometric series with $d = o(n_2^{1/4})$. To show that the test asymptotically holds the level $\alpha$ for growing $n_1, n_2$ and possibly $p$, two assumptions are required. First, $d\le\min\{n_1,n_2,\exp(C_0 p )\}$ has to hold for some positive constant $C_0$. Second, a bound for the maximum over the supremum norm of each of the first and second derivatives of orthonormal functions that grows with $n_2$ is required. The choice of $d$ remains open, and according to \textcite{zhou_two-sample_2017} in practice an optimal choice of $d$ is also not possible. The computation of the multivariate test statistic requires solving an optimization problem with an $\ell_2$-norm constraint. The best optimizer remains unclear.

\paragraph{Test based on ball divergence}\phantomsection\label{pan_ball_2018}
\textcite{pan_ball_2018} introduce a novel measure of the difference between two probability measures in separable Banach spaces, called \textit{Ball Divergence}. The Ball Divergence is defined as the square of the measure difference over a given closed ball collection. It is equal to zero if and only if the probability measures are identical and does not require any moment assumptions. Based on the Ball Divergence, \textcite{pan_ball_2018} propose a metric rank test procedure. Its empirical test statistic is defined based on the difference between averages of the metric ranks. It is robust to outliers and heavy-tail data. The distribution of the test statistic converges to a mixture of $\chi^2$ distributions under the null hypothesis, and it converges to a normal distribution with mean 0 and variance depending on the asymptotic proportion of the sample from the first distribution under the alternative hypothesis. The test does not depend on the ratio of sample sizes and thus can also be applied to imbalanced data. \textcite{pan_ball_2018} state that existing methods do not take extremely imbalanced data into account.\\
The newly proposed test relies on the fact that two Borel probability measures are identical if they agree on all balls in a separable Banach space \parencite{preiss_measures_1991}. It can be applied for data in separable Banach spaces, which overcomes the limitation that many Banach spaces are not of the strong negative type or even of negative type (e.g.\ $\mathbb{R}^p$ with $\ell^q$ metric for $3\le p\le\infty, 2<q\le\infty$) such that e.g.\ the generalized energy distance is not applicable. The square root of the Ball Divergence is a symmetric divergence, but not a metric since it does not satisfy the triangle inequality. The testing procedure can be generalized further to the $k$-sample problem. A connection to the MMD and to the energy statistic is shown through a unified framework of variograms. Consistency against any general alternative can be shown without any additional assumptions and independent of the ratio between the smaller and the larger sample size. \textcite{li_measuring_2022} conclude that the test by \textcite{pan_ball_2018} is  model-free and not constrained by any arguments. The test is implemented in the R package \texttt{Ball} \parencite{ball}.\\

\paragraph{Test based on Jackknife empirical likelihood}\phantomsection\label{wan_empirical_2018}
\textcite{wan_empirical_2018} present a Jackknife Empirical Likelihood (JEL) test that is motivated by the fact that the energy statistic is zero if and only if the two distributions are equal, under the assumption that first moments exist. \textcite{wan_empirical_2018} aim to avoid the problem of an asymptotic distribution that depends on unknown parameters by using the estimated likelihood method to obtain a distribution-free asymptotic behavior. Their test statistic is asymptotically $\chi^2_1$ distributed for any fixed dimension. A Jackknife Empirical Likelihood (EL) is used to circumvent solving nonlinear constraints for a U-statistic as the main obstacle of the EL method. The resulting test statistic is the nonparametric jackknife empirical log-likelihood ratio. To derive its asymptotic distribution, it is assumed that second moments for $\|X-Y\|$ and for the conditional expectations of $\|X-Y\|$, $\|X-X^{\prime}\|$, $\|Y-Y^{\prime}\|$ exist. Under these assumptions, it can also be shown that the resulting asymptotic test is consistent against all fixed alternatives. Under additional assumptions on expectations and on the covariance matrices of $X$ and $Y$, the test is also shown to be consistent against contiguous alternatives $H_1: F_1 = (1 - \delta_{n_1, n_2}) F_2 + \delta_{n_1, n_2} Q$, where $Q$ is a disturbance distribution and $\delta_{n_1, n_2} = \mathcal{O}(N^{-1/2})$.

\paragraph{Test based on projection averaging for Cramér-von Mises statistic}\phantomsection\label{kim_robust_2020}
\textcite{kim_robust_2020} introduce a generalization of the Cramér-von Mises test to the multivariate two-sample problem via projection averaging. They show that the test is consistent against all fixed alternatives and minimax rate optimal against a certain class of alternatives. Moreover, it is robust to heavy-tailed data, free of tuning parameters, and computationally efficient even in high dimensions. The test is shown to have comparable power to existing high-dimensional mean tests under certain location models for $p\to \infty$. \textcite{kim_robust_2020} propose a new metric called \textit{angular distance} as a robust alternative to the Euclidean distance. This solves the problem of the energy statistic that requires that first moments exist, which might be violated for high-dimensional data where outlying observations occur frequently. By introducing the angular distance, a connection to the RKHS approach can be made. The newly proposed test statistic is an unbiased estimate of the squared multivariate Cramér-von Mises statistic and has a simple closed-form expression. It is invariant to orthogonal transformations, nonnegative, and equal to zero if and only if the distributions are equal. Based on this statistic, a permutation test can be performed. It has the same asymptotic power as the oracle test and asymptotic tests that assume knowledge of the underlying distributions, for fixed and contiguous alternatives. \textcite{kim_robust_2020} show that the new test has acceptable power in the contamination model while the energy statistic has very low power. They analyze the finite-sample power and prove minimax rate optimality against a class of alternatives that differ from the null in terms of the CvM-distance. They show that the energy test is not optimal in that context. Moreover, they show consistency in the HDLSS setting under certain conditions. It is also shown that the multivariate CvM-distance is a special case of the generalized energy statistic \parencite{sejdinovic_equivalence_2013} and that it is equal to the MMD associated with the newly introduced angular distance.\\
Throughout their analysis, \textcite{kim_robust_2020} make the assumption of $n_1, n_2 \ge 2$. The CvM statistic averages over $K$ projections to approximate the integral over the unit sphere involved in the calculation of the CvM statistic. A resulting problem is that in high dimensions exponentially many projections may be required to achieve a certain accuracy. Instead, \textcite{kim_robust_2020} give a closed-form expression for the squared multivariate CvM-distance that depends on the expected angles between the differences of $X$ and $Y$ under the assumption that $\beta^TX$ and $\beta^TY$ have continuous distribution functions for $\lambda$-almost all $\beta$ that lie in the $p$-dimensional unit sphere, where $\lambda$ is the uniform probability measure on the $p$-dimensional unit sphere. The asymptotic null distribution of the test statistic is derived, but it is not applicable for calculating critical values. Therefore a permutation test is used instead. \textcite{kim_robust_2020} show that the test is consistent under fixed alternatives if second moments of conditional expectations of the test statistic are assumed. If the distance between the distributions diminishes as the sample grows, the additional assumption of quadratic mean differentiable families and an assumption on eigenvalues is required to achieve power greater than $\alpha$. The new test is more robust than the energy distance test since both can be represented as ${L^2}$-type differences between distribution functions but the energy distance gives uniform weight to the whole real line while the CvM statistic gives most weight on high-density regions. Moreover, the CvM distance is well-defined without moment assumptions in contrast to the energy distance that requires existing first moments. \textcite{kim_robust_2020} prove that the permutation test is minimax rate optimal against a class of alternatives associated with the CvM-distance (CvM-distance of at least $\varepsilon$) and that the energy test is not minimax rate optimal in that context. Additionally, consistency under the HDLSS setting is shown under assumptions on the first and second moments. Under the HDLSS setting and additional moment assumptions (equal covariances, different means) and assumptions on the bandwidth parameter of the Gaussian kernel, they also show the equivalence of CvM, energy statistic, and MMD statistic with the Gaussian kernel. 
The projection averaging approach can also be used for other one-dimensional test statistics like the sign test, Wilcoxon test, and Kendall's tau. \textcite{li_projective_2020} note that the test by \textcite{kim_robust_2020} has cubic computational cost $\mathcal{O}(N^3)$.

\paragraph{Test based on projective ensemble}\phantomsection\label{li_projective_2020}
\textcite{li_projective_2020} construct a robust test through a projective ensemble. The proposed test statistic is a generalization of the Cramér-von Mises statistic that has a simple closed-form expression without tuning parameters. It can be computed in quadratic time and is insensitive to the dimension. \textcite{li_projective_2020} show that the test based on a permutation procedure for approximating critical values is consistent against all fixed alternatives with rate $\sqrt{N}$. Their test does not require a moment assumption and is robust to outliers. The test is a generalization of the robust projection averaging test by \textcite{kim_robust_2020} that does not need the continuity assumption. It is also a member of the class of MMD tests. The test statistic is nonnegative and equal to zero, if and only if the distributions are equal. Its limiting distribution is intractable since it depends on the unknown distributions.

\paragraph{Weighted log-rank-type test}\phantomsection\label{liu_log-rank-type_2022}
\textcite{liu_log-rank-type_2022} present a weighted log-rank-type test for the two- and $k$-sample problem using class of intensity centered score processes. Their idea is to convert multivariate data into survival data to make use of the powerful weighted log-rank test. The transformation can be viewed as a statistic examining the arrival pattern of data at a certain point in space. The test is computationally simple and applicable to high-dimensional data. \textcite{liu_log-rank-type_2022} show consistency against any fixed alternative for Kolmogorov-Smirnov-type and Cramér-von Mises-type statistics. Critical values for the tests are obtained by permutations or with a simulation-based resampling method. A regularity condition on the weight function is required as well as the existence of a bounded density for the first distribution. The choice of the weight function and the test set are left open. Three heuristic strategies are presented to choose the test set. Moreover, a (dis)similarity measure for points must be chosen. Typically, the Euclidean distance is used for that.

\paragraph{Clustering-based $k$-sample tests}\phantomsection\label{paul_clustering-based_2022}
\textcite{paul_clustering-based_2022} propose different distribution-free $k$-sample tests intended for the high dimension low sample size (HDLSS) setting based on clustering the pooled sample.
For the tests, first, the pooled sample is clustered using some clustering algorithm suitable for high-dimensional data, and then a contingency table of the cluster and dataset membership is created. 
The idea behind both tests is that if the datasets come from the same distribution, the cluster and dataset membership are independent while if the datasets come from different distributions, the clustering depends on the true dataset membership. 
For the first test, the Rand index of the clustering is used as a test statistic (RI test). 
It is zero when the clustering is perfect, i.e.\ when the cluster membership is a permutation of the true dataset membership. 
The Rand index should take higher values when all clusters have similar distributions of class labels. 
Therefore, $H_0: F_1 = \dots = F_k$ is rejected for large values. 
The critical value can be calculated using a generalized hypergeometric distribution. 
Due to the discreteness of the Rand index, \textcite{paul_clustering-based_2022} propose to use a randomized test. 
For the second test, the generalized Fisher's test statistic for $k\times\ell$ contingency tables is used (FS test). 
It is intended to assess whether there is a dependence between the dataset membership and the cluster.
Again, a randomized test using the generalized hypergeometric distribution to find the critical values is proposed.
As a clustering algorithm, \textcite{paul_clustering-based_2022} suggest using $K$-means based on the generalized version of the \textit{Mean Absolute Difference of Distances (MADD)}
\[
\rho_{h,\varphi}(z_i, z_j) = \frac{1}{N-2} \sum_{m\in \{1,\dots, N\}\setminus\{i,j\}} \left| \varphi_{h,\psi}(z_i, z_m) - \varphi_{h,\psi}(z_j, z_m)\right|,
\]
as proposed by \textcite{sarkar_perfect_2020} for the HDLSS setting. Here, $z_i, i = 1,\dots,N$, denote points from the pooled sample and \[\varphi_{h,\psi}(z_i, z_j) = h\left(\frac{1}{p}\sum_{l=1}^p\psi|z_{il} - z_{jl}|\right),\] where $h:\mathbb{R}^{+} \to\mathbb{R}^{+}$ and $\psi:\mathbb{R}^{+} \to\mathbb{R}^{+}$ are continous and strictly increasing functions. 
\textcite{paul_clustering-based_2022} consider $h(t) = t$ and $\psi(t) = 1 - \exp(-t)$ for their examples. 
The number of clusters has to be chosen in advance for the RI and FS tests. 
A natural choice is to set the number of clusters to $k$.
\textcite{paul_clustering-based_2022} also present modified versions of the test where the number of clusters is estimated from the data using the Dunn index (MRI, MFS test). 
Setting the number of clusters to $k$ might fail in the case of multimodal distributions. 
In that case, a larger number of clusters might be required where then multiple clusters can correspond to one dataset. 
Moreover, multiscale versions of the tests are presented (MSRI, MSFS test) for the case where the number of clusters is unclear.
The RI or FS tests are then performed for different numbers of clusters and the results are aggregated using a Bonferroni adjustment for the individual tests. 
An upper limit for the number of clusters to be considered must be chosen. 
Under certain moment assumptions and assumptions on the functions $h$ and $\psi$ and on the sample size, consistency of the tests under the HDLSS setting (i.e.\ $p\to\infty$) is shown. The sample size requirements can already be fulfilled for very low sample sizes like $n_i = 4$, depending on the $\alpha$ level and the balance of the sample sizes. Slightly different assumptions are required for the RI, FS / MRI, MFS / MSRI, and MSFS tests. 
All presented tests are implemented in the \texttt{R} package \texttt{HDLSSkST} \parencite{HDLSSkST}.

\section{Summary of data similarity methods}\label{sec:summ.methods}
In the following, we give a brief summary of each of the ten classes that we divided the methods into, i.e.\ (i) comparison of cumulative distribution functions, density functions, or characteristic functions, (ii) methods based on multivariate ranks, (iii) discrepancy measures for distributions, (iv) graph-based methods, (v) methods based on inter-point distances, (vi) kernel-based methods, (vii) methods based on binary classification, (viii) distance and similarity measures for datasets, (ix) comparison based on summary statistics, and (x) different testing approaches. 

\subsection{Comparison of cumulative distribution functions, density functions or characteristic functions}
Since each of the cumulative distribution function, the density function (if it exists), and the characteristic function fully characterizes a distribution, it is natural to compare distributions by one of these functions. Given two datasets for which it is of interest to compare the underlying distributions, empirical versions of the functions can be used. 

For univariate distributions, methods of the Kolmogorov-Smirnov (KS) type are particularly popular. They compare the maximal absolute difference of the respective cumulative distribution functions of the two datasets to be compared. The extension of KS-type methods to multivariate distributions is not straightforward. This class includes two generalizations, one that uses permutations \parencite{bickel_distribution_1969} and one that uses partitioning of the sample space \parencite{biau_asymptotic_2005}.  

For the comparison of datasets based on their empirical density functions, different approaches to density estimation are utilized (e.g.\ kernel density estimation in \textcite{ahmad_goodness_1993, anderson_two-sample_1994, cao_empirical_2006} or estimation of densities based on partitions in \textcite{ntoutsi_general_2008, ganti_framework_1999, roederer_probability_2001, wang_random_2005}) and the resulting estimates of both samples are then compared using different statistics, e.g.\ the $L^2$-norm between the estimates. 

For comparison of distributions by characteristic functions, usually some type of distance, e.g.\ the $L^2$-norm, between the empirical characteristic functions is used \parencite{alba-fernandez_bootstrap_2004, alba-fernandez_test_2008, li_measuring_2022}.

\subsection{Methods based on multivariate ranks}
In the univariate two-sample problem, nonparametric tests based on ranks are popular choices. Since $\mathbb{R}^p$ does not have a natural ordering, the generalization of these methods to the multivariate problem is not straightforward. For the multivariate case, rank-based methods are based either on projecting the multivariate observations to one-dimensional statistics and ranking those \parencite{ghosh_distribution-free_2016} or on multivariate generalizations of ranks based on optimal transport \parencite{ghosal_multivariate_2021, deb_efficiency_2021}. Yet another generalization uses graphs to define ranks for multivariate data \parencite{zhou_new_2023}.

\subsection{Discrepancy measures for distributions} 
There exist various approaches to measure the discrepancy of two distributions. So-called probability metrics are metrics in the mathematical sense (i.e.\ they are positive definite, symmetric, and fulfill the triangle inequality) while discrepancy measures that do not fulfill the triangle inequality are usually called semimetrics or pseudometrics. In general, discrepancy measures that may not fulfill all metric properties are known as divergences. The best-known class of probability metrics are integral probability metrics (IPM), also called probability metrics with a $\xi$-structure \parencite{zolotarev_metric_1976, zolotarev_probability_1984}, as introduced by \textcite{muller_integral_1997}. If two distributions are equal, any function has the same expectation under both distributions. Based on this idea, the supremum difference of the integrals under both distributions over functions belonging to a prespecified set of functions is evaluated. The choice of this set of functions determines the IPM. Divergences include the large class of $f$-Divergences, which are also known as Ali-Silvey distances going back to \textcite{ali_general_1966} or as Csiszár's $\Phi$-divergences going back to \textcite{csiszar_informationstheoretische_1963}. $f$-divergences use the idea that equal distributions assign the same likelihood to each point. Therefore, they measure how far the likelihood ratio of the distributions is from one by using a convex continuous function $f$ that maps a ratio of one to the value zero. The expectation under the first distribution of this function $f$ applied to the likelihood ratio of the two distributions to be compared is evaluated. The choice of the function $f$ specifies the $f$-divergence. There are several other subclasses of probability metrics and divergences following a diverse set of approaches \parencite[e.g.][]{renyi_measures_1961, zolotarev_probability_1984, rachev_probability_1991, munoz_new_2012, zhao_comparing_2021}. 

\subsection{Graph-based methods}
Graph-based methods for comparing distributions are particularly popular in two-sample testing. Most of these fit in the general framework presented for example by \textcite{arias-castro_consistency_2016} or \textcite{mukhopadhyay_nonparametric_2020}. The pooled sample consisting of both datasets is used to construct a graph where each data point corresponds to one node. The methods differ in how edges between these nodes are inserted. Then, in most cases, the number of edges that connect points from different datasets, that is the number of adjacent nodes in the graph from different datasets, is counted. One particularly frequently used example is the $K$-nearest neighbor graph \parencite{weiss_two-sample_1960, friedman_nonparametric_1973, schilling_multivariate_1986, henze_multivariate_1988, nettleton_testing_2001, hall_permutation_2002, chen_weighted_2018, mondal_high_2015}, where each point in the pooled sample corresponds to one node. An edge connects one node to another if the data point corresponding to the second node is one of the $K$ nearest neighbors of the data point corresponding to the first node, with respect to some distance measure for the data points. If the number of edges connecting points from different datasets is high, points of both datasets are mixed well, so the datasets are similar. If the number is low, the datasets are separated well, so they are not similar. Since it is unclear in general what constitutes a high or low number, this is usually determined via a permutation approach.

\subsection{Methods based on inter-point distances}
Many methods for comparing datasets are based on analyzing the distributions of inter-point distances within and between the datasets. A theoretical justification for methods based on inter-point comparisons using a univariate function (e.g.\ a distance) is given by \textcite{maa_reducing_1996}. For two datasets, they consider the two distributions of the in-sample comparisons (i.e.\ $\|X - X^{\prime}\|$ and $\|Y - Y^{\prime}\|$ for all pairs of $X$, $X^{\prime}$ points from the first dataset and $Y$, $Y^{\prime}$ points from the second dataset) and the distribution of the between-sample comparisons (i.e.\ $\|X - Y\|$). They show that the equality of these three distributions is equivalent to the equality of the distributions of the datasets. This holds in general for discrete distributions. For the continuous case, some restrictions on the density function are needed. These include the existence of expectations and a second condition that is for example fulfilled if one of the densities is bounded or continuous. Based on this theorem, many approaches compare the distributions of the within-sample and between-sample distances. The most popular statistic based on this idea is the energy statistic \parencite{zech_new_2003}, which compares the expectations of the distance distributions within and between samples. 

\subsection{Methods based on kernel (mean) embeddings}
Kernel mean embeddings are a standard tool in machine learning. They map probability distributions to functions in so-called reproducing kernel Hilbert spaces (RKHS). Similarity between distributions can then be measured in this RKHS. More precisely, kernel mean embeddings extend feature maps $\phi$ as used by other kernel methods (e.g.\ in the context of kernel support vector machines) to the space of probability distributions by representing each distribution $F$ on the feature space $\mathcal{X}$ as a so-called mean function 
\begin{equation}
\mu_{F}(\cdot) := \int_\mathcal{X} K(x, \cdot) \dif F(x) = \E_{F}(K(X, \cdot)), \nonumber
\end{equation}
where $K:\mathcal{X}\times\mathcal{X}\to\mathbb{R}$ is a symmetric and positive definite kernel function and $X\sim F$ a random variable defined on $\mathcal{X}$. When well-defined, the kernel mean embedding is essentially a transformation of the distribution $F$ to an element in the reproducing kernel Hilbert space (RKHS) $\mathcal{H}$ corresponding to the kernel $K$ \parencite{muandet_kernel_2017}. For characteristic kernels, this representation captures all information about the distribution $F$. This implies that the distance of the kernel mean embeddings of two distributions, measured in the metric that the RKHS is endowed with, is equal to zero if and only if the distributions coincide \parencite{fukumizu_dimensionality_2004, sriperumbudur_injective_2008, sriperumbudur_hilbert_2010}. Therefore, kernel mean embeddings can be used for comparing distributions. The difference of the kernel mean embeddings measured in the RKHS metric is called Maximum Mean Discrepancy (MMD) \parencite{gretton_kernel_2006}, which is also the most popular method of this class. 

\subsection{Methods based on binary classification}
The idea behind methods in this class is to perform a binary classification of the data points from two given datasets and to evaluate the quality of this classification. More detailed,  the data points are labeled with their membership to the first or second dataset, respectively, and then some binary classification rule is fitted to the dataset labels on the pooled sample. If this classification rule performs well (e.g.\ measured by the classification error) the datasets are considered to be different in some sense, while for datasets that come from the same distribution, it is expected that the classification rule does not perform better than random guessing. To learn the classification rule, various classification methods like random forests or neural networks can be utilized and there are also different proposals on how to evaluate their performance \parencite{lopez-paz_revisiting_2017, kim_global_2019, cheng_classification_2022, yu_two-sample_2007, hediger_use_2021}. Alternatively, some approaches compare the whole (one-dimensional) distributions of scores, e.g.\ predicted probabilities, obtained from the classification of the data points \parencite{friedman_multivariate_2004}.

\subsection{Distance and similarity measures for datasets}
In contrast to defining a distance or similarity measure of the underlying distribution, some methods directly define the distance or similarity of the datasets themselves, using characteristics that are only indirectly connected to the underlying distributions. These methods are in part defined in the context of meta-learning. They use for example the correlation between meta-features like the number of variables in the datasets or other descriptive statistics \parencite{feurer_initializing_2015}, or the agreement of the performance of different learning algorithms on the datasets \parencite{leite_selecting_2012, leite_exploiting_2021}. Moreover, there are approaches to define distances between datasets by viewing them as metric measure spaces \parencite{memoli_distances_2017} or by using optimal transport \parencite{alvarez-melis_geometric_2020}.

\subsection{Comparison based on summary statistics}
The idea behind the methods of this class is to first summarize a dataset using different summary statistics. Then, a distance between these summary statistics is used as the distance between the datasets. This approach is less complex than using potentially complicated distances directly on the datasets, and it can also lead to simpler interpretations.

\subsection{Different testing approaches}
Comparing two datasets can be seen as a two-sample problem, i.e.\ testing for equality of their distributions. Many of the methods across all classes are introduced as test statistics for two- or $k$-sample testing. In this class, more two- and $k$-sample tests that do not fit in any of the other classes are collected.

\section{Approach for comparison of data similarity methods}\label{sec:comp}
So far, we classified more than 100 different methods for quantifying the similarity of datasets into ten groups described above. Now, we rate these methods with regard to their applicability, interpretability, and theoretical properties, in order to be able to compare them with each other. This comparison can then facilitate the choice of an appropriate method for the data that researchers have at hand. For this comparison, we introduce 22 different criteria which are explained in the following Section~\ref{sec:criteria}. The procedure for the comparison of the methods is then described in Section~\ref{meth-comp}. 
Note that the criteria do not include the performance of the methods, e.g.\ type I error rates and power for two- and $k$-sample tests, as this is hard to formalize, and for many methods, there are no empirical results yet. Moreover, to our knowledge, there are no neutral comparison studies of the methods yet. Rather, when comparisons are provided, they are usually presented in the context of an article proposing a new method \parencite[e.g.][]{biswas_nonparametric_2014,chen_new_2017,chwialkowski_fast_2015,jitkrittum_interpretable_2016,liu_classifier_2018,liu_learning_2020,lopez-paz_revisiting_2017,mondal_high_2015,petrie_graph-theoretic_2016,sarkar_graph-based_2020}.
Therefore, we focus on criteria that can be judged without performing extensive simulations and leave a neutral comparison of the method performance open for further research.

\subsection{Criteria for the comparison of data similarity measures}\label{sec:criteria}
\paragraph{Applicability}
Favorable are methods that can be used for general applications. To judge the applicability, we introduce the following criteria. \\\\
\textbf{Does the method allow incorporation of a target variable in a meaningful way?} \\
Many datasets consist of influencing (independent) variables and a target (dependent) variable. Presumably, in most contexts, it is not reasonable to treat this target variable in the same way as the influencing variables. Therefore, dataset similarity measures should also take into account the different role of the target variable. This criterion is counted as fulfilled if the method explicitly accounts for a target variable in the datasets. \\\\
\textbf{Does the method work on numeric data? Does the method work on categorical data?} \\
Numeric and categorical data are often treated differently. Ideally, dataset similarity measures should be able to handle both kinds of data. Each of these criteria is counted as fulfilled if the method is defined for the respective type of data. \\\\
\textbf{Does the method work for datasets that have different numbers of observations?} \\
Some of the methods might not be able to handle different sample sizes. It is desirable that a method can handle differently-sized datasets. The criterion is counted as fulfilled if the method is explicitly defined for datasets of different sizes. \\\\
\textbf{Does the method work if the number of variables exceeds the number of observations?} \\ The case of more variables than observations might be hard to handle due to identifiability as well as the curse of dimensionality. However, since it is a common case in applications like the analysis of high-dimensional gene expression data, data similarity measures that work even for numbers of variables larger than the number of observations might be needed. This criterion is counted as fulfilled if the method can be applied to data where the number of variables is larger than the number of observations. We do not evaluate how well the method works in that case, but only if the measure can be applied at all. \\\\
\textbf{Can the method be used to compare more than two datasets at a time?} \\
In some applications, researchers might be faced with more than two datasets. In that case, it is useful if multiple datasets can be compared at once. In general, it is always possible to extend methods comparing two datasets to the $k$-sample case for $k>2$ by aggregating the pairwise comparisons. For this criterion, we check if the method is explicitly defined for more than two datasets. \\\\
\textbf{Can the method be used without a separate training dataset?} \\
In some applications, data can be scarce, e.g.\ data derived from expensive experiments. In that case, it is undesirable or even impossible to hold out data for training a model involved in the dataset similarity measure. This criterion is counted as fulfilled if the method does not require holding out training data. \\\\
\textbf{Is the method independent of further assumptions?} \\ Further assumptions like continuity of distributions or the existence of certain moments reduce the applicability of a method and are therefore unwanted. This criterion is fulfilled if there are no explicit or implicit assumptions made that are not covered by the other criteria. For example, if the method requires numerical data, this criterion is fulfilled, while it is unfulfilled if continuous data is required. \\\\
\textbf{Is the method free of parameters that need to be chosen or tuned?} \\ 
Choosing good parameter values often requires good knowledge of the method and of the datasets at hand and is therefore often a hard task. Thus, for ease of application, it is desirable for users that they do not need to choose parameters prior to applying the method. Default parameters or suggestions on how to choose the parameters are very helpful but do still leave some uncertainty for the parameter choice. Therefore, the criterion is only counted as fulfilled if the method has no free tuning parameters to choose.\\\\
\textbf{Is the method implemented?} \\
Implementation of a method highly increases its applicability for practitioners. This criterion is counted as fulfilled if an implementation in any software is publicly available, e.g.\ via an R package or as code (e.g.\ in R, matlab, python, or others) in any publicly available repository. Otherwise, we count the criterion as unknown since we cannot guarantee that there is no implementation if we find none. We searched the publications introducing or reviewing the respective methods themselves as well as CRAN (\url{https://cran.r-project.org/}) and Bioconductor (\url{https://bioconductor.org/}) for implementations of the methods.\\\\
\textbf{What is the computational complexity of the method?} \\
In times of big data, methods with high-cost complexity might be inapplicable. Therefore, a low complexity of the method is desirable. A value for this criterion is given if cost complexity is mentioned in the publications introducing or reviewing the respective methods. For this criterion, we do not decide whether it is fulfilled or not but simply report the complexity if known since in general it is unclear which complexities can be counted as ``good''. Moreover, usually only the complexity with regard to the number of observations is given while in some applications the number of features might be of higher interest.

\paragraph{Interpretability}
To judge the result of a dataset comparison, the interpretability of the used measure is very helpful. To rate the interpretability of each measure, we use the following criteria.\\\\
\textbf{Does the measure have interpretable units?} \\ Interpretable units allow the user to judge what an increase in the measure by one unit means. For example, for accuracies given as percentages, one unit increase can be interpreted as classifying one additional observation in 100 correctly, or a one unit increase in many graph-based methods can be interpreted as one additional edge that connects points from different samples. In contrast, for example, one unit increase in the $L^q$ metric of the density functions is not interpretable. This criterion is fulfilled if it is intuitively interpretable what an increase of the measure by one unit means.\\\\
\textbf{Is the measure upper bounded? Is the measure lower bound?} \\
Bounds allow us to set the observed value of a measure into context and thus to judge if the observed value represents a low or high similarity or distance, respectively. These criteria are fulfilled if the measure is bounded. If known, the concrete bounds are provided. 

\paragraph{Theoretical properties}
There are several desirable theoretical properties that a data similarity measure might have. \\\\
\textbf{Is the measure invariant to rotation/ location change/ homogeneous scale transformations?} \\
Invariance under certain transformations can be useful since it might for example allow to rescale or shift both datasets in the same way without influencing the similarity values. The criteria are counted as fulfilled if the respective transformation of the datasets does not change the value of the measure.\\\\
\textbf{Does the measure fulfill the metric properties, i.e.\ is it positive definite, symmetric, and does it fulfill the triangle inequality?} \\
Metrics are well-known in mathematics and used in many different contexts. The requirement of positive definiteness ensures that a value of zero is attained if and only if the datasets, respectively their distributions, coincide. Symmetry makes sure that the ordering of the datasets, i.e.\ which one is defined to be the first or second, does not change their similarity. The triangle inequality holds if the sum of the distance of one dataset to a second plus the distance of this second dataset to a third dataset cannot be smaller than the distance directly between the first and third dataset. Again, each of these criteria is fulfilled if the measure fulfills the respective property. For symmetry, this is often obvious even if not explicitly mentioned by the authors. Positive definiteness and the triangle inequality are counted as unknown if they are not explicitly mentioned in the publications introducing or reviewing the respective methods. If the measure is defined for more than two distributions, symmetry and the triangle inequality are checked for the special case of $k = 2$ distributions.\\\\
\textbf{Is the two- or $k$-sample test based on the data similarity measure consistent?} \\
This criterion is only applicable to methods for which a two- or $k$-sample test is defined. As such tests are defined for many of the presented methods, the testing performance is of interest. As direct power comparisons of the methods are infeasible, only consistency of the test is considered as it can be assessed without simulations. Following the presented literature, we distinguish between consistency under the usual limiting regime, i.e.\ $n_i\to\infty$, $n_i/N\to\pi_i\in(0,1), i = 1,\dots,k$, and $p$ fixed, and high dimension low sample size setup (HDLSS) consistency, i.e.\ $n_i$ fixed and $p\to\infty$. In almost all cases, some additional assumptions on the distributions or on parts of the test statistic like the graph in graph-based tests or the kernel in kernel-based tests are required. As these differ fundamentally from test to test, we only check whether there is some proof of consistency under certain assumptions. In that case, the criterion counts as fulfilled. If there is only proof for the test to not be consistent under the respective limiting regime, it is counted as unfulfilled. If there are known conditions under which it is consistent and known conditions under which it is not, it is counted as conditionally fulfilled. It is also counted as conditionally fulfilled if consistency is only shown for a certain variant or special case of the test. If there are no statements regarding the consistency of the test in the literature, the criterion is counted as unknown. For methods for which no test is defined, the criterion is counted as inapplicable.

\subsection{Method comparison procedure}\label{meth-comp}
The comparison of all presented methods is performed as follows. For each method, each of the criteria explained above is checked and the results are tabulated. If a criterion is fulfilled, the method gets a checkmark in the corresponding row of the criterion. If it is not fulfilled, the method gets a cross for that criterion. If it is neither described in the literature nor obvious whether the criterion is fulfilled, the field is left empty (referring to unknown). If a method has free parameters and a criterion is only fulfilled for certain choices of these parameters, the check is given in parentheses. For the lower and upper bounds and the complexity, concrete values are given if known.

In the end, to evaluate how good each method is with regard to our criteria, we count how many criteria are fulfilled, how many are fulfilled conditionally on some free parameters, how many are unfulfilled, and for how many it is unclear. The complexity is not considered in these numbers as it is unclear what a good complexity is in general. The distinction between criteria that are always fulfilled and criteria that are fulfilled for certain parameters allows down-weighting the latter in the comparison. This might be of interest since in many cases there is no single parameter setting that fulfills all properties that can (in principle) be fulfilled by some setting. We analyze which of the methods fulfill most of the criteria as these might be the most promising methods in a general setting. For concrete data at hand, some of the criteria might be irrelevant. To facilitate finding the best suiting method we complement this article with an online tool (\url{https://shiny.statistik.tu-dortmund.de/data-similarity}), which allows filtering by certain criteria that are relevant to the problem at hand. 

\section{Results of comparison of data similarity methods}\label{sec:results}
In the following, we present the results of the method comparison. First, we demonstrate the criteria for one example method. Then, we give an overview of the results for all methods. Finally, we present a detailed comparison. All figures presented are created using R \parencite{R_4_1_2}. 

\subsection{Example for criteria evaluation: Cross-match test}
In the following, we check the criteria for one example method, namely the cross-match test statistic \parencite{rosenbaum_exact_2005}. 
The cross-match test is a graph-based method that uses the optimal non-bipartite matching. The optimal non-bipartite matching is the graph where each data point in the pooled sample is connected to exactly one other data point such that the sum over the edge lengths, i.e.\ the distances between the corresponding points, is minimal. In the case of an odd number of data points, one observation is left out such that the resulting matching has the lowest sum of edge lengths.

\begin{figure}[!t]
\centering
\includegraphics[width = 0.7\linewidth]{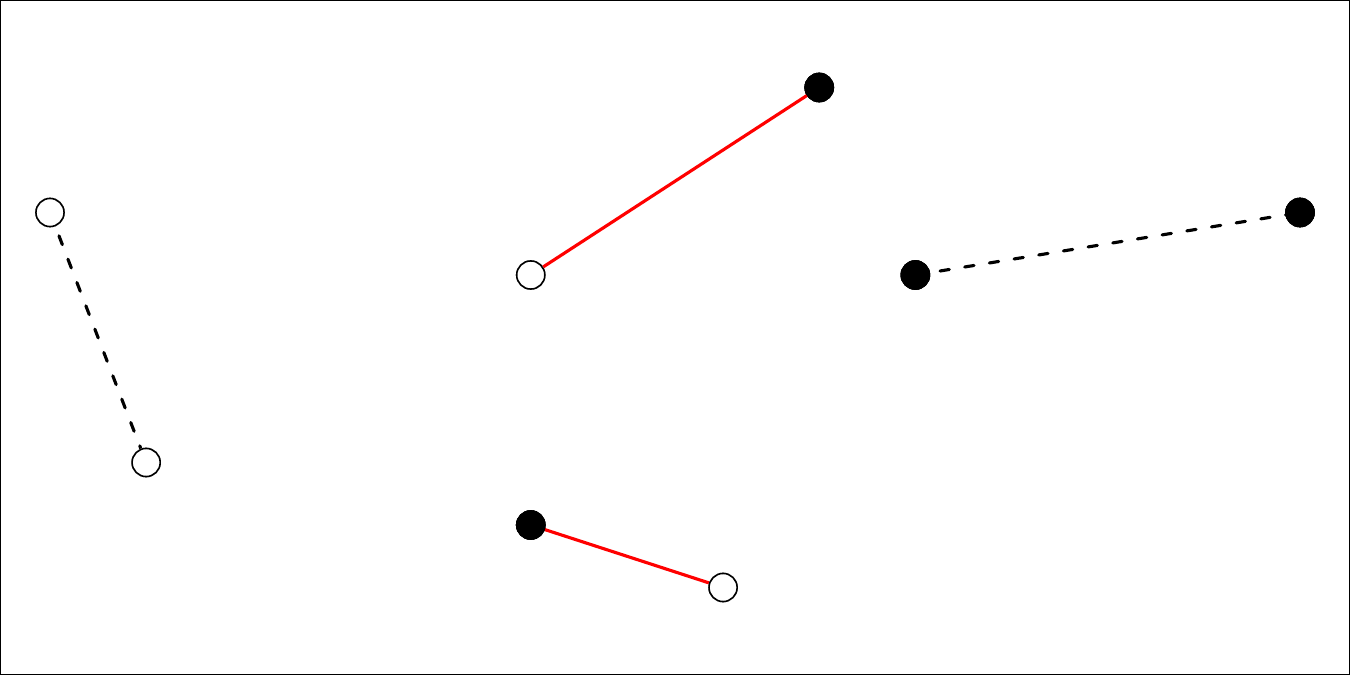}
\caption{Optimal non-bipartite matching for the pooled sample of two example datasets. White points correspond to the first dataset, black points correspond to the second dataset. Lines between points indicate edges. Edges between points from different datasets are indicated by red solid lines, and edges between points from the same dataset by black dashed lines.}
\label{fig:ex.matching}
\end{figure}

Figure \ref{fig:ex.matching} shows the optimal non-bipartite matching for an example dataset. The cross-match statistic is given as the number of edges that connect points from different datasets. For testing, the edge count standardized by the expectation and standard deviation under the null is used. For example, in Figure \ref{fig:ex.matching}, the edges connecting points from different datasets are indicated by red and solid lines. The edge count statistic takes the value two.

We now check the criteria described in Section~\ref{sec:criteria} for the cross-match test statistic.
\paragraph{Applicability:}
\begin{itemize}
\item \textbf{Sensible inclusion of target variable?} Since the distances of the observations are taken, all variables are treated the same. $\Rightarrow$ Unfulfilled
\item \textbf{Numeric variables?} The test is intended for numeric data. $\Rightarrow$ Fulfilled
\item \textbf{Categorical variables?} Categorical data can lead to ties for which the statistic is not uniquely defined. $\Rightarrow$ Unfulfilled 
\item \textbf{Unequal sample sizes permitted?} The datasets are pooled, so the sample sizes do not play a role in calculating the statistic. $\Rightarrow$ Fulfilled
\item \textbf{$p>n_i$ permitted?} Data is transformed into distances. $\Rightarrow$ Fulfilled \parencite[see also][]{biswas_nonparametric_2014}
\item \textbf{Applicable to more than two datasets at a time ($k > 2$)?} The statistic is defined for exactly two datasets. $\Rightarrow$ Unfulfilled
\item \textbf{No additional training data / train test split required?} The calculation requires no training step. $\Rightarrow$ Fulfilled
\item \textbf{No further assumptions on distributions required?} The calculation indirectly requires the uniqueness of the optimal non-bipartite matching, so no ties are allowed. $\Rightarrow$ Unfulfilled
\item \textbf{No tuning / choice of additional parameters required?} There are no additional parameters. $\Rightarrow$ Fulfilled
\item \textbf{Implemented in any software?} The cross-match test is implemented in the R~\parencite{R_4_1_2} package \texttt{crossmatch}~\parencite{crossmatch}. $\Rightarrow$ Fulfilled
\item \textbf{Computational complexity?} The complexity for calculating the optimal non-bipartite matching is $\mathcal{O}(N^3)$, where $N$ denotes the total sample size, i.e.\ the size of the pooled sample \parencite{rosenbaum_exact_2005}.
\end{itemize}
\FloatBarrier
\paragraph{Interpretability:}
\begin{itemize}
\item \textbf{Interpretable units?} An increase of one unit for the cross-match statistic can be interpreted as one additional edge in the optimal non-bipartite matching that connects two points from different datasets. $\Rightarrow$ Fulfilled
\item \textbf{Lower bound?} If each observation is connected to another observation from the same dataset, the minimum value of zero is attained.
\item \textbf{Upper bound?} If each observation from the smaller of the two datasets is connected to an observation from the other dataset, the maximum value of $\min\{n_1, n_2\}$ is attained, where $n_1$ and $n_2$ are the numbers of observations in the first and second dataset, respectively.
\end{itemize}

\paragraph{Theoretical properties:}
\begin{itemize}
\item \textbf{Rotation invariant?} Distances are rotation invariant, so the optimal non-bipartite matching and therefore the edge count statistic stays the same under rotation. $\Rightarrow$ Fulfilled
\item \textbf{Location change invariant?} Distances are location change invariant, so the optimal non-bipartite matching and therefore the edge count statistic stays the same under location change. $\Rightarrow$ Fulfilled
\item \textbf{Scale invariant?} For a change in scale, all distances change by a constant factor, so the optimal non-bipartite matching and therefore the edge count statistic stays the same under scale transformations. $\Rightarrow$ Fulfilled
\item \textbf{Positive definite?} For more similar datasets, higher values are expected. $\Rightarrow$ Unfulfilled
\item \textbf{Symmetric?} The roles of the first and the second dataset are interchangeable since the data is pooled. $\Rightarrow$ Fulfilled
\item \textbf{Triangle inequality?} It is not known whether the triangle inequality is fulfilled. 
\end{itemize}

\subsection{General insights from overall results}
\begin{figure}[H]
	\centering
	\includegraphics[width = \textwidth, trim = {0 0 12cm 13.5cm}, clip]{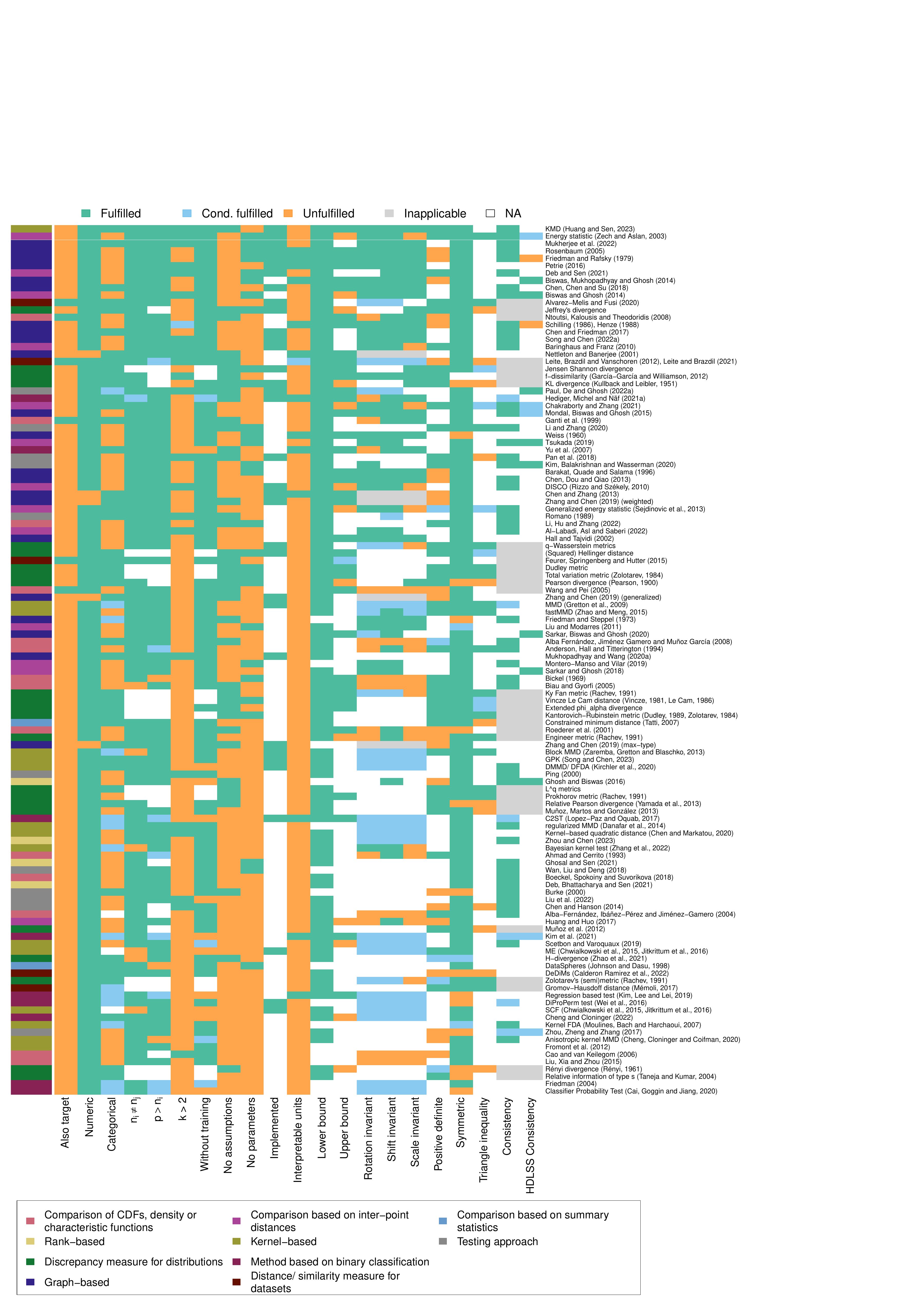}
	\caption{Comparison of all methods regarding the theoretical criteria. $n_i,~i=1,\dots,k$ denote the sample sizes, $k$ denotes the number of datasets to compare, $p$ denotes the number of features per dataset.}
	\label{fig:heatmap}
\end{figure}

Figure \ref{fig:heatmap} shows a heatmap of all methods and criteria, where the color of each field indicates whether a criterion is fulfilled for the respective method. The methods are ordered first by the highest proportion of fulfilled criteria, then by the highest proportion of conditionally fulfilled criteria, and then by the lowest proportion of unfulfilled criteria. We take into account the proportions instead of absolute numbers of fulfilled criteria since we do not want to give a structural advantage to methods that define a test or are applicable to numeric data, since more criteria can be applied to such methods.

There are many graph-based methods at the top. Apart from this, overall the classes are mostly mixed up. The best method according to the ordering is the nonparametric (kernel) measure of multi-sample dissimilarity (KMD) of \textcite{huang_kernel_2022} which fulfills 16 out of 21 criteria. It uses the association between the features and the sample membership to quantify the dissimilarity of multiple distributions. The estimator for KMD is based on a graph in which two points of the pooled sample are connected by an edge if they are close in distance, e.g.\ the $K$-nearest neighbor graph. The second best method is the Energy statistic \parencite{zech_new_2003, szekely_energy_2017}, which is based on inter-point distances and compares the mean of between-sample distances to the means of within-sample distances and fulfills 14 out of 21 criteria and 1 conditionally. 
Following this method, there are three methods that each fulfill 14 out of 21 criteria but none conditionally.
These are all graph-based tests, namely the Friedman-Rafksy test
\noindent \parencite{friedman_multivariate_1979} that uses the minimum spanning tree, the cross-match test \parencite{rosenbaum_exact_2005} that uses the optimal non-bipartite matching, and the graph-based test of \textcite{mukherjee_distribution-free_2022} based on optimal non-bipartite matchings that generalizes the cross-match test to categorical data and multiple datasets by using the Mahalanobis distance of a matrix that consists of the pairwise cross-match statistics of all pairs of datasets.
\FloatBarrier

We can see that certain criteria are fulfilled by most of the methods, such as applicability to numeric data, unequal sample sizes, that there is a lower bound and symmetry. On the other hand, certain other criteria are unfulfilled for most methods, such as the sensible inclusion of a target variable, applicability to more than two datasets at a time, no further assumptions, no tuning parameters, and interpretable units. In many cases, it is unknown if a method is implemented, if it has an upper bound, or if the triangle inequality holds.

\begin{figure}[!b]
\centering
\includegraphics[width = \linewidth]{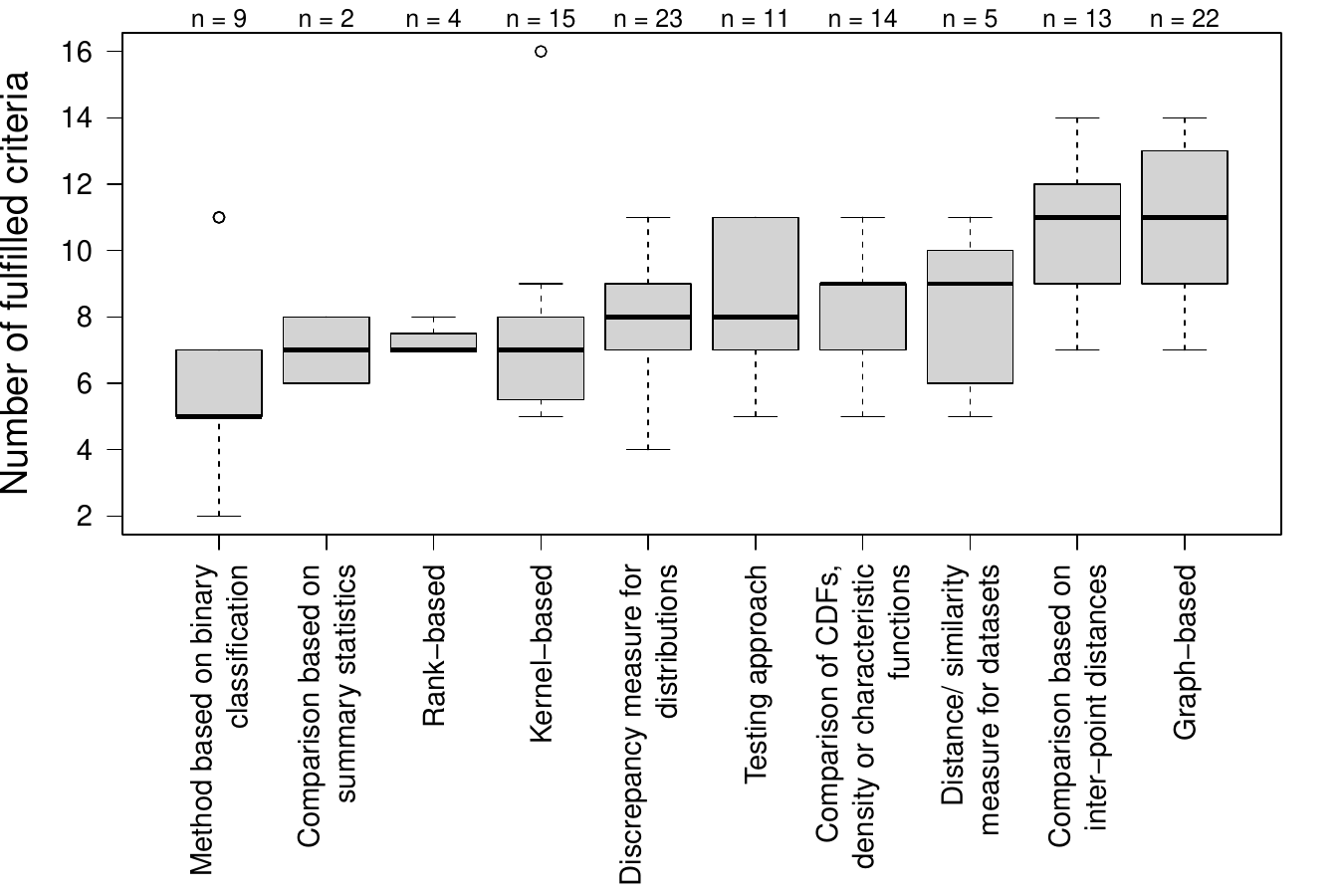}
\caption{Comparison of methods regarding the number of fulfilled criteria, grouped by classes. The classes are ordered by the median number of fulfilled criteria. n denotes the number of methods in the respective group.}
\label{fig:boxplots}
\end{figure}

Figure \ref{fig:boxplots} shows boxplots of the number of fulfilled criteria for each method, grouped by classes. The median number of fulfilled criteria ranges from five, for methods based on binary classification, to eleven for graph-based methods and methods based on inter-point distances. The number of fulfilled criteria also varies notably within the classes.

The number of fulfilled criteria on its own gives a first idea of the overall performance of the method with regard to our criteria. However, depending on the application, the criteria are not equally important, since some criteria might be mandatory and others negligible. For example, for a dataset comparison where some of the variables are numeric and others are categorical, a method that can handle both types of data is required. Further, if the data is not transformed, the invariance properties of a method might not be of interest. 

Therefore, in the following section, a detailed list of criteria is given for each method. To facilitate the choice and comparison of suitable methods for a dataset comparison, the online tool (\url{https://shiny.statistik.tu-dortmund.de/data-similarity}) can be used, which allows filtering and sorting of the tables of Section~\ref{sec:comp.tables} by different criteria. In addition, the tool makes it easy to search for specific methods and it allows users to hide criteria that are not relevant to their application of interest, in order to make the comparison results more concise.

\FloatBarrier
\subsection{Detailed method comparison}\label{sec:comp.tables}
Tables \ref{tab:comp1} to \ref{tab:comp10} show which of the methods fulfill which of our criteria and summarize how many of the criteria are fulfilled or unfulfilled for each method. The cells in the table are filled as explained in Section~\ref{sec:criteria}. For upper and lower bounds the criterion is fulfilled if a bound is given, all other criteria are fulfilled if they have a checkmark or a checkmark within parentheses. Parentheses around checkmarks mean that parameters can be chosen such that the criterion is fulfilled. Crosses in parentheses mean that for all but single choices the criterion is not fulfilled. Empty fields mean that it is neither described in literature nor obvious whether the criterion is fulfilled. The complexity is not considered in the calculation of the score so the maximum number of fulfilled criteria is 21. For methods that are inapplicable to numeric data, the transformations of rotating, shifting, or scaling the data are not meaningful. Therefore, the invariance criteria are inapplicable for such methods. This is denoted by a dash. Similarly, consistency does not apply as a criterion to methods that do not define any two- or $k$-sample procedure. $n_i,\: i = 1,\dots,k$ denote the sample sizes, $N = \sum n_i$ denotes the total sample size of the pooled sample, $p$ the number of features, and $k$ the number of datasets. 
\FloatBarrier
\begin{table}[!hb]
\caption{Comparison of approaches based on \textbf{cumulative distribution functions, density or characteristic functions} regarding applicability, interpretability, and theoretical properties.}
\label{tab:comp1}
\resizebox{\textwidth}{!}{
\begin{tabular}{lccc|ccccccc|cccc}
		Method/ Article & \rotatebox{90}{\textcite{bickel_distribution_1969}} & \rotatebox{90}{\textcite{biau_asymptotic_2005}} & \rotatebox{90}{\textcite{boeckel_multivariate_2018}} & \rotatebox{90}{\textcite{ntoutsi_general_2008}} & \rotatebox{90}{\textcite{ganti_framework_1999}} & \rotatebox{90}{\textcite{roederer_probability_2001}} & \rotatebox{90}{\textcite{wang_random_2005}} & \rotatebox{90}{\textcite{ahmad_goodness_1993}} & \rotatebox{90}{\textcite{anderson_two-sample_1994}} & \rotatebox{90}{\textcite{cao_empirical_2006}} & \rotatebox{90}{\textcite{alba-fernandez_bootstrap_2004}} & \rotatebox{90}{\textcite{alba-fernandez_test_2008}} & \rotatebox{90}{\textcite{liu_test_2015}} & \rotatebox{90}{\textcite{li_measuring_2022}}\\ 
		\midrule
		Also target? & \mytimes & \mytimes & \mytimes & \mycheck & \mycheck & \mytimes & \mycheck & \mytimes & \mytimes & \mytimes & \mytimes & \mytimes & \mytimes & \mytimes\\ 
		Numeric? & \mycheck & \mycheck & \mycheck & \mycheck & \mycheck & \mycheck & \mycheck & \mycheck & \mycheck & \mycheck & \mycheck & \mycheck & \mycheck & \mycheck \\ 
		Ca\-te\-go\-ri\-cal? & \mytimes & \mytimes & \mytimes & \mytimes & \mycheck & \mytimes & \mytimes & \mytimes & \mytimes & \mytimes &  & \mycheck & \mytimes & \mytimes \\
		$n_i\ne n_j$? & \mycheck & \mytimes & \mycheck & \mycheck & \mycheck & \mycheck & \mycheck & \mycheck & \mycheck & \mycheck & \mycheck & \mycheck & \mycheck & \mycheck\\ 
		$p>n_i$? & \mytimes & \mycheck & \mycheck & \mycheck & \mycheck & \mycheck & \mycheck & \condcheck & \condcheck &  &  & \mycheck & \mycheck & \mycheck \\ 
		$k > 2$? & \mytimes & \mytimes & \mytimes & \mytimes & \mytimes & \mytimes & \mytimes & \mytimes & \mytimes & \mycheck & \mytimes & \mytimes & \mytimes & \mytimes \\ 
		Without training? & \mycheck & \mycheck & \mycheck & \mycheck & \mycheck & \mycheck & \mycheck & \mycheck & \mycheck & \mycheck & \mycheck & \mycheck & \mycheck & \mycheck \\
\bottomrule
	\end{tabular}
}

\end{table}

\newpage
\begin{table}
\caption{Comparison of approaches based on \textbf{multivariate ranks} and \textbf{probability metrics} regarding applicability, interpretability, and theoretical properties. $^\ast$ assumptions only needed to show consistency for high dimension low sample size (HDLSS) setting.}
\label{tab:comp2}
\centering
\resizebox{\textwidth}{!}{
	\begin{tabular}{lcccc|ccccccccc}
		Method/ Article & \rotatebox{90}{\textcite{ghosh_distribution-free_2016}} & \rotatebox{90}{\textcite{ghosal_multivariate_2021}} & \rotatebox{90}{\textcite{deb_efficiency_2021}} & \rotatebox{90}{\textcite{zhou_new_2023}} & \rotatebox{90}{Engineer metric \parencite{rachev_probability_1991}} & \rotatebox{90}{Zolotarev's (semi)metric \parencite{rachev_probability_1991}} & \rotatebox{90}{Ky Fan metric \parencite{rachev_probability_1991}} & \rotatebox{90}{Prokhorov metric \parencite{rachev_probability_1991}} & \rotatebox{90}{Dudley metric} &  \rotatebox{90}{Total variation metric \parencite{zolotarev_probability_1984}} & \rotatebox{90}{Kantorovich-Rubinstein metric \parencite{zolotarev_probability_1984,dudley_real_1989}} & \rotatebox{90}{$L^q$ metrics} & \rotatebox{90}{$q$-Wasserstein metrics}\\
		\midrule
		Also target? & \mytimes & \mytimes & \mytimes & \mytimes & \mytimes & \mytimes & \mytimes & \mytimes & \mytimes & \mytimes & \mytimes & \mytimes  & \mytimes\\ 
		Numeric? & \mycheck & \mycheck & \mycheck & \mycheck & \mycheck & \mycheck & \mycheck & \mycheck & \mycheck  & \mycheck & \mycheck & \mycheck & \mycheck\\ 
		Ca\-te\-go\-ri\-cal? & \mytimes & \mytimes & \mytimes & \mytimes & \mytimes & \mycheck & \mycheck & \mytimes & \mycheck &  \mycheck & \mycheck & \mycheck & \mytimes\\
		$n_i\ne n_j$?  & \mycheck & \mycheck & \mycheck & \mycheck & \mycheck &  &  &  &   &  &  & &\mycheck\\ 
		$p>n_i$?  & \mycheck &  & \mycheck & \mycheck & \mycheck &  &  &  &   &  &  & & \mycheck\\ 
		$k > 2$? & \mytimes & \mytimes & \mytimes & \mytimes & \mytimes & \mytimes & \mytimes & \mytimes & \mytimes & \mytimes & \mytimes & \mytimes & \mytimes\\ 
		Without training?  & \mytimes & \mycheck & \mycheck & \mycheck & \mycheck &  & &  &  &  &  & & \mycheck\\
\bottomrule
\end{tabular}}
\end{table}

\newpage
\begin{table}
\caption{Comparison of \textbf{divergences} regarding applicability, interpretability, and theoretical properties. $^\ast$ holds for square root transformation. $^{\ast\ast}$ holds only in case of $\alpha = 1/2$, resp. $s = 1/2$. $^{\ast\ast\ast}$ values calculated using the general formula given in \textcite{vajda_metric_2009}.}
\label{tab:comp3}
\centering
\resizebox{\textwidth}{!}{
	\begin{tabular}{lcccccccccccccc}
		Method/ Article & \rotatebox{90}{(Squared) Hellinger distance} & \rotatebox{90}{Vincze Le Cam distance \parencite{vincze_concept_1981, cam_asymptotic_1986}} & \rotatebox{90}{KL divergence \parencite{kullback_information_1951}} & \rotatebox{90}{Jeffrey's divergence} & \rotatebox{90}{Extended $\phi_{\alpha}$ divergence} & \rotatebox{90}{Jensen Shannon divergence} & \rotatebox{90}{Pearson divergence \parencite{pearson_criterion_1900}} & \rotatebox{90}{Relative Pearson divergence \parencite{yamada_relative_2013}} &  \rotatebox{90}{$f$-dissimilarity \parencite{gyorfi_f-dissimilarity_1975, garcia-garcia_divergences_2012}} & \rotatebox{90}{Rényi divergence \parencite{renyi_measures_1961}} & \rotatebox{90}{Relative information of type s \parencite{taneja_relative_2004}} & \rotatebox{90}{H-divergence \parencite{zhao_comparing_2021}}  & \rotatebox{90}{\textcite{munoz_new_2012}} &  \rotatebox{90}{\textcite{munoz_new_2013}}\\
		\midrule
		Also target? & \mytimes & \mytimes & \mytimes & \mytimes & \mytimes & \mytimes & \mytimes & \mytimes & \mytimes & \mytimes & \mytimes &  \mytimes & \mytimes & \mytimes\\ 
		Numeric? &  \mycheck & \mycheck & \mycheck & \mycheck & \mycheck & \mycheck & \mycheck & \mycheck & \mycheck & \mycheck & \mycheck &  \mycheck & \mycheck  & \mycheck\\ 
		Ca\-te\-go\-ri\-cal? & \mycheck & \mycheck & \mycheck & \mycheck & \mycheck & \mycheck & \mycheck & \mycheck & \mycheck & \mycheck & \mycheck &  \mycheck & \mytimes & \mytimes\\
		$n_i\ne n_j$?  &  &  & \mycheck & \mycheck & & & \mycheck & \mycheck & \mycheck &  &  & \mytimes & \mycheck & \mycheck\\ 
		$p>n_i$?  &  &  &  &  & & & &  &  &  &  & \mycheck & & \mycheck \\ 
		$k > 2$?  & \mytimes & \mytimes & \mytimes & \mytimes & \mytimes & \mytimes & \mytimes & \mytimes & \mycheck & \mytimes & \mytimes &  \mytimes & \mytimes & \mytimes\\ 
		Without training?  &  &  & \mycheck & \mycheck & \mycheck &  & \mycheck & \mycheck &  &  &  & \mycheck & \mycheck & \mycheck \\
\bottomrule
\end{tabular}}
\end{table}

\newpage
\begin{table}
\caption{Comparison of \textbf{graph-based methods} regarding applicability, interpretability, and theoretical properties. $^\ast$ no assumptions mentioned in the original articles, but the method implicitly requires uniqueness of the constructed graph \parencite{chen_graph-based_2013}.
	$^{\ast\ast}$ $K$ minimum number of categories, $M$ number of minimum spanning trees. $^{\ast\ast\ast}$ for unstandardized statistic. }
\label{tab:comp4}
\centering
\resizebox{\textwidth}{!}{
	\begin{tabular}{lcccccccccccccc}
		Method/ Article & \rotatebox{90}{\textcite{friedman_multivariate_1979}} & \rotatebox{90}{\textcite{rosenbaum_exact_2005}} & \rotatebox{90}{\textcite{chen_graph-based_2013}} & \rotatebox{90}{\textcite{biswas_distribution-free_2014}} & \rotatebox{90}{\textcite{petrie_graph-theoretic_2016}} & \rotatebox{90}{\textcite{chen_new_2017}} & \rotatebox{90}{\textcite{chen_weighted_2018}} & \rotatebox{90}{\textcite{zhang_graph-based_2019} (weighted)} & \rotatebox{90}{\textcite{zhang_graph-based_2019} (generalized)} & \rotatebox{90}{\textcite{zhang_graph-based_2019} (max-type)} & \rotatebox{90}{\textcite{sarkar_graph-based_2020}} & \rotatebox{90}{\textcite{mukhopadhyay_nonparametric_2020}} & \rotatebox{90}{\textcite{mukherjee_distribution-free_2022}} & \rotatebox{90}{\textcite{song_new_2022}}\\ 
		\midrule	
		Also target? & \mytimes & \mytimes & \mytimes & \mytimes & \mytimes & \mytimes & \mytimes & \mytimes & \mytimes & \mytimes & \mytimes & \mytimes & \mytimes & \mytimes \\ 
		Numeric? & \mycheck & \mycheck & \mytimes & \mycheck & \mycheck & \mycheck & \mycheck & \mytimes & \mytimes & \mytimes & \mycheck & \mycheck  & \mycheck & \mycheck \\ 
		Ca\-te\-go\-ri\-cal? & \mytimes & \mytimes & \mycheck & \mytimes & \mytimes & \mycheck & \mycheck & \mycheck & \mycheck & \mycheck & \mytimes & \mycheck &  \mycheck & \mytimes \\
		$n_i\ne n_j$?  & \mycheck & \mycheck & \mycheck & \mycheck & \mycheck & \mycheck & \mycheck & \mycheck & \mycheck & \mycheck & \mycheck & \mycheck & \mycheck & \mycheck \\ 
		$p>n_i$?  & \mycheck & \mycheck & \mycheck &  \mycheck & \mycheck & \mycheck & \mycheck & \mycheck & \mycheck & \mycheck & \mycheck & \mycheck & \mycheck & \mycheck \\ 
		$k > 2$?  & \mytimes & \mytimes & \mytimes & \mytimes & \mycheck & \mytimes & \mytimes & \mytimes & \mytimes & \mytimes & \mytimes & \mycheck & \mycheck & \mycheck \\
\bottomrule
\end{tabular}}
\end{table}

\newpage
\begin{table}
\caption{Comparison of methods based on \textbf{nearest neighbors} regarding applicability, interpretability, and theoretical properties. $^\ast$ only in combination with numerical data.}
\label{tab:comp5}
\centering
\resizebox{\textwidth}{!}{
	\begin{tabular}{lcccccccc}
		Method/ Article & \rotatebox{90}{\textcite{weiss_two-sample_1960}} & \rotatebox{90}{\textcite{friedman_nonparametric_1973}} & \rotatebox{90}{\textcite{schilling_multivariate_1986, henze_multivariate_1988}} & \rotatebox{90}{\textcite{barakat_multivariate_1996}} & \rotatebox{90}{\textcite{nettleton_testing_2001}} & \rotatebox{90}{\textcite{hall_permutation_2002}} & \rotatebox{90}{\textcite{chen_ensemble_2013}} & \rotatebox{90}{\textcite{mondal_high_2015}}\\ 
		\midrule	
		Also target? & \mytimes & \mytimes & \mytimes & \mytimes & \mytimes & \mytimes & \mytimes & \mytimes \\ 
		Numeric? & \mycheck & \mycheck & \mycheck & \mycheck & \mytimes & \mycheck & \mycheck & \mycheck \\ 
		Ca\-te\-go\-ri\-cal? & \mytimes & ~\condcheck$^\ast$ & \mytimes & \mytimes & \mycheck & \mytimes & \mytimes & \mytimes \\
		$n_i\ne n_j$?  & \mycheck & \mycheck & \mycheck & \mycheck & \mycheck & \mycheck & \mycheck & \mycheck \\ 
		$p>n_i$?  & \mycheck & \mycheck & \mycheck & \mycheck & \mycheck & \mycheck & \mycheck & \mycheck \\ 
		$k > 2$?  & \mytimes & \mytimes & \condcheck & \mytimes & \mycheck & \mytimes & \mytimes & \mytimes \\ 
		Without training?  & \mycheck & \mycheck & \mycheck & \mycheck & \mycheck & \mycheck & \mycheck & \mycheck \\
\bottomrule
\end{tabular}}

\end{table}

\newpage
\begin{table}
\caption{Comparison of methods based on \textbf{inter-point distances} regarding applicability, interpretability, and theoretical properties. $^{\ast}$ $m$ denotes the number of random projections.}
\label{tab:comp6}
\centering
\resizebox{\textwidth}{!}{
	\begin{tabular}{lccccccccccccc}
		Method/ Article & \rotatebox{90}{Energy statistic \parencite{zech_new_2003}} & \rotatebox{90}{Generalized energy statistic \parencite{sejdinovic_equivalence_2013}} & \rotatebox{90}{\textcite{chakraborty_new_2021}} &  \rotatebox{90}{DISCO \parencite{rizzo_disco_2010}} & \rotatebox{90}{\textcite{huang_efficient_2017}} & \rotatebox{90}{\textcite{deb_multivariate_2021}} & \rotatebox{90}{\textcite{al-labadi_bayesian_2022}} & \rotatebox{90}{\textcite{baringhaus_rigid_2010}} & \rotatebox{90}{\textcite{liu_triangle_2011}} & \rotatebox{90}{\textcite{biswas_nonparametric_2014}} & \rotatebox{90}{\textcite{sarkar_high-dimensional_2018}} & \rotatebox{90}{\textcite{montero-manso_two-sample_2019}} & \rotatebox{90}{\textcite{tsukada_high_2019}}\\ 
		\midrule	
		Also target? & \mytimes & \mytimes & \mytimes & \mytimes & \mytimes & \mytimes & \mytimes & \mytimes & \mytimes & \mytimes & \mytimes & \mytimes & \mytimes \\ 
		Numeric? & \mycheck & \mycheck & \mycheck & \mycheck & \mycheck & \mycheck & \mycheck & \mycheck & \mycheck & \mycheck & \mycheck & \mycheck & \mycheck \\ 
		Ca\-te\-go\-ri\-cal? & \mytimes & \mycheck & \mycheck & \mytimes & \mytimes & \mytimes & \mytimes & \mytimes & \mytimes & \mytimes & \mytimes & \mytimes & \mytimes \\
		$n_i\ne n_j$?  & \mycheck & \mycheck & \mycheck & \mycheck & \mycheck & \mycheck & \mycheck & \mycheck & \mycheck & \mycheck & \mycheck & \mycheck & \mycheck \\ 
		$p>n_i$?  & \mycheck & \mycheck & \mycheck & \mycheck & \mycheck & \mycheck & \mycheck & \mycheck & \mycheck & \mycheck & \mycheck & \mycheck & \mycheck \\ 
		$k > 2$?  & \mycheck & \mytimes & \mytimes & \mycheck & \mytimes & \mycheck & \mycheck & \mycheck & \mytimes & \mycheck & \mytimes & \mytimes & \mytimes \\ 
		Without training?  & \mycheck & \mycheck & \mycheck & \mycheck & \mycheck & \mycheck & \mycheck & \mycheck & \mycheck & \mycheck & \mycheck & \mycheck & \mycheck \\
\bottomrule
\end{tabular}}
\end{table}

\newpage
\begin{table}
\caption{Comparison of methods based on \textbf{variations of MMD} regarding applicability, interpretability, and theoretical properties. $^{\ast}$ for block size $B = [n^{\gamma}]$. $^{\ast\ast}$ $L$ denotes the number of basis functions for approximating kernels. $^{\ast\ast\ast}$ $N_R$ denotes the number of reference points.}
\label{tab:comp7}
\centering
\resizebox{\textwidth}{!}{
	\begin{tabular}{lccccccccc}
		Method/ Article & \rotatebox{90}{(Linear) $MMD^2$ \parencite{gretton_fast_2009, muandet_kernel_2017, gretton_kernel_2012}} & \rotatebox{90}{Block MMD \parencite{zaremba_b-test_2013}} & \rotatebox{90}{fastMMD \parencite{zhao_fastmmd_2015}} & \rotatebox{90}{ME \parencite{chwialkowski_fast_2015, jitkrittum_interpretable_2016}} & \rotatebox{90}{SCF \parencite{chwialkowski_fast_2015, jitkrittum_interpretable_2016}} & \rotatebox{90}{regularized MMD \parencite{danafar_testing_2014}} & \rotatebox{90}{Anisotropic kernel MMD \parencite{cheng_two-sample_2020}} & \rotatebox{90}{DMMD/ DFDA \parencite{kirchler_two-sample_2020}} &   \rotatebox{90}{GPK \parencite{song_generalized_2021}}\\ 
		\midrule	
		Also target? & \mytimes & \mytimes & \mytimes & \mytimes & \mytimes & \mytimes & \mytimes & \mytimes & \mytimes \\ 
		Numeric? & \mycheck & \mycheck & \mycheck & \mycheck & \mycheck & \mycheck & \mycheck & \mycheck & \mycheck \\ 
		Ca\-te\-go\-ri\-cal? & \condcheck & \condcheck & \mytimes &  &  & \condcheck & \mytimes & \mycheck & \mycheck \\
		$n_i\ne n_j$?  & \mycheck & \mytimes & \mycheck & \mytimes & \mytimes & \mycheck & \mycheck & \mycheck & \mycheck \\ 
		$p>n_i$?  & \mycheck & \mycheck & \mycheck & \mycheck & \mycheck & \mycheck & \mytimes & \mycheck & \mycheck \\ 
		$k > 2$?  & \mytimes & \mytimes & \mytimes & \mytimes & \mytimes & \mycheck & \mytimes & \mytimes & \mytimes \\ 
		Without training?  & \mycheck & \mycheck & \mycheck & \mytimes & \mytimes & \mycheck & \condcheck & \mytimes & \mycheck \\
\bottomrule
\end{tabular}}
\end{table}

\newpage
\begin{table}
\caption{Comparison of \textbf{kernel-based methods other than MMD} and of methods based on \textbf{binary classification} regarding applicability, interpretability, and theoretical properties. $^{\ast}$ assumptions are only needed for showing properties of the estimator. $^{\ast\ast}$ for $K$-nearest neighbor graph. $^{\ast\ast\ast}$ for permutation version.}
\label{tab:comp8}
\centering
\resizebox{\textwidth}{!}{
	\begin{tabular}{lcccccc|ccccccccc}
		Method/ Article & \rotatebox{90}{Kernel FDA \parencite{eric_testing_2007}} & \rotatebox{90}{\textcite{fromont_kernels_2012}} & \rotatebox{90}{\textcite{scetbon_comparing_2019}} & \rotatebox{90}{Kernel-based quadratic distance \parencite{chen_kernel_2020}} & \rotatebox{90}{Bayesian kernel test \parencite{zhang_bayesian_2022}} & \rotatebox{90}{KMD \parencite{huang_kernel_2022}} & \rotatebox{90}{\textcite{friedman_multivariate_2004}} & \rotatebox{90}{C2ST \parencite{lopez-paz_revisiting_2017}} & \rotatebox{90}{Regression based test \parencite{kim_global_2019}} & \rotatebox{90}{\textcite{cheng_classification_2022}} & \rotatebox{90}{\textcite{yu_two-sample_2007}} & \rotatebox{90}{DiProPerm test \parencite{wei_direction-projection-permutation_2016}} & \rotatebox{90}{Classifier Probability Test \parencite{cai_two-sample_2020}} & \rotatebox{90}{\textcite{kim_classification_2021}} & \rotatebox{90}{\textcite{hediger_use_2021}}\\ 
		\midrule	
		Also target? & \mytimes & \mytimes & \mytimes & \mytimes & \mytimes & \mytimes & \mytimes & \mytimes & \mytimes & \mytimes & \mytimes & \mytimes & \mytimes & \mytimes & \mytimes \\ 
		Numeric? & \mycheck & \mycheck & \mycheck & \mycheck & \mycheck & \mycheck & \mycheck & \mycheck & \mycheck & \mycheck & \mycheck & \mycheck & \mycheck & \mycheck & \mycheck \\ 
		Ca\-te\-go\-ri\-cal? & \condcheck & \mytimes & \mytimes & \mytimes & \condcheck & \mycheck & \condcheck & \condcheck & \condcheck & \mycheck & \mycheck & \condcheck & \condcheck & \condcheck & \mycheck \\
		$n_i\ne n_j$?  & \mycheck & \mycheck & \mycheck & \mycheck & \mytimes & \mycheck & \mycheck & \mycheck & \mycheck & \mycheck & \mycheck & \mycheck & \mycheck & \mycheck & \condcheck \\ 
		$p>n_i$?  & \mycheck & \mycheck & \mycheck & \mycheck & \mycheck & \mycheck & \condcheck & \condcheck & \condcheck & \mycheck & \mycheck & \mycheck & \condcheck & \condcheck & \mycheck \\
\bottomrule
\end{tabular}}
\end{table}

\newpage
\begin{table}
\caption{Comparison of \textbf{distance and similarity measures for datasets} and \textbf{comparison based on summary statistics} regarding applicability, interpretability, and theoretical properties. $^\ast$ combination of numeric and categorical data required. $^{\ast\ast}$ finite sample space required.}
\label{tab:comp9}
\centering
\resizebox{0.7\textwidth}{!}{
	\begin{tabular}{lccccc|cc}
		Method/ Article & \rotatebox{90}{\textcite{feurer_initializing_2015}} & \rotatebox{90}{Gromov-Hausdoff distance \parencite{memoli_distances_2017}} & \rotatebox{90}{\textcite{leite_selecting_2012, leite_exploiting_2021}} & \rotatebox{90}{DeDiMs \parencite{calderon_ramirez_dataset_2022}} & \rotatebox{90}{\textcite{alvarez-melis_geometric_2020}} & \rotatebox{90}{DataSpheres \parencite{johnson_comparing_1998}} & \rotatebox{90}{Constrained minimum distance \parencite{tatti_distances_2007}}\\ 
		\midrule	
		Also target? & \mycheck & \mytimes & \mycheck & \mytimes & \mycheck & \mytimes &  \mytimes\\ 
		Numeric? & \mycheck & \mycheck & \mycheck & \mycheck & \mycheck & ~\mycheck$^\ast$ & ~~\mycheck$^{\ast\ast}$ \\ 
		Ca\-te\-go\-ri\-cal? & \mycheck & \condcheck & \mycheck & \mycheck & \mycheck & ~\mycheck$^\ast$ & \mycheck \\
		$n_i\ne n_j$?  & \mycheck &  & \mycheck & \mycheck & \mycheck & \mycheck &  \mycheck \\ 
		$p>n_i$?  & \mycheck &  & \condcheck & \mycheck & \mycheck & \mytimes & \mycheck \\ 
		$k > 2$?  & \mytimes & \mytimes & \mycheck & \mytimes & \mytimes & \mytimes & \mytimes \\ 
		Without training?  & \mycheck &  & \mycheck & \mycheck & \mycheck & \mycheck & \mycheck \\
\bottomrule
\end{tabular}}
\end{table}

\newpage
\begin{table}
\caption{Comparison of \textbf{different testing approaches} regarding applicability, interpretability, and theoretical properties.$^\ast$ for RI version }
\label{tab:comp10}
\centering
\resizebox{\textwidth}{!}{
	\begin{tabular}{lccccccccccc}
		Method/ Article & \rotatebox{90}{\textcite{romano_bootstrap_1989}} & \rotatebox{90}{\textcite{burke_multivariate_2000}} & \rotatebox{90}{\textcite{ping_bootstrap_2000}} & \rotatebox{90}{\textcite{chen_bayesian_2014}} & \rotatebox{90}{\textcite{zhou_two-sample_2017}} & \rotatebox{90}{\textcite{pan_ball_2018}} & \rotatebox{90}{\textcite{wan_empirical_2018}} & \rotatebox{90}{\textcite{kim_robust_2020}} & \rotatebox{90}{\textcite{li_projective_2020}} & \rotatebox{90}{\textcite{liu_log-rank-type_2022}} & \rotatebox{90}{\textcite{paul_clustering-based_2022}}\\ 
		\midrule	
		Also target? & \mytimes & \mytimes & \mytimes & \mytimes & \mytimes & \mytimes & \mytimes & \mytimes & \mytimes & \mytimes & \mytimes \\ 
		Numeric? & \mycheck & \mycheck & \mycheck & \mycheck & \mycheck & \mycheck & \mycheck & \mycheck & \mycheck & \mycheck & \mycheck\\ 
		Ca\-te\-go\-ri\-cal? & \mycheck & \mycheck & \mytimes & \mytimes & \mytimes & \mytimes & \mytimes & \mytimes & \mytimes & \mytimes &\condcheck   \\
		$n_i\ne n_j$?  & \mycheck & \mycheck & \mycheck & \mycheck & \mycheck & \mycheck & \mycheck & \mycheck & \mycheck & \mycheck & \mycheck\\ 
		$p>n_i$?  & \mycheck & \mycheck & \mycheck &  & \mycheck & \mycheck & \mycheck & \mycheck & \mycheck & \mycheck &\mycheck\\ 
		$k > 2$?  & \mycheck & \mytimes & \mycheck & \mycheck & \mytimes & \mycheck & \mytimes & \mytimes & \mytimes & \mycheck &\mycheck\\ 
		Without training?  & \mycheck & \mycheck & \mycheck & \mycheck & \mycheck & \mycheck & \mycheck & \mycheck & \mycheck & \mytimes & \mycheck\\
\bottomrule
\end{tabular}}
\end{table}

\FloatBarrier	

\section{Conclusion}\label{sec:summary}
In statistics and machine learning, measuring the similarity between two or more datasets has widespread applications. Extremely many approaches for quantifying dataset similarity have been proposed in the literature. We examined more than 100 methods for quantifying the similarity of datasets.
The methods were selected from an extensive literature search by using the following criteria: 
\begin{itemize}
\item The method is applicable for multivariate datasets.
\item The method requires no specific parametric or distributional assumptions on the underlying distributions of the datasets (e.g.\ normal distribution).
\item The method does not focus on a particular property of the data (e.g.\ means), but on the entire dataset or its entire distribution. 
\end{itemize}
We classified the methods into ten classes based on their main ideas, including 
\begin{enumerate}
\item Comparison of cumulative distribution functions, density functions, or characteristic functions
\item Methods based on multivariate ranks 
\item Discrepancy measures for distributions
\item Graph-based methods
\item Methods based on inter-point distances 
\item Kernel-based methods
\item Methods based on binary classification 
\item Distance and similarity measures for datasets
\item Comparison based on summary statistics
\item Different testing approaches.
\end{enumerate}
We presented an extensive review of these methods. For each method, we introduced the underlying ideas, formal definitions, and important properties. An overview of the methods can be found in Table \ref{tab:method.list} and a summary of the classes can be found in Section \ref{sec:summ.methods}.

Moreover, we compared all these methods with respect to 22 criteria that can be divided into the three categories applicability (e.g.\ is the method applicable to numeric or categorical data), interpretability (e.g.\ is the statistic bounded), and theoretical properties (e.g.\ metric properties). The criteria can be used to judge which methods are best suited for quantifying the similarity of given datasets.
Overall, we found that graph-based methods had the highest numbers of fulfilled criteria.

To facilitate the choice of an appropriate data similarity measure for a concrete application, we provided detailed comparisons of the methods. Moreover, we designed an online tool (\url{https://shiny.statistik.tu-dortmund.de/data-similarity}) that allows for custom filtering of the criteria and sorting of the methods. Therefore, the online tool can provide more specific guidance for the choice of a suitable dataset similarity method for concrete data at hand in addition to the overall comparison presented in this paper.
We intend to expand this online tool over time. Suggestions for new methods to be included, as well as additional entries for criteria not yet marked as fulfilled or unfulfilled, are welcome. These can be added as an issue in the GitHub repository (\url{https://github.com/MariekeStolte/ComparisonToolDatasetSimilarity.git}).

Note that the comparison so far does not include the performance of the methods, e.g.\ type I error rates and power for two- and $k$-sample tests. Therefore, no statements can be made as to whether the methods that perform well in this theoretical comparison also perform well in practice. There are limited simulation results on the performance of the methods available in some of the respective articles.

Moreover, the discussion is restricted to datasets with the same number of variables since the aspect of comparing datasets with different dimensions is very rarely discussed in the literature. Further, in the applications we have in mind the comparison of datasets with different dimensions is also not relevant. 

For future research, we plan to incorporate the best-performing methods into a comparison of parametric and Plasmode simulation studies. Within this comparison, a critical step is to quantify how far assumptions of the simulations deviate from a true data-generating process.
It is desirable to quantify this deviation in terms of a dataset similarity or distance rather than in terms of specific parameters that are changing. 
Moreover, we plan to conduct an empirical comparison of the methods to evaluate how well the methods perform in practice and to provide a fair comparison of the method performance. 

\section*{Funding}
This work has been supported (in part) by the Research Training Group ”Biostatistical Methods for High-Dimensional Data in Toxicology” (RTG 2624, Project P1) funded by the Deutsche Forschungsgemeinschaft (DFG, German Research Foundation - Project Number 427806116).

\begin{sloppypar}
\printbibliography[title = References]
\end{sloppypar}
\end{document}